\RequirePackage{amsmath}
\let\oldvec\vec%
\RequirePackage{fix-cm}
\documentclass{svjour3}                     %
\let\vec\oldvec%
\smartqed  %

\usepackage{graphicx}
\usepackage{cite}
\usepackage{mathptmx}      %
\usepackage{amsmath}
\usepackage{mathtools}
\usepackage{amssymb} %
\usepackage{mathrsfs} %
\usepackage{fp}   %

\usepackage[subnum]{cases} %
\usepackage{cancel}

\usepackage{xstring} %

\def\u2#1{\underline{\underline{#1}}} %
\newcommand{\bs}[1]{\boldsymbol{#1}} %

\newcommand{\fpdv}[1]{\partial_{#1}}
\newcommand{\ffrac}[2]{#1/#2}

\usepackage[shortlabels]{enumitem} %
\usepackage{pifont}     %

\usepackage{parskip} %
\usepackage[normalem]{ulem} %
\usepackage{accents}
\usepackage{csquotes} %
\def\colorize<#1>{\temporal<#1>%
{\color{black!30}}%
{\color{red}}%
{\color{black}}%
}
\usepackage{varwidth,lipsum} %
\usepackage{tabto} %
\usepackage{tabstackengine} %

\usepackage{geometry}
\usepackage[percent]{overpic} %
\usepackage[rightcaption]{sidecap}
\usepackage{caption}
\usepackage{subfig} %
\usepackage{graphicx} %
\usepackage[export]{adjustbox} %
\usepackage{placeins} %
\usepackage{epstopdf} %
\usepackage{latexsym}%
\usepackage{keyval}%
\usepackage{ifthen}%
\usepackage{moreverb}%
\usepackage[subfolder]{gnuplottex}%

\usepackage{tikz}
\usetikzlibrary{decorations.markings}

\DeclareRobustCommand{\fillcircle}[1]{
\begin{tikzpicture}[baseline=-0.5ex]
	\draw[#1 ,fill= #1 ] (0,0) circle (.5ex);
\end{tikzpicture}
}
\DeclareRobustCommand{\lineplain}[1]{
\begin{tikzpicture}[baseline=-0.5ex]
	\draw[line width=0.5mm, #1] (0,0) -- (0.5,0);
\end{tikzpicture}
}
\DeclareRobustCommand{\linedot}[1]{
\begin{tikzpicture}[baseline=-0.5ex]
	\draw[line width= 0.5mm, dash pattern=on 2pt off 1pt, #1] (0,0) -- (0.5,0);
\end{tikzpicture}
}

\graphicspath{{./}}

\usepackage{color}
\usepackage{xcolor}
\usepackage{transparent} %

\definecolor{dark_blue}{RGB}{0,9,129}
\definecolor{dark_green}{RGB}{18,172,88} 
\definecolor{dark_blue_cite}{RGB}{0,84,168}
\definecolor{light_grey}{RGB}{190,190,190}

\definecolor{RedDark}{RGB}{139,0,0}
\definecolor{Red}{RGB}{255,0,0}
\definecolor{GreenDark}{RGB}{34,139,34} %
\definecolor{Green}{RGB}{50,205,50} %

\definecolor{Blueone}{RGB}{0,96,173} %
\definecolor{Redone}{RGB}{173,77,0} %

\usepackage{xifthen}%
\usepackage{xparse}
\RequirePackage{physics}
\newcommand{\vari}[1]{\widetilde{#1}}

\newcommand{\fontscal}[1]{#1} %
\newcommand{\fontscalemp}[1]{\mathsf{#1}} %
\newcommand{\fontvec}[1]{\bs{\mathrm{#1}}} %
\newcommand{\fontmat}[1]{\bs{\mathcal{#1}}} %
\NewDocumentCommand{\Time}{o o }{%
  \IfNoValueTF{#1}
    {\ensuremath{t}} %
    {%
        \IfNoValueTF{#2}
          {\IfEqCase{#1}{
	        {period}{\ensuremath{T}}%
	        {sim}{\ensuremath{t_{sim}}}%
    			{conv}{\ensuremath{t_{conv}}}%
          }[\PackageError{\time}{Undefined option to time: #1}{}]
          }%
        {\IfEqCase{#1}{
	        {sim}{\ensuremath{t_{sim}(#2)}}%
          }[\PackageError{\time}{Undefined option to time: #1 + #2}{}]
        }
      }
}

\DeclareRobustCommand{\card}[1][Default]{
    \IfEqCase{#1}{%
      {Default}{\ensuremath{N}}%
        {spe}{\ensuremath{m}}%
        {fluid}{\ensuremath{n_{f}}}%
        {fspe}{\ensuremath{m_k}}%
        {lfspe}{\ensuremath{m_l}}%
        {k}{\ensuremath{k}}%
        {k}{\ensuremath{k^{\prime}}}%
        {j}{\ensuremath{j}}%
        {cell}{\ensuremath{N_{cell}}}%
        {face}{\ensuremath{N_{face}}}%
		{fcelli}{\ensuremath{N_\mathcal{S}^{i}}}%
        {microstate}{\ensuremath{\Omega}}%
        {class}{\ensuremath{n_{c}}}%
    }[\PackageError{card}{Undefined option to cardinal: #1}{}]%
}%
\DeclareRobustCommand{\ite}[1][Default]{
    \IfEqCase{#1}{%
      {Default}{\ensuremath{N}}%
      {spe}{\ensuremath{j}}%
      {speall}{\ensuremath{(j)}}%
      {fspeall}{\ensuremath{(j,k)}}%
      {lfspeall}{\ensuremath{(j,l)}}%
      {lfall}{\ensuremath{(l)}}%
      {fall}{\ensuremath{(k)}}%
      {diff}{\ensuremath{d}}%
      {tspe}{\ensuremath{z}}%
      {fspe}{\ensuremath{j,k}}%
      {fspei}{\ensuremath{i,k}}%
      {lspe}{\ensuremath{j,l}}%
      {fnspe}{\ensuremath{\card[fspe],k}}%
      {f1spe}{\ensuremath{1,k}}%
      {f1spe}{\ensuremath{1,k}}%
      {1spe}{\ensuremath{j,1}}%
      {2spe}{\ensuremath{j,2}}%
      {nspe}{\ensuremath{\card[spe]}}%
      {fluid}{\ensuremath{k}}%
      {fluidp}{\ensuremath{k^{\prime}}}%
      {lfluidp}{\ensuremath{l^{\prime}}}%
      {lfluid}{\ensuremath{l}}%
      {mfluid}{\ensuremath{m}}%
      {fluidl}{\ensuremath{l}}%
      {nfluid}{\ensuremath{\card[fluid]}}%
      {H2O}{\ensuremath{H_{2}O}}%
      {sk}{\ensuremath{k}}%
      {s1}{\ensuremath{1}}%
      {s2}{\ensuremath{2}}%
      {sint}{\ensuremath{\mathrm{I}}}%
      {int}{\ensuremath{\mathrm{I}}}%
      {euler}{\ensuremath{\mathcal{E}}}%
      {capi}{\ensuremath{\mathcal{C}}}%
      {var}{\ensuremath{\lambda}}%
      {left}{\ensuremath{L}}%
      {right}{\ensuremath{R}}%
      {middle}{\ensuremath{M}}%
      {normal}{\ensuremath{n}}%
      {cap}{\ensuremath{c}}%
      {pm}{\ensuremath{\pm}}%
      {plus}{\ensuremath{-}}%
      {minus}{\ensuremath{+}}%
      {cell}{i}%
      {face}{j}%
    }[\ensuremath{#1}]%
}%

\NewDocumentCommand{\setvar}{o o }{%
  \IfNoValueTF{#1}
    {\ensuremath{\xi}} %
    {%
        \IfNoValueTF{#2}
          {\IfEqCase{#1}{
              {entropy}{\ensuremath{\eta}}%
              {entropyvar}{\ensuremath{\eta_{\lambda}}}%
              {natural}{\ensuremath{\zeta}}%
              {naturalk}{\ensuremath{\zeta_{k}}}%
              {pressure}{\ensuremath{\pi}}%
              {pressurek}{\ensuremath{\pi_{k}}}%
              {thermo}{\ensuremath{\xi}}%
              {thermok}{\ensuremath{\xi_{k}}}%
              {thermoext}{\ensuremath{\Xi_{k}}}%
              {pdf}{\ensuremath{\xi_{f}}}%
          }[\PackageError{\setvar}{Undefined option to setvar: #1}{}]
          }%
        {\IfEqCase{#1}{
            {natural}{\IfEqCase{#2}{%
            {var}{\ensuremath{\zeta^{\tau}}}%
            {vart}{\ensuremath{\zeta^{\tau '}}}%
            {vartt}{\ensuremath{\zeta^{\tau ''}}}%
            	{fluidiso}{\ensuremath{\overline{\zeta_{\ite[fluid]}}}}%
            	}[\ensuremath{\zeta_{\ite[#2]}}]%
            	}
            {pressure}{\IfEqCase{#2}{%
            	{fluidiso}{\ensuremath{\overline{\pi_{\ite[fluid]}}}}%
            	}[\ensuremath{\pi_{\ite[#2]}}]%
				}%
            {thermo}{\IfEqCase{#2}{%
            	{fluidiso}{\ensuremath{\overline{\xi_{\ite[fluid]}}}}%
            	}[\ensuremath{\xi_{\ite[#2]}}]%
            	}%
            {thermoext}{\ensuremath{\Xi_{\ite[#2]}}}%
          }
        }
      }
}

\NewDocumentCommand{\speed}{o o o }{%
  \IfNoValueTF{#1}
    {\ensuremath{\bs{v}}}%
    {%
        \IfNoValueTF{#2}%
          {\IfEqCase{#1}{%
        {Default}{\ensuremath{\bs{v}}}%
        {vk}{\bs{v}_{k}}%
        {sk}{v_{k}}%
        {sint}{v_{I}}%
        {surface}{\fontvec{v}_{s}}%
        {int}{\bs{v}_{I}}%
        {moy}{\overline{\bs{v}}}%
        {delta}{\delta \bs{v}}%
        {diff}{\bs{v}_{d}}%
        {mean}{\bs{v}_{h}}%
        {v}{\bs{v}}%
        {v1}{\bs{v}_{1}}%
        {v2}{\bs{v}_{2}}%
        {x}{v_{x}}%
        {y}{v_{y}}%
        {z}{v_{z}}%
        {s}{v}%
        {sx1}{v_{x_{1}}}%
        {sx2}{v_{x_{2}}}%
        {sx3}{v_{x_{3}}}%
        {sxi}{v_{x_{i}}}%
        {sxj}{v_{x_{j}}}%
        {sdiff}{v_{d}}%
        {s1}{v_{1}}%
        {s2}{v_{2}}%
        {sair}{V_{air}}%
        {sH2O}{V_{H_{2}O}}%
        {sl}{V_{L}}%
        {sg}{V_{G}}%
        {sound}{a}%
        {sound1}{a_{1}}%
        {sound2}{a_{2}}%
        {var}{\ensuremath{\bs{v}_{\lambda}}}%
        {normal}{\ensuremath{v_{\ite[normal]}}}%
          }[\ensuremath{\bs{v}_{\ite[#1]}}]%
          }%
          {\IfNoValueTF{#3}
            {\IfEqCase{#1}{%
            {wave}{\ensuremath{S_{\ite[#2]}}}%
            {normal}{\ensuremath{v_{n,\ite[#2]}}}%
            {sound}{\IfEqCase{#2}{%
                   {wallis}{\ensuremath{a_{\scriptscriptstyle{W}}}}%
                   {frozen}{\ensuremath{a_{\scriptscriptstyle{F}}}}%
                   }[\ensuremath{a_{\ite[#2]}}]}%
            {v}{\ensuremath{\bs{v}_{\ite[#2]}}}%
            {s}{\IfEqCase{#2}{%
            		{fluctu}{\ensuremath{v^{\prime}}}%
            		}[\ensuremath{v_{\ite[#2]}}]}%
            {S}{\IfEqCase{#2}{%
            		{fluctu}{\ensuremath{V^{\prime}}}%
            		}[\ensuremath{V_{\ite[#2]}}]}%
            {i}{\IfEqCase{#2}{%
            		{fluctu}{\ensuremath{v^{\prime}_{i}}}%
            		}[\PackageError{speed}{Undefined option to speed: 0  = #2}{}]}%
            {j}{\IfEqCase{#2}{%
            		{fluctu}{\ensuremath{v^{\prime}_{j}}}%
            		}[\PackageError{speed}{Undefined option to speed: 0  = #2}{}]}%
            {x}{\IfEqCase{#2}{%
            		{moy}{\ensuremath{V}}%
            		{delta}{\ensuremath{ v_{x}}}%
            		{fluctu}{\ensuremath{v^{\prime}_{x}}}%
            		}[\PackageError{speed}{Undefined option to speed: 0  = #2}{}]}%
            {y}{\IfEqCase{#2}{%
            		{delta}{\ensuremath{ v_{y}}}%
            		{fluctu}{\ensuremath{v^{\prime}_{y}}}%
            		}[\PackageError{speed}{Undefined option to speed: 0  = #2}{}]}%
            {z}{\IfEqCase{#2}{%
            		{delta}{\ensuremath{ v_{z}}}%
            		{fluctu}{\ensuremath{v^{\prime}_{z}}}%
            		}[\PackageError{speed}{Undefined option to speed: 0  = #2}{}]}%
            {fluid}{\ensuremath{\qty(\bs{v}_{\ite[#1]})_{\ite[#2]}}}%
            {int}{\ensuremath{\bs{v}_{\ite[#1],\ite[#2]}}}%
            {cell}{\IfEqCase{#2}{%
            	{face}{\ensuremath{v}_{\ite[#1] \ite[#2]}}}}%
            }[\PackageError{speed}{Undefined option to speed: #2}{}]%
            }%
            {\IfEqCase{#1}{%
            	  {x}{\IfEqCase{#2}{%
            		{moy}{\IfEqCase{#3}{%
            			{cst}{\ensuremath{V}_{0}}%
            			}[\PackageError{speed}{Undefined option to speed: 0  = #3}{}]}%
            		}[\PackageError{speed}{Undefined option to speed: 0  = #2}{}]}%
              {vol}{\ensuremath{V_{\ite[#2],\ite[#3]}}}%
              {spec}{\ensuremath{v_{\ite[#2],\ite[#3]}}}%
              {frac}{\ensuremath{\alpha_{\ite[#2],\ite[#3]}}}%
			 {sound}{\IfEqCase{#2}{%
                          {fluid}{\IfEqCase{#3}{%
                          	{K}{\ensuremath{(a_{\ite[#2]})_{\ite[#3]}}}%
                          	}[\PackageError{speed}{Undefined option to speed: 0  = #3}{}]}%
                      }[{\IfEqCase{#3}{%
                      		{wallis}{\ensuremath{a_{\scriptscriptstyle{W}}}}%
                      		}[\PackageError{speed}{Undefined option to speed: 0  = #3}{}]]}]}
              }[\PackageError{speed}{Undefined option with 3 entries to speed: #1}{}]
            }%
          }%
    }%
}
\NewDocumentCommand{\pressure}{o o }{%
  \IfNoValueTF{#1}
    {\ensuremath{\ensuremath{p}}} %
    {%
        \IfNoValueTF{#2}
          {\IfEqCase{#1}{
        	{fluid}{p_{\ite[fluid]}}%
        	{var}{\ensuremath{p_{\lambda}}}%
        	{delta}{\ensuremath{\delta p}}%
        	{moy}{\ensuremath{\overline{p}}}%
        	{smel}{\ensuremath{\Delta p}}%
        	{extend}{\ensuremath{\mathcal{P}}}%
        	{contact}{\ensuremath{p_{c}}}%
        	{infty}{\ensuremath{p^{\infty}}}%
        	{error}{\ensuremath{\xi_{p}}}%
          }[\ensuremath{p_{\ite[#1]}}]%
          }%
          {\IfEqCase{#1}{%
        		{fluid}{\IfEqCase{#2}{%
        			{PG}{\ensuremath{p^{PG}_{\ite[fluid]}}}%
        			{iso}{\ensuremath{p^{o}_{\ite[fluid]}}}%
        			{inf}{\ensuremath{p^{\infty}_{\ite[fluid]}}}%
        			}[\ensuremath{\qty(p_{\ite[fluid]})_{\ite[#2]}}]%
        			}
      			{cell}{\ensuremath{(p_{\ite[#2]})_{\ite[#1]}}}%
        		{int}{\ensuremath{p_{\ite[#1],\ite[#2]}}}%
        		{PG}{\ensuremath{p^{PG}_{\ite[#2]}}}%
       			{inf}{\ensuremath{p^{\infty}_{\ite[#2]}}}%
        		{iso}{\ensuremath{p^{o}_{\ite[#2]}}}%
          	{var}{\ensuremath{p_{\ite[#2], \ite[var]}}}%
       	 	{delta}{\ensuremath{\delta p_{\ite[#2]}}}%
          	{extend}{\IfEqCase{#2}{%
          		{delta}{\delta \mathcal{P}}%
          		}[\ensuremath{\mathcal{P}_{\ite[#2]}}]%
          	}%
          	{smel}{\ensuremath{\Delta p_{\ite[#2]}}}%
          	}[\PackageError{pressure}{Undefined option with 2 entries to pressure: #1}{}]
          }
    }
}
\NewDocumentCommand{\density}{o o }{%
  \IfNoValueTF{#1}
    {\ensuremath{\ensuremath{\rho}}} %
    {%
        \IfNoValueTF{#2}
          {\IfEqCase{#1}{
        	{fluid}{\rho_{\ite[fluid]}}%
        	{var}{\ensuremath{\rho_{\lambda}}}%
        	{delta}{\ensuremath{\delta \rho}}%
        	{moy}{\ensuremath{\overline{\rho}}}%
          }[\ensuremath{\rho_{\ite[#1]}}]
          }%
          {\IfEqCase{#1}{
          {fluid}
            {\IfEqCase{#2}{
            	{moy}{\ensuremath{\overline{\rho}_{\ite[#1]}}}%
            	}[\ensuremath{\qty(\rho_{\ite[fluid]})_{\ite[#2]}}]
            	}%
            {var}{\ensuremath{\rho_{\ite[#2], \ite[var]}}}%
            }[\PackageError{density}{Undefined option with 2 entries to density: #1}{}]
          }
    }
}
\DeclareRobustCommand{\volume}[1][Default]{
    \IfEqCase{#1}{%
    {Default}{\ensuremath{V}}%
    {sk}{\ensuremath{V_{k}}}%
    {s1}{\ensuremath{V_{1}}}%
    {s2}{\ensuremath{V_{2}}}%
    {sp}{\ensuremath{V^{\prime}}}%
    {fluid}{\ensuremath{V_{\ite[fluid]}}}%
    {cell}{\ensuremath{\mathcal{V}_{\ite[cell]}}}%
    }[\ensuremath{V_{\ite[#1]}}]%
}%

\DeclareRobustCommand{\volspec}[1][Default]{
    \IfEqCase{#1}{%
    {Default}{\ensuremath{v}}%
    {sk}{\ensuremath{v_{k}}}%
    {fluid}{\ensuremath{v_{\ite[fluid]}}}%
    {s1}{\ensuremath{v_{1}}}%
    {s2}{\ensuremath{v_{2}}}%
    {sp}{\ensuremath{v^{\prime}}}%
    }[\PackageError{volspec}{Undefined option to vol spec: #1}{}]%
}%

\NewDocumentCommand{\vol}{o o o }{%
  \IfNoValueTF{#1}
    {\ensuremath{V}} %
    {%
        \IfNoValueTF{#2}
          {\IfEqCase{#1}{
            {vol}{\ensuremath{V}}   %
            {fluid}{\ensuremath{V_{\ite[fluid]}}}   %
            {fluidall}{\ensuremath{V_{(\ite[fluid])}}}    %
              {spec}{\ensuremath{v}}    %
              {specp}{\ensuremath{v^{\prime}}}    %
              {specpp}{\ensuremath{v^{\prime \prime}}}    %
              {frac}{\ensuremath{\alpha}} %
              {mel}{\ensuremath{\Delta V}}%
          }[\PackageError{vol}{Undefined option to volume: #1}{}]
          }%
          {\IfNoValueTF{#3}
            {\IfEqCase{#1}{
              {vol}{\ensuremath{V_{\ite[#2]}}}
                {frac}{\IfEqCase{#2}{
                	{fluctu}{\ensuremath{\tilde{\alpha}}}%
                	}[\ensuremath{\alpha_{\ite[#2]}}]}%
                {fracall}{\ensuremath{\alpha_{(\ite[#2])}}}%
                {fract}{\ensuremath{\tilde{\alpha}_{\ite[#2]}}}%
                {spec}{\ensuremath{v_{\ite[#2]}}}%
            }[\PackageError{vol}{Undefined option to volume: #2}{}]
            }%
            {\IfEqCase{#1}{
              {vol}{\ensuremath{V_{\ite[#2],\ite[#3]}}}   %
              {spec}{\ensuremath{v_{\ite[#2],\ite[#3]}}}
              {frac}{\IfEqCase{#2}{%
	               	{fluid}{\ensuremath{(\alpha_{\ite[#2]})_{\ite[#3]}}}%
	               	{cell}{\ensuremath{(\alpha_{\ite[#2]})_{\ite[#3]}}}%
               		}[{\IfEqCase{#3}{%
				    {fluctu}{\ensuremath{\tilde{\alpha_{\ite[#2]}}}}%
               		{cell}{(\alpha_{\ite[#2]})_{\ite[#3]}}}}]%
                	}
                	}[\PackageError{vol}{Undefined option with 3 entries to volume: #1}{}]
            }
    }
}
}

\NewDocumentCommand{\mass}{o o o }{%
  \IfNoValueTF{#1}
    {\ensuremath{M}} %
    {%
        \IfNoValueTF{#2}
          {\IfEqCase{#1}{
            {vol}{\ensuremath{M}}
            {fluid}{\ensuremath{M_{\ite[fluid]}}}
            {fluidall}{\ensuremath{M_{(\ite[fluid])}}}
            {fspeall}{\ensuremath{M_{(\ite[spe],\ite[fluid])}}}
            {frac}{\ensuremath{Y}}%
            {sk}{\ensuremath{M_{k}}}%
            {transfer}{\ensuremath{\Gamma}}%
          }[\PackageError{mass}{Undefined option to mass: #1}{}]
          }%
          {\IfNoValueTF{#3}
            {\IfEqCase{#1}{
              {vol}{\ensuremath{M_{\ite[#2]}}}
              {frac}{\IfEqCase{#2}{%
              	{moy}{\ensuremath{\overline{Y}}}%
              	{delta}{\ensuremath{\delta Y}}%
              	}[\ensuremath{\ensuremath{y_{\ite[#2]}}}]}%
              {fracb}{\ensuremath{\overline{y}_{\ite[#2]}}}%
              {fract}{\ensuremath{\tilde{y}_{\ite[#2]}}}%
              {fracall}{\ensuremath{y_{(\ite[#2])}}}%
            {transfer}{\ensuremath{\Gamma_{\ite[#2]}}}%
            }[\PackageError{mass}{Undefined option to mass: #2}{}]
            }%
            {\IfEqCase{#1}{
              {vol}{\ensuremath{M_{\ite[#2],\ite[#3]}}}   %
                {frac}{\IfEqCase{#2}{
                	{fluid}{\ensuremath{(y_{\ite[#2]})_{\ite[#3]}}}%
                	}[\ensuremath{y_{\ite[#2],\ite[#3]}}]%
                	}%
                {fracb}{\ensuremath{\overline{y}_{\ite[#2],\ite[#3]}}}%
                }[\PackageError{mass}{Undefined option with 3 entries to mass: #1}{}]
            }
          }
    }
}

\DeclareRobustCommand{\massfrac}[1][Default]{
    \IfEqCase{#1}{%
        {Default}{\ensuremath{Y}}%
        {sk}{\ensuremath{y_{k}}}%
        {s}{\ensuremath{y}}%
        {s1}{\ensuremath{y_{1}}}%
        {s2}{\ensuremath{y_{2}}}%
        {sN}{\ensuremath{y_{N}}}%
        {sNj}{\ensuremath{y_{N_{j}}}}%
        {var}{\ensuremath{\massfrac_{\lambda}}}
        {skp}{\ensuremath{y_{k^{\prime}}}}%
    }[\PackageError{massfrac}{Undefined option to massfrac: #1}{}]%
}%

\NewDocumentCommand{\tem}{o o }{%
  \IfNoValueTF{#1}
    {\ensuremath{T}} %
    {\IfNoValueTF{#2}
  		{\IfEqCase{#1}{
		    	{fluid}{\ensuremath{T_{\ite[fluid]}}}%
		    	{int}{\ensuremath{T_{\ite[int]}}}%
	    }[\ensuremath{T_{\ite[#1]}}]%
        }%
        {\IfEqCase{#1}{
          {fluid}{\IfEqCase{#2}{
          	{fluidp}{\ensuremath{T_{\ite[#1],\ite[#2]}}}%
          	{lfluid}{\ensuremath{T_{\ite[#1],\ite[#2]}}}%
          		}[\ensuremath{T_{\ite[#1],\ite[#2]}}]%
          	}%
          }
        }
      }
}

\NewDocumentCommand{\gibbs}{o o }{%
  \IfNoValueTF{#1}
    {\ensuremath{g}} %
    {\IfNoValueTF{#2}
  		{\IfEqCase{#1}{
      		{smel}{\ensuremath{\Delta g}}%
      		{mel}{\ensuremath{\Delta g}}%
      	}[\ensuremath{g_{\ite[#1]}}]%
        }%
        {\IfEqCase{#1}{
          {fspe}{\IfEqCase{#2}{%
          	{iso}{\ensuremath{g^{o}_{\ite[fspe]}}}%
          		}[\PackageError{gibbs}{Undefined option to gibbs fspe iso: #2}{}]%
          	}%
          {fluid}{\IfEqCase{#2}{
          	{PG}{\ensuremath{g^{PG}_{\ite[fluid]}}}%
          	{iso}{\ensuremath{g^{o}_{\ite[fluid]}}}%
          		}[\ensuremath{g_{\ite[#2]}}]%
          	}%
          {mel}{\ensuremath{\Delta g_{\ite[#2]}}}%
          {smel}{\ensuremath{\Delta g_{\ite[#2]}}}%
          {mix}{\ensuremath{G_{\ite[#2]}}}%
          }
        }
      }
}

\NewDocumentCommand{\entropy}{o o }{%
  \IfNoValueTF{#1}
    {\ensuremath{s}} %
    {%
        \IfNoValueTF{#2}
          {\IfEqCase{#1}{
            {flux}{\ensuremath{\fontvec{G}}}%
            {vol}{\ensuremath{S}}%
              {fluid}{\ensuremath{s_{\ite[fluid]}}}%
              {sk}{\ensuremath{s_{k}}}%
              {svolk}{\ensuremath{S_{k}}}%
              {svol}{\ensuremath{S}}%
              {s}{\ensuremath{s}}%
              {s1}{\ensuremath{s_{1}}}%
              {s2}{\ensuremath{s_{2}}}%
              {smel}{\ensuremath{\Delta s}}%
              {PG}{\ensuremath{s^{PG}}}%
          }[\ensuremath{s_{\ite[#1]}}]%
          }%
        {\IfEqCase{#1}{
          {fluid}{\IfEqCase{#2}{
          	{PG}{\ensuremath{s_{\ite[fluid]}^{PG}}}%
          	{iso}{\ensuremath{s_{\ite[fluid]}^{o}}}%
          	}[\ensuremath{\qty(s_{\ite[fluid]})_{\ite[#2]}}]%
          	}%
          {vol}{\ensuremath{S_{\ite[#2]}}}%
          {frac}{\ensuremath{s_{\ite[#2]}}}%
          {var}{\ensuremath{s_{\ite[#2], \ite[var]}}}%
          {smel}{\ensuremath{\Delta s_{\ite[#2]}}}%
          }
        }
      }
}
\NewDocumentCommand{\enthalpy}{o o }{%
  \IfNoValueTF{#1}
    {\ensuremath{s}} %
    {%
        \IfNoValueTF{#2}
          {\IfEqCase{#1}{
            {vol}{\ensuremath{H}}%
              {fluid}{\ensuremath{h_{\ite[fluid]}}}%
              {sk}{\ensuremath{h_{k}}}%
              {svolk}{\ensuremath{H_{k}}}%
              {svol}{\ensuremath{H}}%
              {s}{\ensuremath{h}}%
              {s1}{\ensuremath{h_{1}}}%
              {s2}{\ensuremath{h_{2}}}%
              {smel}{\ensuremath{\Delta h}}%
              {tot}{\ensuremath{H}}%
          }[\ensuremath{h_{\ite[#1]}}]
          }%
        {\IfEqCase{#1}{
          {fluid}{\ensuremath{\qty(h_{\ite[fluid]})_{\ite[#2]}}}%
          {vol}{\ensuremath{H_{\ite[#2]}}}%
          {frac}{\ensuremath{h_{\ite[#2]}}}%
          {var}{\ensuremath{h_{\ite[#2], \ite[var]}}}%
          }
        }
      }
}

\NewDocumentCommand{\fpot}{o o o }{%
  \IfNoValueTF{#1}
    {\ensuremath{f}} %
    {%
        \IfNoValueTF{#2}
          {\IfEqCase{#1}{%
            {volfrac}{\ensuremath{f_{\alpha}}}%
            {interf}{\ensuremath{F^{\Sigma}}}%
          	}[\PackageError{fpot}{Undefined option to function potential: #1}{}]
          }%
          {\IfNoValueTF{#3}
            {\IfEqCase{#1}{%
         	 	{volfrac}{\ensuremath{f_{\alpha_{\ite[#2]}}}}%
            		{interf}{\ensuremath{f^{\Sigma}_{\ite[#2]}}}%
          		}[\PackageError{fpot}{Undefined option to fpot: #2}{}]%
            }%
            {\IfEqCase{#1}{%
          		{volfrac}{\IfEqCase{#2}{%
            			{mix}{\ensuremath{F_{\alpha_{\ite[#3]}}}}%
            			{mel}{\ensuremath{\Delta f_{\alpha_{\ite[#3]}}}}%
          			}[\PackageError{fpot}{Undefined option to fpot: #1, #2, #3 with volfrac}{}]
          			}
            		{interf}{\ensuremath{F^{\Sigma}_{\ite[#2],\ite[#3]}}}%
              }[\PackageError{fpot}{Undefined option with 3 entries to fpot: #1}{}]
            }
          }
    }
}
\NewDocumentCommand{\energy}{o o }{%
  \IfNoValueTF{#1}
    {\ensuremath{e}} %
    {%
        \IfNoValueTF{#2}
          {\IfEqCase{#1}{
            {vol}{\ensuremath{E}}
            {svol}{\ensuremath{E}}
            {frac}{\ensuremath{e}}%
            {sk}{\ensuremath{e_{k}}}%
            {fluid}{\ensuremath{e_{\ite[fluid]}}}%
            {fluidp}{\ensuremath{e_{\ite[fluidp]}}}%
            {s1}{\ensuremath{e_{1}}}%
            {s2}{\ensuremath{e_{2}}}%
              {smel}{\ensuremath{\Delta e}}%
              {PG}{\ensuremath{e^{PG}}}%
              {tot}{\ensuremath{E}}%
              {diff}{\ensuremath{E_{d}}}%
              {kinetic}{\ensuremath{\mathscr{K}}}%
              {internal}{\ensuremath{\epsilon}}%
              {kineticsmall}{\ensuremath{\mathscr{K}_i}}%
              {interfacial}{\ensuremath{\mathscr{U}_i}}%
              {potential}{\ensuremath{\mathscr{U}}}%
              {capl}{\ensuremath{\mathscr{E}_{c}^{l}}}%
              {caps}{\ensuremath{\mathscr{E}_{c}^{s}}}%
              {free}{\ensuremath{f}}%
          }[\ensuremath{e_{\ite[#1]}}]%
          }%
        {\IfEqCase{#1}{
          {fluid}{\IfEqCase{#2}{%
          	{PG}{\ensuremath{e_{\ite[fluid]}^{PG}}}%
          	{iso}{\ensuremath{e_{\ite[fluid]}^{o}}}%
          	}[\ensuremath{\qty(e_{\ite[fluid]})_{\ite[#2]}}]%
          	}%
          {free}{\IfEqCase{#2}{%
          	{smel}{\ensuremath{\Delta f}}%
          	}[\ensuremath{f_{\ite[#2]}}]%
          	}%
          {vol}{\IfEqCase{#2}{%
          	{kinetic}{\ensuremath{E_{k}}}%
          	{potential}{\ensuremath{E}_{p}}%
          	}[\ensuremath{E_{\ite[#2]}}]%
          	}%
          {frac}{\ensuremath{e_{\ite[#2]}}}%
          {tot}{\ensuremath{E_{\ite[#2]}}}%
          {internal}{\ensuremath{\epsilon_{\ite[#2]}}}%
          {free}{\ensuremath{f}_{\ite[#2]}}%
          {smel}{\ensuremath{\Delta e_{\ite[#2]}}}%
          }
        }
      }
}
\NewDocumentCommand{\force}{o o }{%
  \IfNoValueTF{#1}
    {\ensuremath{\bs{f}}} %
    {%
        \IfNoValueTF{#2}
          {\IfEqCase{#1}{
            {gravity}{\ensuremath{\bs{g}}}%
            {body}{\ensuremath{\bs{f}^{b}}}%
            {surface}{\ensuremath{\bs{f}^{s}}}%
            {drag}{\ensuremath{\bs{F}_{d}}}%
          }[\PackageError{force}{Undefined option to force: #1}{}]
          }%
        {\IfEqCase{#1}{
          {gravity}{\ensuremath{\bs{g}_{\ite[#2]}}}%
          {body}{\ensuremath{\bs{f}^{b}_{\ite[#2]}}}%
          {drag}{\ensuremath{\bs{F}_{d,\ite[#2]}}}%
          }
        }
      }
}
\NewDocumentCommand{\heat}{o o }{%
  \IfNoValueTF{#1}
    {\ensuremath{\bs{q}}}%
    {%
        \IfNoValueTF{#2}
          {\IfEqCase{#1}{
            {flux}{\ensuremath{\bs{q}}}%
            {source}{\ensuremath{\bs{\dot{q}}}}%
          }[\PackageError{heat}{Undefined option to heat: #1}{}]
          }%
        {\IfEqCase{#1}{
          {flux}{\ensuremath{\bs{q}_{\ite[#2]}}}
          {source}{\ensuremath{\bs{\dot{q}}_{\ite[#2]}}}
         }
        }
      }
}
\NewDocumentCommand{\coeff}{o o o }{%
  \IfNoValueTF{#1}
    {\ensuremath{\lambda}}%
    {%
        \IfNoValueTF{#2}
          {\IfEqCase{#1}{
            {relax}{\ensuremath{\lambda}}%
            {surften}{\ensuremath{\sigma}}%
			{isentrop}{\ensuremath{\gamma}}%
			{diffusion}{\ensuremath{D}}%
          }[\PackageError{coeff}{Undefined option to coeff: #1}{}]
          }%
          {\IfNoValueTF{#3}
            {\IfEqCase{#1}{%
          		{relax}{\IfEqCase{#2}{%
      				{puls}{\ensuremath{\epsilon}}%
      				{speed}{\ensuremath{\lambda}}%
      				{pressure}{\ensuremath{\mu}}%
          			}[\PackageError{coeff}{Undefined option to coeff - relax: #2}{}]}         		
       		   	{isentrop}{\ensuremath{\gamma_{\ite[#2]}}}%
		        }[\PackageError{coeff}{Undefined option to coeff #1 with a second #2}{}]
            }%
            {\IfEqCase{#1}{%
          		{relax}{\IfEqCase{#2}{
            		{speed}{\IfEqCase{#3}{%
            			{inv}{\ensuremath{\epsilon^{\speed[s]}}}%
            			}[\PackageError{\coeff}{Undefined option to coeff: relax, speed, #3 with coeff}{}]}%
            		{pressure}{\IfEqCase{#3}{%
            			{inv}{\ensuremath{\epsilon^{\pressure}}}%
            			}[\PackageError{\coeff}{Undefined option to coeff: relax, pressure, #3 with coeff}{}]}%
          			}[\PackageError{\coeff}{Undefined option to coeff: #1, #2, #3 with coeff}{}]
          			}
              }[\PackageError{\coeff}{Undefined option with 3 entries to coeff: #1}{}]
            }
          }
    }
}

\NewDocumentCommand{\gas}{o o }{%
  \IfNoValueTF{#1}
    {\ensuremath{r}} %
	{%
        \IfNoValueTF{#2}
          {\IfEqCase{#1}{
		      {rspec}{\ensuremath{\text{r}}}%
      		  {cv}{\ensuremath{\text{c}_{v}}}%
      		  {gamma}{\ensuremath{\gamma}}%
          }[\PackageError{\gas}{Undefined option to gas: #1}{}]%
          }%
        {\IfEqCase{#1}{
	      {rspec}{\ensuremath{\text{r}_{\ite[#2]}}}%
   		  {cv}{\ensuremath{\text{c}_{v,\ite[#2]}}}%
   		  {gamma}{\ensuremath{\gamma_{\ite[#2]}}}%
         }[\PackageError{\gas}{Undefined option to gas: #2}{}]%
        }
      }
}

\NewDocumentCommand{\source}{o o o }{%
  \IfNoValueTF{#1}
    {\ensuremath{\mathcal{T}}}%
    {%
        \IfNoValueTF{#2}
          {\IfEqCase{#1}{
            {stress}{\ensuremath{\fontvec{T}}}%
            {viscous}{\ensuremath{\fontvec{T}}}%
            {diffusion}{\ensuremath{\fontvec{D}}}%
            {relax}{\ensuremath{\fontvec{r}}}%
            {interfarea}{\ensuremath{\fontvec{S}_{\Sigma}}}%
            {meancurv}{\ensuremath{\fontvec{S}_{\meancurv}}}%
            {gausscurv}{\ensuremath{\fontvec{S}_{\gausscurv}}}%
           }[\PackageError{source}{Undefined option to source: #1}{}]%
          }%
          {\IfNoValueTF{#3}
            {\IfEqCase{#1}{%
          		{stress}{\ensuremath{\fontvec{T}_{\ite[#2]}}}%
          		{viscous}{\ensuremath{\fontvec{D}_{\ite[#2]}}}%
          		{relax}{\IfEqCase{#2}{%
            			{speed}{\ensuremath{\fontvec{r}^{\speed[s]}}}%
            			{pressure}{\ensuremath{\fontvec{r}^{\pressure}}}%
          			}[\PackageError{source}{Undefined option to source: #1, #2 with relax}{}]%
          			}%
          		}[\PackageError{source}{Undefined option to source: #2}{}]%
            }%
            {\IfEqCase{#1}{%
          		{relax}{\IfEqCase{#2}{
            			{speed}{\ensuremath{\fontvec{r}^{\speed[s]}_{\ite[#3]}}}%
            			{pressure}{\ensuremath{\fontvec{r}^{\pressure}_{\ite[#3]}}}%
          			}[\PackageError{source}{Undefined option to source: #1, #2, #3 with relax}{}]
          			}
              }[\PackageError{speed}{Undefined option with 3 entries to speed: #1}{}]
            }
          }
    }
}
\NewDocumentCommand{\exchange}{o o }{%
  \IfNoValueTF{#1}
    {\ensuremath{m}}%
    {%
        \IfNoValueTF{#2}
          {\IfEqCase{#1}{
            {mass}{\ensuremath{c^{+}}}%
            {momentum}{\ensuremath{\fontvec{m^{+}}}}%
            {energy}{\ensuremath{e^{+}}}%
          }[\PackageError{source}{Undefined option to source: #1}{}]
          }%
        {\IfEqCase{#1}{
            {mass}{\ensuremath{c^{+}_{\ite[#2]}}}%
            {momentum}{\ensuremath{\fontvec{m^{+}}_{\ite[#2]}}}%
            {energy}{\ensuremath{e^{+}_{\ite[#2]}}}%
         }
        }
      }
}

\DeclareRobustCommand{\rhovar}[1][Default]{
    \IfEqCase{#1}{%
        {Default}{\vari{\rho}}%
    }[\PackageError{rhovar}{Undefined option to rhovar: #1}{}]%
}%

\NewDocumentCommand{\pulsation}{o }{%
  \IfNoValueTF{#1}
    {\ensuremath{\omega}} %
  {\IfEqCase{#1}{
      {suppl}{\ensuremath{\Delta \omega}}%
      {delta}{\ensuremath{\delta \omega}}%
	  {interfarea}{\ensuremath{\omega_{\scriptscriptstyle{\Sigma}}}}%
	  {volfrac}{\ensuremath{\omega_{\scriptscriptstyle{\alpha}}}}%
	  }
      [\ensuremath{\omega_{\ite[#1]}}]%
    }
}

\NewDocumentCommand{\micro}{o o o }{%
  \IfNoValueTF{#1}
    {\ensuremath{\mu}} %
    {%
        \IfNoValueTF{#2}
          {\IfEqCase{#1}{
             {inertia}{\ensuremath{\nu}}%
             {viscosity}{\ensuremath{\fontscalemp{\epsilon}}}%
          }[\ensuremath{\mu_{\ite[#1]}}]%
          }%
        	  {\IfNoValueTF{#3}
            {\IfEqCase{#1}{
             {inertia}{\IfEqCase{#2}{
             		{volfrac}{\ensuremath{\nu_{\scriptscriptstyle{\alpha}}}}%
             		{interf}{\ensuremath{\nu_{\scriptscriptstyle{\Sigma}}}}%
             		}[\ensuremath{\nu_{\ite[#2]}}]%
          		  }%
             {viscosity}{\IfEqCase{#2}{
             		{volfrac}{\ensuremath{\fontscalemp{\epsilon_{\scriptscriptstyle{\alpha}}}}}%
             		{interf}{\ensuremath{\fontscalemp{\epsilon_{\scriptscriptstyle{\Sigma}}}}}%
             		}[\ensuremath{\fontscalemp{\epsilon_{\ite[#2]}}}]%
          		  }%
            }[\PackageError{micro}{Undefined option micro to #1 + #2}{}]
            }%
            {\IfEqCase{#1}{
             	{inertia}{to be defined}%
          		}[\PackageError{micro}{Undefined option micro to #1=lag + #2 + #3}{}]%
            }
          }
    }
}

\NewDocumentCommand{\Lagr}{o o }{%
  \IfNoValueTF{#1}
    {\ensuremath{L}} %
    {%
        \IfNoValueTF{#2}
          {\IfEqCase{#1}{
      		{filter}{\ensuremath{L_{c}}}%
      		{star}{\ensuremath{L^{\star}}}%
      		{var}{\accentset{\circ}{\ensuremath{L}}}%
      		{fluid}{\ensuremath{L_{\ite[fluid]}}}%
          }[\ensuremath{L_{\ite[#1]}}]%
          }%
        {\IfEqCase{#1}{
             {star}{\ensuremath{L^{\star}_{\ite[#2]}}}%
        		}[\PackageError{Lagr}{Undefined option to Lagrangian #1 + #2}{}]%
      	}
     }
}
\DeclareRobustCommand{\Lagrdiff}[1][Default]{
    \IfEqCase{#1}{%
      {Default}{\ensuremath{\dd L}}%
      {u}{\ensuremath{\bs{K}}}%
      {gradalpha}{\ensuremath{\bs{D}}}%
      {dtalpha}{\ensuremath{M}}%
      {dtsigma}{\ensuremath{N}}%
    }[\PackageError{Lagrdiff}{Undefined option to Lagrdiff: #1}{}]%
}%
\NewDocumentCommand{\LAPvar}{o }{%
  \IfNoValueTF{#1}%
    {\ensuremath{\lambda}}%
  {\IfEqCase{#1}{%
        {0}{\ensuremath{\lambda}}%
        {1}{\ensuremath{\lambda_{1}}}%
        {2}{\ensuremath{\lambda_{2}}}%
        {kp}{\lambda_{k^{\prime}}}%
        {fluidp}{\lambda_{\ite[fluidp]}}%
		{t}{\ensuremath{\lambda}}%
		{vect}{\bs{\chi}} %
		{fluid}{\lambda_{\ite[fluid]}} %
      }[\ensuremath{\lambda_{\ite[#1]}}]%
    }%
}%

\NewDocumentCommand{\pathM}{o o }{%
  \IfNoValueTF{#1}
    {\ensuremath{\gamma}} %
    {%
        \IfNoValueTF{#2}
          {\IfEqCase{#1}{
             {var}{\ensuremath{\Gamma}}%
        		{0}{\ensuremath{\gamma_{\lambda}}}%
        		{1}{\ensuremath{\gamma_{\lambda_{1}}}}%
        		{2}{\ensuremath{\gamma_{\lambda_{2}}}}%
        		{kp}{\gamma_{\lambda_{k^{\prime}}}}%
        		{fluidp}{\gamma_{\lambda_{\ite[fluidp]}}}%
			{t}{\gamma_{\ensuremath{\lambda}}}%
			{vect}{\gamma_{\bs{\chi}}} %
			{fluid}{\gamma_{\lambda_{\ite[fluid]}}} %
          }[\ensuremath{\gamma_{\ite[#1]}}]%
          }%
        {\IfEqCase{#1}{
             {var}{\ensuremath{\Gamma_{\ite[#2]}}}%
        		}[\PackageError{pathM}{Undefined option pathM to #1 + #2}{}]%
      	}
     }
}

\NewDocumentCommand{\xlag}{o }{%
  \IfNoValueTF{#1}
    {\ensuremath{\bs{X}}} %
  {\IfEqCase{#1}{%
      {tobedefined}{\ensuremath{\bs{X}}}%
      }%
      [\ensuremath{\bs{X}_{\ite[#1]}}]%
    }%
}%
\NewDocumentCommand{\xeul}{o }{%
  \IfNoValueTF{#1}
    {\ensuremath{\bs{x}}} %
  {\IfEqCase{#1}{%
      {tobedefined}{\ensuremath{\bs{x}}}%
      }%
      [\ensuremath{\bs{x}_{\ite[#1]}}]%
    }%
}%

\NewDocumentCommand{\pathx}{o o o }{%
  \IfNoValueTF{#1}
    {\ensuremath{\varphi}} %
    {%
        \IfNoValueTF{#2}
          {\IfEqCase{#1}{
             {eul}{\ensuremath{\vb*{\varphi}}}%
             {lag}{\ensuremath{\vb*{\varphi}^L}}%
             {var}{\ensuremath{\hat{\varphi}}}%
          }[\ensuremath{\varphi_{\ite[#1]}}]%
          }%
        	  {\IfNoValueTF{#3}
            {\IfEqCase{#1}{
             {var}{\ensuremath{\hat{\varphi}_{\ite[#2]}}}%
             {lag}{\IfEqCase{#2}{
             		{var}{\ensuremath{\vb*{\varphi}^L_{\LAPvar}}}%
             		}[\ensuremath{\vb*{\varphi}^L_{#2}}]%
          		  }%
             {eul}{\IfEqCase{#2}{
             		{var}{\ensuremath{\vb*{\varphi}_{\LAPvar}}}%
             		}[\PackageError{pathx}{Undefined option to #1=eul + #2}{}]%
          		  }%
            }[\PackageError{pathx}{Undefined option to #1 + #2}{}]
            }%
            {\IfEqCase{#1}{
             	{lag}{\IfEqCase{#2}{
             		{var}{\ensuremath{\vb*{\varphi}^L_{\LAPvar[#3]}}}%
             		}[\PackageError{pathx}{Undefined option to #1=lag + #2 + #3}{}]%
          		  }%
                }[\PackageError{pathx}{Undefined option with 3 entries to pathx: #1}{}]
            }
          }
    }
}

\NewDocumentCommand{\SV}{o }{%
  \IfNoValueTF{#1}
    {\ensuremath{\fontvec{u}}} %
  {\IfEqCase{#1}{
      {face}{\ensuremath{\fontvec{u}_{\ite[face]}}}%
      {int}{\ensuremath{\fontvec{u}_{\ite[int]}}}%
      {t}{\ensuremath{\tilde{\fontvec{u}}}}%
      }
      [\ensuremath{\fontvec{u}_{\ite[#1]}}]%
    }
}

\NewDocumentCommand{\CV}{o o }{%
  \IfNoValueTF{#1}
    {\ensuremath{\fontvec{q}}} %
    {%
        \IfNoValueTF{#2}
          {\IfEqCase{#1}{
             {scal}{\ensuremath{q}}%
          }[\ensuremath{\fontvec{q}_{\ite[#1]}}]%
          }%
        {\IfEqCase{#1}{
            {scal}{\ensuremath{q_{\ite[#2]}}}%
          }
        }
      }
}

\NewDocumentCommand{\CF}{o o }{%
  \IfNoValueTF{#1}
    {\ensuremath{\fontvec{f}}} %
    {%
        \IfNoValueTF{#2}
          {\IfEqCase{#1}{
      		{face}{\ensuremath{\fontvec{f}_{\ite[face]}}}%
      		{int}{\ensuremath{\fontvec{f}_{\ite[int]}}}%
      		{euler}{\ensuremath{\fontvec{f}^{e}}}%
      		{capi}{\ensuremath{\fontvec{f}^{c}}}%
          }[\ensuremath{\fontvec{f}_{\ite[#1]}}]%
          }%
        {\IfEqCase{#1}{
      		{euler}{\IfEqCase{#2}{
      			{num}{\ensuremath{\fontvec{\phi}^{e}}}%
      			}[\ensuremath{\fontvec{f}^{e}_{\ite[#2]}}]}%
      		{capi}{\IfEqCase{#2}{
      			{num}{\ensuremath{\fontvec{\phi}^{c}}}%
      			}[\ensuremath{\fontvec{f}^{c}_{\ite[#2]}}]}%
      		}[\PackageError{\CF}{Undefined option to CF: #2}{}]%
        }
      }
}
\NewDocumentCommand{\NCF}{o o }{%
  \IfNoValueTF{#1}
    {\ensuremath{\fontmat{N}}} %
    {%
        \IfNoValueTF{#2}
          {\IfEqCase{#1}{
      		{v}{\ensuremath{\fontvec{n}}}%
          }[\ensuremath{\fontmat{N}_{\ite[#1]}}]%
          }%
        {\IfEqCase{#1}{
            {v}{\ensuremath{\fontvec{n}_{\ite[#2]}}}%
          }
        }
      }
}

\newcommand{\transvector}{\fontvec{t}}
\DeclareRobustCommand{\transfer}[1]{%
    \IfEqCase{#1}{%
        {salpha}{\fontscal{t_{\alpha}}} %
        {s1}{\fontscal{t_{1}}} %
        {s2}{\fontscal{t_{2}}} %
        {sk}{\fontscal{t_{k}}} %
        {v}{\fontvec{t}} %
        {m}{\fontmat{T}}%
    }[\PackageError{transfer}{Undefined option to transfer: #1}{}]%
}%

\NewDocumentCommand{\jacobian}{o }{%
  \IfNoValueTF{#1}
    {\ensuremath{\ensuremath{\fontmat{A}}}}%
  {\IfEqCase{#1}{
      {add}{\ensuremath{\fontmat{A}_{add}}}%
      }
      [\ensuremath{\fontmat{A}_{\ite[#1]}}]%
    }
}

\newcommand{\x}[1][Default]{
    \IfEqCase{#1}{%
        {Default}{\ensuremath{\bs{x}}} %
    }[\PackageError{x}{Undefined option to x: #1}{}]%
}%

\NewDocumentCommand{\radius}{o }{%
  \IfNoValueTF{#1}
    {\ensuremath{\ensuremath{R}}}%
  {\IfEqCase{#1}{
      {add}{\ensuremath{r_{add}}}%
      }
      [\ensuremath{R_{\ite[#1]}}]%
    }
}
\NewDocumentCommand{\diam}{o }{%
  \IfNoValueTF{#1}
    {\ensuremath{\ensuremath{d}}}%
  {\IfEqCase{#1}{
      {zl}{\ensuremath{Z_{l}}}%
      {zg}{\ensuremath{Z_{g}}}%
      {inj}{\ensuremath{d_{inj}}}%
      }
      [\ensuremath{d_{\ite[#1]}}]%
    }
}

\newcommand{\charlen}[1][Default]{
    \IfEqCase{#1}{%
        {Default}{\ensuremath{\ell}} %
        {c}{\ensuremath{\ell}_{c}} %
        {filter}{\ensuremath{\ell}_{c}} %
        {DNS}{\ensuremath{\ell_{DNS}}} %
        {mfp}{\ensuremath{\ell_{fr}}} %
        {fp}{\ensuremath{\ell_{p}}} %
        {macro}{\ensuremath{L}} %
        {KH}{\ensuremath{d}} %
    }[\PackageError{charlen}{Undefined option to charlen: #1}{}]%
}%

\NewDocumentCommand{\wave}{o o }{%
  \IfNoValueTF{#1}
    {\ensuremath{wave}} %
    {%
        \IfNoValueTF{#2}
          {\IfEqCase{#1}{
        {len}{\ensuremath{\lambda}}%
        {num}{\ensuremath{k}}%
        {puls}{\ensuremath{n}}%
          }[\ensuremath{\vb{n}_{\ite[#1]}}]
          }%
          {\IfEqCase{#1}{
        {len}{\ensuremath{\lambda}}%
        {num}{\ensuremath{k}}%
        {puls}{\IfEqCase{#2}{%
        		{tot}{\ensuremath{N}}%
          	}[\PackageError{wave}{Undefined option with 2 entries to wave: #1}{}]
          }
        }[\PackageError{wave}{Undefined option with 2 entries to wave: #1}{}]}
    }
}

\newcommand{\pdf}[1][Default]{
  \IfEqCase{#1}{%
      {Default}{\ensuremath{f}}           %
        {DNS}{\ensuremath{f}_{DNS}}           %
        {filter}{\ensuremath{f}_{c}}    %
    }[\PackageError{pdf}{Undefined option to pdf: #1}{}]%
}

\newcommand{\lset}[1][Default]{
    \IfEqCase{#1}{%
        {Default}{\ensuremath{\varphi}}         %
        {DNS}{\ensuremath{\varphi}_{DNS}} 
        {prime}{\ensuremath{\varphi^{\prime}}}      %
        {filter}{\ensuremath{\varphi}_{c}}  %
        {fluctu}{\ensuremath{\tilde{\varphi}}}  %
    }[\PackageError{lset}{Undefined option to lset: #1}{}]%
}%

\newcommand{\meancurv}[1][Default]{%
    \IfEqCase{#1}{%
        {Default}{\ensuremath{H}}         %
        {DNS}{H_{DNS}}      %
        {prime}{\ensuremath{H^{\prime}}}      %
        {max}{\ensuremath{H_{\max}}} 			%
        {fluctu}{\ensuremath{\tilde{H}}}  %
        {filter}{\ensuremath{H_{c}}}%
        {var}{\ensuremath{H_{ \lambda}}}%
        {delta}{\ensuremath{\delta H}}%
        {moy}{\ensuremath{\overline{H}}}%
        {K}{\ensuremath{H_{K}}}%
    }[\ensuremath{\meancurv_{\ite[#1]}}]%
}%

\newcommand{\gausscurv}[1][Default]{%
    \IfEqCase{#1}{%
        {Default}{\ensuremath{G}}%
        {DNS}{\ensuremath{G_{DNS}}}%
        {prime}{\ensuremath{G^{\prime}}}%
        {filter}{\ensuremath{G}_{c}}%
    }[\ensuremath{G_{\ite[#1]}}]%
}%
\newcommand{\invar}[1][Default]{%
    \IfEqCase{#1}{%
        {Default}{\ensuremath{I}}%
        {DNS}{\ensuremath{I_{DNS}}}%
    }[\ensuremath{I_{\ite[#1]}}]%
}%

\NewDocumentCommand{\curv}{o o }{%
  \IfNoValueTF{#1}
    {\ensuremath{\kappa}} %
    {%
        \IfNoValueTF{#2}
          {\IfEqCase{#1}{
        		{local}{\ensuremath{\kappa}}%
          }[\PackageError{\curv}{Undefined option to curv with entry : #1}{}]%
          }%
          {\IfEqCase{#1}{%
        		{local}{\IfEqCase{#2}{%
        			{filter}{\ensuremath{\kappa_{c}}}%
        			}[\PackageError{\curv}{Undefined option to curv with entry : #2}{}]%
        			}%
          	}[\ensuremath{\kappa_{\ite[#1],\ite[#2]}}]%
          }
    }
}
\NewDocumentCommand{\interfarea}{o o }{%
  \IfNoValueTF{#1}
    {\ensuremath{\Sigma}} %
    {%
        \IfNoValueTF{#2}
          {\IfEqCase{#1}{
        		{Default}{\ensuremath{\Sigma}}        %
        		{DNS}{\ensuremath{\Sigma_{DNS}}}      %
        		{prime}{\ensuremath{\Sigma^{\prime}}}     %
        		{filter}{\ensuremath{\Sigma_{c}}}   %
        		{fluctu}{\ensuremath{\tilde{\Sigma}}}   %
        		{ref}{\ensuremath{\Sigma_{0}}}   %
        		{c}{\ensuremath{\Sigma_{c}}}  %
        		{var}{\ensuremath{\Sigma_{\lambda}}}  %
        		{s}{\ensuremath{\Sigma}}  %
       		{K}{\ensuremath{\Sigma_{K}}}  %
        		{min}{\ensuremath{\Sigma_{\min}}}  %
        		{spot}{\ensuremath{f_{\Sigma}}}  %
        		{ext}{\ensuremath{\mathscr{S}}}%
          }[\ensuremath{\Sigma_{\ite[#1]}}]%
          }%
          {\IfEqCase{#1}{%
        		{fluid}{\ensuremath{\Sigma_{\ite[#1],\ite[#2]}}}%
        		{fluctu}{\IfEqCase{#2}{%
        			{var}{\ensuremath{\tilde{\Sigma}_{\lambda}}}%
        			{delta}{\ensuremath{\delta \tilde{\Sigma}}}  %
				{K}{\ensuremath{\tilde{\Sigma}_{\ite[#2]}}}%
        			}[\PackageError{interfarea}{Undefined option with entry fluctu to interfarea: #1}{}]%
        			}%
          	}[\ensuremath{\Sigma_{\ite[#1],\ite[#2]}}]%
          }
    }
}
\NewDocumentCommand{\normal}{o o }{%
  \IfNoValueTF{#1}
    {\ensuremath{\vb{n}}} %
    {%
        \IfNoValueTF{#2}
          {\IfEqCase{#1}{
         {s}{\ensuremath{n}}%
        {Default}{\ensuremath{\bs{n}}}%
        {DNS}{\ensuremath{\vb{n}_{DNS}}}%
        {filter}{\ensuremath{\vb{n}}_{c}}%
        {face}{\ensuremath{\vb{n}}_{\ite[face]}}%
          }[\ensuremath{\vb{n}_{\ite[#1]}}]
          }%
          {\IfEqCase{#1}{
        		{fluid}{\ensuremath{\qty(\vb{n}_{\ite[fluid]})_{\ite[#2]}}}%
          	{var}{\ensuremath{\vb{n}_{\ite[#2], \ite[var]}}}%
	        {face}{\ensuremath{\vb{n}}_{\ite[face]\ite[#2]}}%
	        {cell}{\ensuremath{\vb{n}}_{\ite[cell]\ite[#2]}}%
          	}[\PackageError{normal}{Undefined option with 2 entries to normal: #1}{}]
          }
    }
}

\NewDocumentCommand{\volfrac}{o o }{%
  \IfNoValueTF{#1}
    {\ensuremath{\ensuremath{\alpha}}} %
    {%
        \IfNoValueTF{#2}
          {\IfEqCase{#1}{
        	{fluid}{\alpha_{\ite[fluid]}}%
        	{var}{\ensuremath{\alpha_{\lambda}}}%
        	{moy}{\ensuremath{\overline{\alpha}}}%
        	{delta}{\ensuremath{\delta \alpha}}%
        	{smel}{\ensuremath{\Delta \alpha}}%
        	{filter}{\alpha_{c}}%
		    {fluctu}{\tilde{\alpha}}%
	        {skt}{\tilde{\alpha}_{k}}%
        	{skpot}{f_{k}}%
        	{s1pot}{f_{1}}%
        	{s2pot}{f_{2}}%
	        {x1}{\ensuremath{\alpha_{x_{1}}}}%
    	    {x2}{\ensuremath{\alpha_{x_{2}}}}%
        	{x3}{\ensuremath{\alpha_{x_{3}}}}%
	        {xi}{\ensuremath{\alpha_{x_{i}}}}%
          }[\ensuremath{\alpha_{\ite[#1]}}]
          }%
          {\IfEqCase{#1}{
        	{moy}{\ensuremath{\overline{\alpha}_{\ite[#2]}}}%
        	{fluid}{\IfEqCase{#2}{
        		{moy}{\ensuremath{\overline{\alpha}_{\ite[#1]}}}%
        		}[\ensuremath{\qty(\alpha_{\ite[fluid]})_{\ite[#2]}}]%
        		}%
          	{var}{\ensuremath{\alpha_{\ite[#2], \ite[var]}}}%
          	}[\PackageError{volfrac}{Undefined option with 2 entries to volfrac: #1}{}]
          }
    }
}
\DeclareRobustCommand{\ball}[1][Default]{
    \IfEqCase{#1}{%
      {Default}{\mathcal{B}}%
      {c}{\mathcal{B}_{c}}%
      {filter}{\mathcal{B}_{c}}%
    }[\PackageError{ball}{Undefined option to ball: #1}{}]%
}%

\NewDocumentCommand{\surface}{o o }{%
  \IfNoValueTF{#1}
    {\ensuremath{\mathcal{S}}}%
    {%
        \IfNoValueTF{#2}
          {\IfEqCase{#1}{
     		{face}{\ensuremath{\mathcal{S}_{\ite[face]}}}%
     		{cell}{\ensuremath{\mathcal{S}_{\ite[cell]}}}%
      		{maths}{\ensuremath{\mathscr{S}}}%
          }[\ensuremath{\mathcal{S}_{\ite[#1]}}]%
          }%
          {\IfEqCase{#1}{
     		{face}{\ensuremath{\mathcal{S}_{\ite[face]\ite[#2]}}}%
     		{cell}{\ensuremath{\mathcal{S}_{\ite[cell]\ite[#2]}}}%
      		{maths}{\ensuremath{\mathscr{S}_{\ite[#2]}}}%
          	}[\PackageError{surface}{Undefined option with 2 entries to surface: #2}{}]
          }
    }
}

\NewDocumentCommand{\spaces}{o o o }{%
  \IfNoValueTF{#1}
    {\ensuremath{\mathbb{R}}}%
    {%
        \IfNoValueTF{#2}
          {\IfEqCase{#1}{
        	{M}{\mathcal{M}}%
        	{TM}{\mathcal{TM}}%
        	{TqM}{\mathcal{T}_{\pathM}\mathcal{M}}%
       	 	{TR}{\mathcal{TR}}%
        	{TR7}{\mathcal{TR}^{7}}%
        	{R7}{\mathbb{R}^{7}}%
        	{R77}{\mathbb{R}^{7\times 7}}%
        	{R2}{\mathbb{R}^{2}}%
        	{R3}{\mathbb{R}^{3}}%
        	{R5}{\mathbb{R}^{5}}%
        	{R55}{\mathbb{R}^{5\times 5}}%
        	{R4}{\mathbb{R}^{4}}%
        	{R}{\mathbb{R}}%
        	{Rp}{\mathbb{R}^{p}}%
        	{Rd}{\mathbb{R}^{d}}%
        	{Rpp}{\mathbb{R}^{p\times p}}%
        	{Rplus}{\mathbb{R}_{+}}%
       	 	{up}{\Omega} %
        	{upelem}{\omega} %
        	{fluid}{\mathcal{N}_{f}}%
        	{fspe}{\mathcal{N}_{s}(\ite[fluid])}%
        	{f1spe}{\mathcal{N}_{s}(1)}%
        	{f2spe}{\mathcal{N}_{s}(2)}%
        	{lfspe}{\mathcal{N}_{s}(\ite[lfluid])}%
        	{spe}{\mathcal{N}_{s}}%
        	{setvar}{\mathcal{O}}%
          	}[\ensuremath{\mathcal{R}^{\ite[#1]}}]
          }%
          {\IfNoValueTF{#3}
            {\IfEqCase{#1}{%
        	{fspe}{\mathcal{N}_{s}(\ite[#2])}%
				{fluid}{\ensuremath{\qty(\vb{n}_{\ite[fluid]})_{\ite[#2]}}}%
	        	{setvar}{\IfEqCase{#2}{
	        		{entropy}{\ensuremath{\mathcal{O}_{\zeta_{\eta}}}}%
              		{natural}{\ensuremath{\mathcal{O}_{\zeta}}}%
              		{naturalk}{\ensuremath{\mathcal{O}_{\zeta_{k}}}}%
              		{pressure}{\ensuremath{\mathcal{O}_{\pi}}}%
              		{pressurek}{\ensuremath{\mathcal{O}_{\pi_{k}}}}%
             		{thermo}{\ensuremath{\mathcal{O}_{\xi}}}%
              		{thermok}{\ensuremath{\mathcal{O}_{\xi_{k}}}}%
              		{thermoext}{\ensuremath{\mathcal{O}_{\Xi_{k}}}}%
              		{entropyvar}{\ensuremath{\mathcal{O}_{\eta_{\lambda}}}}%
					}[\PackageError{spaces}{Undefined option with 2 entries to setvar in spaces: #2}{}]}%
          	}[\PackageError{spaces}{Undefined option with 2 entries to spaces: #1}{}]}%
            {\IfEqCase{#1}{%
	        	{setvar}{\IfEqCase{#2}{
					{natural}{\IfEqCase{#3}{%
            			{fluidiso}{\ensuremath{\mathcal{O}_{\overline{\zeta_{\ite[fluid]}}}}}%
            			}[\ensuremath{\mathcal{O}_{\zeta_{\ite[#3]}}}]%
						}%
					{pressure}{\IfEqCase{#3}{%
            			{fluidiso}{\ensuremath{\mathcal{O}_{\overline{\pi_{\ite[fluid]}}}}}%
            			}[\ensuremath{\mathcal{O}_{\pi_{\ite[#3]}}}]%
						}%
					{thermo}{\IfEqCase{#3}{%
            			{fluidiso}{\ensuremath{\mathcal{O}_{\overline{\xi_{\ite[fluid]}}}}}%
            			}[\ensuremath{\mathcal{O}_{\xi_{\ite[#3]}}}]%
						}%
              		{thermoext}{\ensuremath{\mathcal{O}_{\Xi_{\ite[#3]}}}}%
					}[\PackageError{spaces}{Undefined option with 3 entries to setvar in spaces: #2}{}]}%
              }[\PackageError{speed}{Undefined option with 3 entries to speed: #1}{}]}
          }
    }
}

\newcommand{\spacetx}[1]{%
    \IfEqCase{#1}{%
        {0}{\mathcal{V}}%
        {t}{\mathcal{V}_{t}}%
        {t0}{\mathcal{V}_{t_{0}}}%
        {t1}{\mathcal{V}_{t_{1}}}%
    }[\PackageError{spacetx}{Undefined option to spacetx: #1}{}]%
}%

\newcommand{\capillarity}[1][Default]{
	\IfEqCase{#1}{%
		{Default}{\ensuremath{\sigma}}         %
		{filter}{\ensuremath{\sigma_{c}}}       %
		{fluctu}{\ensuremath{\tilde{\sigma}}}      %
	}[\PackageError{capillarity}{Undefined option to capillarity: #1}{}]%
}%
\NewDocumentCommand{\viscosity}{o }{%
  \IfNoValueTF{#1}
    {\ensuremath{\ensuremath{\mu}}}%
  {\IfEqCase{#1}{
      {add}{\ensuremath{\mu_{add}}}%
      }
      [\ensuremath{\mu_{\ite[#1]}}]%
    }
}

\NewDocumentCommand{\Idmat}{o }{%
  \IfNoValueTF{#1}
    {\ensuremath{\fontmat{I}_{d}}}%
  {\IfEqCase{#1}{
      {7}{\ensuremath{\fontmat{I}_{7}}}%
      {3}{\ensuremath{\fontmat{I}_{3}}}%
      }
      [\ensuremath{\fontmat{I}_{\ite[#1]}}]%
    }
}
\newcommand{\average}[1]{\ensuremath{\left\langle #1 \right\rangle}}

\newcommand{\ddmat}[1]{%
    \IfEqCase{#1}{%
        {1}{D_{1,t}}%
        {2}{D_{2,t}}%
        {k}{D_{k,t}}%
        {kp}{D_{k^{\prime},t}}%
    }[\PackageError{ddmat}{Undefined option to ddmat: #1}{}]%
}%

\NewDocumentCommand{\eigen}{o o }{%
  \IfNoValueTF{#1}
    {\ensuremath{\fontvec{r}}} %
    {%
        \IfNoValueTF{#2}
          {\IfEqCase{#1}{
        		{vec}{\ensuremath{\fontvec{r}}}%
        		{rvec}{\ensuremath{\fontvec{r}}}%
        		{lvec}{\ensuremath{\fontvec{l}}}%
        		{val}{\ensuremath{\lambda}}%
          }[\PackageError{eigen}{Undefined option with 1 entries to eigen: #1}{}]
          }%
          {\IfEqCase{#1}{
        		{rvec}{\ensuremath{\fontvec{r}_{\ite[#2]}}}%
        		{lvec}{\ensuremath{\fontvec{l}_{\ite[#2]}}}%
        		{val}{\ensuremath{\lambda_{\ite[#2]}}}%
          	}[\PackageError{eigen}{Undefined option with 2 entries to eigen: #2}{}]
          }
    }
}

\NewDocumentCommand{\reynolds}{o o }{%
  \IfNoValueTF{#1}
    {\ensuremath{\text{Re}}} %
    {%
        \IfNoValueTF{#2}
          {\IfEqCase{#1}{%
        		{add}{\ensuremath{\text{Re}_{add}}}%
          }[\ensuremath{\text{Re}_{\ite[#1]}}]%
          }%
          {\IfEqCase{#1}{%
        		{add}{\ensuremath{\ensuremath{\text{Re}_{add}}}}%
          	}[\PackageError{reynolds}{Undefined option with 2 entries to reynolds: #2}{}]%
          }
    }
}
\NewDocumentCommand{\weber}{o o }{%
  \IfNoValueTF{#1}
    {\ensuremath{\text{We}}} %
    {%
        \IfNoValueTF{#2}
          {\IfEqCase{#1}{%
        		{add}{\ensuremath{\text{We}_{add}}}%
          }[\ensuremath{\text{We}_{\ite[#1]}}]%
          }%
          {\IfEqCase{#1}{%
        		{add}{\ensuremath{\ensuremath{\text{We}_{add}}}}%
          	}[\PackageError{weber}{Undefined option with 2 entries to weber: #2}{}]%
          }
    }
}

\NewDocumentCommand{\momentumratio}{o o }{%
  \IfNoValueTF{#1}
    {\ensuremath{\text{M}}} %
    {%
        \IfNoValueTF{#2}
          {\IfEqCase{#1}{
        		{add}{\ensuremath{\text{M}_{add}}}%
          }[\ensuremath{\text{M}_{\ite[#1]}}]
          }%
          {\IfEqCase{#1}{
        		{add}{\ensuremath{\ensuremath{\text{M}_{add}}}}%
          	}[\PackageError{momentumratio}{Undefined option with 2 entries to momentumratio: #2}{}]
          }
    }
}

\newcommand{\citeay}[1]{\cite{#1}} 
\usepackage[pdftex,breaklinks, urlcolor=blue,pdftitle={FTAC},pdfauthor={Pierre CORDESSE},pdfsubject={LaTeX}]{hyperref}
\hypersetup{
  colorlinks   = true,
  citecolor    = blue,
  linkcolor = blue,
}

\usepackage{todonotes}
\usepackage{empheq}

\usepackage{latexsym}
\journalname{ 
}%
\begin{document}

\title{\large Validation strategy of reduced-order two-fluid flow models based on a hierarchy of direct numerical simulations%
}

\titlerunning{Validation strategy of reduced-order two-fluid flow models}        %

\author{Pierre Cordesse         	\and
        Alberto Remigi   		\and
        Benjamin Duret			\and
        Angelo~Murrone			\and
        Thibaut M\'enard			\and
        Fran\c{c}ois-Xavier Demoulin	\and
        Marc~Massot
}

\authorrunning{P. Cordesse, A. Remigi, B. Duret, A. Murrone, T. M\'enard, F.X. Demoulin and M. Massot} %

\institute{P. Cordesse, A. Murrone \at
              ONERA, DMPE, 8 Chemin de la Huni\`ere, 91120 Palaiseau, France \\
              \email{pierre.cordesse@polytechnique.edu}           %
           \and
           P. Cordesse, M. Massot \at
           CMAP, CNRS, Ecole Polytechnique, Institut Polytechnique de Paris, 
91128 Palaiseau cedex, France
           \and
           A. Remigi, B. Duret, T. M\'enard, F.X. Demoulin \at
            CNRS UMR 6614 - CORIA, 76801, Saint Etienne du Rouvray cedex, France
}

\date{Received: date / Accepted: date}

\maketitle

\section*{Abstract} Whereas direct numerical simulation (DNS) have reached a high level of description in the field of atomization processes, they are not yet able to cope with industrial needs since they lack resolution and are too costly. Predictive simulations relying on reduced order modeling have become mandatory for applications ranging from cryotechnic to aeronautic combustion chamber  liquid injection. Two-fluid models provide a good basis in order to conduct such simulations, even if recent advances allow to refine subscale modeling using geometrical variables in order to reach a unified model including separate phases and disperse phase descriptions based on high order moment methods. The simulation of such models has to rely on dedicated numerical methods and still lacks assessment of its predictive capabilities. The present paper constitutes a building block of the investigation of  a hierarchy of test-cases designed to be amenable to DNS while close enough to industrial configurations, for which we propose a comparison of two-fluid compressible simulations with DNS data-bases. We focus in the present contribution on 
an air-assisted water atomization using a planar liquid sheet injector. Qualitative and quantitative comparisons with incompressible DNS allow us to identify and analyze strength and weaknesses of the reduced-order modeling and numerical approach in this specific configuration and set a framework for more refined models since they already provide a very interesting level of comparison on averaged quantities.
\\[12pt]
\section*{Introduction}
Engines safety and efficiency are two of the main priorities given to the aeronautical and space industries. In the combustion chamber, the multi-scale and multi-physics phenomena are very complex and their interaction still animate the research domain. The primary atomization plays a crucial part in the way the engines work, thus must be thoroughly studied to understand its impact on high frequencies instabilities and on the generation of droplets evaporating and eventually reacting with the gaseous flow field. The former have been encountered in the past and can lead to critical damages of rockets. The latter is nowadays a key concern in the aeronautic industry since facing strong restrictions from the International Civil Aviation Organization norms \citeay{Icao_2019}. Even though experimentations must be conducted to enable simulation validation and to understand the observed physical phenomena, predictive numerical simulations are mandatory, at least as a complementary tool to understand the physics but even more to conceive new combustion chambers and predict instabilities and droplet size distributions they may generate in a given configuration.

In sub-critical condition, downside an air-assisted coaxial injector,  three two-phase flow topologies are to be found. At the injector exit, the two phases are separated by a smooth interface. Downstream a polydisperse spray of droplets is carried by the gaseous phase. In between, shear stress caused by strong velocity gradients tears the liquid core apart and ligaments are formed. This process is called primary atomization. The ligaments get thinner and thinner until they break into droplets during the secondary atomization process. In this \textit{mixed region}, the subscale physics and the topology of the flow are very complex.
\begin{figure}
  \begin{center}
    \includegraphics[width=0.50\textwidth]{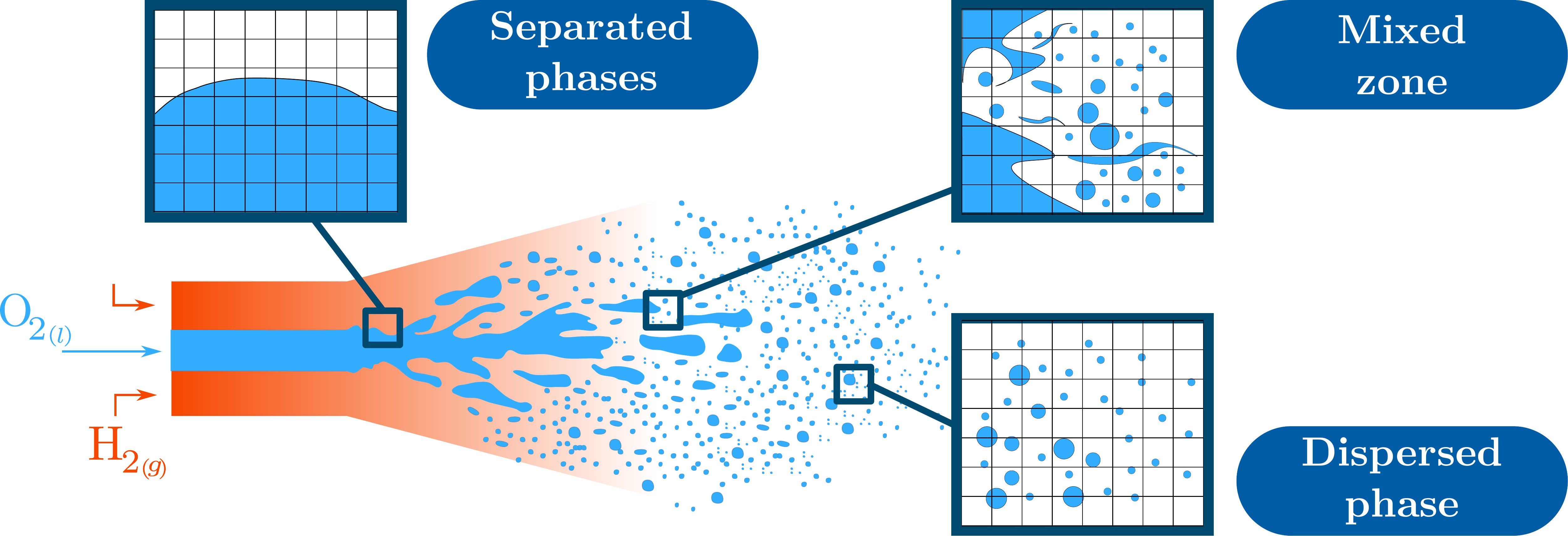}
  \end{center}
  \caption{Description of the various regimes and flow topologies in jet atomization for cryotechnic injectors in subcritical conditions.}
\end{figure}
The typical range of flow numbers in sub-critical cryogenic cylindrical jet in real configuration are, for the liquid Reynolds number, $Re_{L} = 10^{5}$, for the hydrodynamic Weber number, $We_{H} = 10^{5}$. In the aeronautical simplex atomizers employed on mid-range engines, the typical liquid Reynolds number of the liquid sheet flowing out from the atomizer is of the order of $Re_{L} = 10^{3}$, while the hydrodynamic Weber number is $We_{H}=10-100$. Experimental test benches such as Mascotte, operated by the French Aerospace Lab (ONERA) for cryogenic combustion, offer some experimental data but usually at lower flow numbers and are very expensive to operate and to instrument. As a result, numerical simulations are adequate to increase our understanding of the physics and perform early-stage optimizations, thus accelerating developments while reducing the costs.  

Direct numerical simulations (DNS) in real configuration of such engines are still out of reach, CPU needs being too high due to the high Reynolds and Weber numbers. Therefore predictive numerical tools using reduced-order models must be developed. However great care must be taken on the choices of these models in order to both have solid mathematics properties and to lead to predictive simulations after a validation process. 

The strategy retained at ONERA is to perform numerical simulations of the primary atomization from the injection to the combustion of the spray by coupling reduced-order models. One leading course of action consists in coupling Eulerian models, more specifically Kinetic-Based-Moment Methods (KBMM) to describe the disperse flow (see \citeay{Sibra_2017,Essadki_2018,Essadki_IJMF_2019}) and diffuse interface models\footnote{Since the signification of diffuse interface models vary in the scientific community, DIM refer in the present paper to any averaged model which allows locally the presence of each phase. These models can be obtained by a statistical averaging process for instance \citeay{Cordesse_PhD}.} (DIM) to reproduce the separated phases and the \emph{mixed region}. These models are implemented in the industrial CFD code CEDRE \citeay{Gaillard_2016}. While the predictiveness of KBMM solvers in dispersed flow have already been demonstrated, it is not yet the case for the diffuse interface models (DIM) which are meant to handle the challenging \emph{mixed region}.

Among the hierarchy of DIM, well-known models such as the multi-species compressible Navier-Stokes, referred as the \emph{four equation} model have been massively adopted in the industry \citeay{Letouze_2015} but have shown their limitations as they neglect the fluids thermodynamics disequilibrium in the mixing zones. Therefore, recent efforts have been done to increase the disequilibrium description of the phases by implementing and testing the so-called \emph{five equation} model \citeay{Kapila_2001} accounting for two temperatures \citeay{Murrone_2018} or even the Baer-Nunziato model \citeay{Baer_Nunziato_1986}, also called \emph{seven equation model}, whereby two temperatures, two pressures and two velocities are solved. Stemming from rational thermodynamics or Least Action Principle, the complexity of these models are three-folds: 1) the macroscopic set of equations of these models often include non-conservative terms, 2) these models can not be derived from physics at small scale of interface dynamics, thus require closure of interfacial quantities such as the interfacial velocity or the interfacial pressure, 3) the thermodynamics has to be postulated and requires assumptions. The mathematical properties of the Baer-Nunziato model have been studied in \citeay{Embid_Baer_1992, Coquel_2002, Gallouet_2004} for instance and many closure have been proposed for the interfacial terms based on wave-type considerations and the entropy inequality. In \citeay{Cordesse_2019_CMS}, a theoretical framework for the derivation of supplementary conservation laws for systems of partial differential equation including first-order non-conservative terms has been proposed, thus extending the Godunov-Mock standard approach for systems of conservation laws \citeay{Godunov_1961,Mock_1980}. The theory has brought about entropy supplementary conservative equations together with constraints on the interfacial quantities and the definition of the thermodynamics for non-miscible fluids and also when accounting for some level of mixing of the two phases. Nonetheless, this study has also emphasized the risk of inconsistency between the system and the associated postulated thermodynamics when accounting for non-ideal effects. Indeed most multi-fluids thermodynamics approaches are based on the assumption of no interactions between the fluids, resulting in equipping each fluid with its own thermodynamics and then defining mixing thermodynamics quantities by taken arithmetic average of each phase contribution. However in many industrial configurations, such as in jet atomizations, the fluids face non-ideal effects such as compaction or surface tension, questioning the validity of having an isolated phase thermodynamics approach. In \citeay{Cordesse_CMT_2019}, the authors propose to build a multi-fluid thermo- dynamics starting from a single fundamental thermodynamics differential equation of the mixture and a set of pressure laws. The derivation exhibits a natural decomposition of thermodynamics quantities into an isolated phase contribution and mixing terms, expressed in terms of the stated pressure laws. Thus, given any pressure law including non-ideal effects, the method yields the consistent thermodynamics variables as well as the associated extended system of equations. A strongly connected question for such systems is the ability to derive an entropic symmetrization in the sense of Godunov-Mock and the related constraints on the decomposition, as well as the study of the spectrum and hyperbolicity. The proposed framework introduced in \citeay{Cordesse_2019_CMS} allows to shed some light on these questions. Nevertheless, in the case of Baer-Nunziato model, the lack of strict convexity of the proposed entropies prevents its Godunov-Mock symmetrization. Based on the new developments in \citeay{Cordesse_CMT_2019}, we envision equipping the Baer-Nunziato system with an extended thermodynamics closure will lead to a strictly convex entropy and thus to its symmetrization in the sense of Godunov-Mock. This is the subject of current research. Besides, next generation two-phase flow models adding sub-scale effects emerge \citeay{Herrmann_2015,Cordesse_2018_NASATM_LAP} and add again some mathematical challenges. 

As a result, while the mathematical properties of these Eulerian models are still under current investigation, even the most basic element of the hierarchy of models, that is the Bear-Nunziato model, gives rise to numerical challenges. First, the models require numerical schemes that are in adequacy with the closures of the interfacial terms. Second, the presence of non-conservative terms in the system of equations complexifies the definition of weak solutions in the sense of distributions, and Rankine-Hugoniot jump conditions, which allow to solve accurately the local Riemann problems, can not be unambiguously defined. No general theory exists for these reduced-order models, making the derivation of efficient numerical schemes challenging. The most simplest way to handle the non-conservative terms is to assume local constancy of the interfacial pressure and velocity on each interface at each time integration \citeay{Saurel_1999,Zein_2010,Furfaro_Saurel_2015} allowing a simple discretization. Various other numerical strategies exist such as exact Riemann solvers \citeay{Andrianov_2004,Schwendeman_2006,Deledicque_2007} or approximate Riemann solvers \citeay{Saurel_1999,Saurel_2009,Ambroso_2012} including correction techniques based on mixture conservative equations. In addition, exact solver of the linearized Riemann problem have been also developped since they are able to include the non-conservative terms into the discretization \citeay{Sainsaulieu_1995,Gallouet_2004,Pelanti_2014}. Finally, an entropy-satisfying relaxation scheme has been proposed for the Baer-Nunziato model in \citeay{Coquel_2014_Seguin,Coquel_2017} for which a discrete entropy inequality is proven, guarantying the nonlinear stability of the numerical method. Furthermore the applications we are seeking are characterized by strong gradients and discontinuities, thus the numerical schemes employed must be robust and accurate enough. They need to be assessed to highlight their promising predictiveness before being widely deployed in the industry. In this study, we thus focus on the Bear-Nunziato \emph{seven equation model} \citeay{Baer_Nunziato_1986} equipped with a simple numerical strategy as in \citeay{Saurel_1999,Zein_2010,Furfaro_Saurel_2015}, which constitutes a relevant first candidate for numerical strategy formulation and reduced-order simulations before tackling more recent and refined models.

Therefore, a hierarchy of specific test cases aiming at reproducing real engine configurations has been selected and reproduced with DNS in order to build an assessment tool to validate sophisticated reduced-order models such as the \emph{seven equation model}. %
 In \citeay{Cordesse_2018_NASATM_simu}, we have started with an air-assisted water atomization using a coaxial injector, which in addition provides experimental results from the LEGI test bench. The comparison has shown good agreements in terms of liquid core length and important CPU gains between the \emph{seven equation model} implemented in the CEDRE code and the incompressible DNS results obtained with the ARCHER code. It has also shown the limits of diffuse interface models to capture complex liquid structures such as ligaments, rings or deformed droplets and has encouraged to add a sub-scale description of the interface dynamics through geometric variables such as the interfacial area density or the mean and Gaussian curvatures as proposed in \citeay{Cordesse_2018_NASATM_LAP}.

In the present work, we carry on the construction of a hierarchy of DNS test-cases to validate this reduced-order DIM. We propose a complementary second test case, an air-assisted water atomization using this time a planar injector rather than a co-axial injector. This planar injector reproduces in terms of Weber and Reynolds number the liquid sheet flowing out a swirling atomizer used in agricultural applications \citeay{Belhadef_2012}, $Re_{L} =1.5 \, 10^{3}$ and $We_{R} = 4.0\, 10^{2}$ and $p=1 \, bar$. The Reynolds and Weber numbers are also typical from the mid-range aeronautical engines. As analyzed in the present contribution, this test case offers also an atomization regime with different flow features, which makes this second test case complementary with the first one in order to eventually validate our reduced-order models on a cryogenic coaxial injection.

This paper is organized as followed. In the first Section~\ref{sec:math_modelling}, we introduce the mathematical properties of the reduced-order model and the governing equations solved in the DNS. Then in Section~\ref{sec:num_methods}, the numerical methods implemented in the two codes, CEDRE and ARCHER are presented. In Section~\ref{sec:config}, the investigated configuration is described. Finally in Section~\ref{sec:result}, we present the results obtained in both simulations and  provide a qualitative and quantitative comparisons. We finally conclude on the reduced order model assessment. %
\section{Mathematical modelling}\label{sec:math_modelling}
At the top of the hierarchy of diffuse interface models \citeay{Drui_JFM_2019} stands the Baer-Nunziato model \citeay{Baer_Nunziato_1986}, also called the \textit{seven equation model}, accounting for full disequilibrium of the phases. This first-order non-linear non-conservative system of partial differential equations is composed of a mass, momentum and energy equation for each phase and a seventh equation on the volume fraction $\volfrac$ to reconstruct the interface. The extended form proposed in \citeay{Saurel_1999} introduces an interfacial pressure, $\pressure[sint]$, and an interfacial velocity, $\speed[sint]$, that need to be closed. The quasi-linear form of the system in the one-dimensional case takes the form
\begin{subequations}\label{eq:chap1_BNZ_model_anal_maths_with_relax_terms}
\renewcommand\arraystretch{1}
\setlength\arraycolsep{8pt}
\begin{align}
\fpdv{t} \CV + \fontmat{A}(\CV) \fpdv{x} \CV =\source[relax](\CV), 
\end{align}
with $\fontmat{A}(\CV)  = \fpdv{\CV} \CF(\CV) + \NCF(\CV)$ and 
\begin{align}
\fpdv{\CV} \CF(\CV) =
	\begin{pmatrix}
        0 & \fontvec{0} & \fontvec{0} \\
        \fontvec{0} 	& \partial_{\CV[2]} \CF[2](\CV[2]) & \fontvec{0}  \\
    		\fontvec{0}	& \fontvec{0} & \partial_{\CV[1]} \CF[1](\CV[1])
    \end{pmatrix}, 
&&
\NCF(\CV) = 
	\begin{pmatrix}
        \speed[s][int] & \fontvec{0} & \fontvec{0}\\
        \NCF[v][2] & \fontvec{0} & \fontvec{0}\\
    		\NCF[v][1] & \fontvec{0} & \fontvec{0}
    \end{pmatrix},
\end{align}
\vspace{-1em}
\begin{align}
\CV[fluid] =
	\begin{pmatrix}
		\volfrac[fluid] \density[fluid] \\
		\volfrac[fluid] \density[fluid] \speed[s][fluid] \\
		\volfrac[fluid] \density[fluid] \energy[tot][fluid]
	\end{pmatrix}, \	 
&&
\CF[fluid](\CV[fluid]) = 
	\begin{pmatrix}
 		\volfrac[fluid] \density[fluid] \speed[s][fluid] \\
 		\volfrac[fluid](\density[fluid] \speed[s][fluid]^{2}+\pressure[fluid]) \\
 		\volfrac[fluid] ( \density[fluid] \energy[tot][fluid]+\pressure[fluid])\speed[s][fluid]
 	\end{pmatrix}, \
&&
\NCF[v][fluid](\CV) = \epsilon_{\ite[fluid]}
	\begin{pmatrix}
		 0 \\
		 \pressure[int] \\
		 \pressure[int] \speed[s][int]
	\end{pmatrix},
&&
\end{align}
\end{subequations}
where the column vector $\CV \in \spaces[R7]$ is defined by $\CV^{\transvector} = \left(\volfrac[2],\, \CV[2]^{\transvector},\, \CV[1]^{\transvector} \right)$, the conservative flux $\CF: \CV \in \Omega \mapsto \spaces[R7]$ reads $\CF(\CV)^{\transvector} = (0,\, \CF[2](\CV[2])^{\transvector},\, \CF[1](\CV[1])^{\transvector})$ and $\NCF: \CV \in \Omega \mapsto \spaces[R77]$ is the matrix containing the non-conservative terms. Furthermore we have $\epsilon_{\ite[fluid]} = (-1)^{\ite[fluid]+1}$. Then, $\vol[frac][fluid]$ is the volume fraction of phase $k \in \left[ 1,2 \right]$, $\density[fluid]$ the partial density, $\speed[s][fluid]$ the phase velocity, $\pressure[fluid]$ the phase pressure, $\energy[tot][fluid]=\energy[fluid] + \allowbreak \speed[s][fluid]^{2}/2$ the total energy per unit of mass and $\energy[fluid]$ the internal energy. Concerning the relaxation source term, $\source[relax](\CV)$, it decomposes into
\begin{subequations}\label{eq:chap1_BNZ_model_anal_maths_relax_term}
\renewcommand\arraystretch{1}
\begin{align}
	\source[relax] = \coeff[relax][speed] \source[relax][speed] + \coeff[relax][pressure] \source[relax][pressure], \
\source[relax][speed] = 
\begin{pmatrix}
	0\\
	\source[relax][speed][2] \\
	\source[relax][speed][1]
\end{pmatrix}, \
\source[relax][pressure] = 
\begin{pmatrix}
	\pressure[diff] \\
	\source[relax][pressure][2] \\
	\source[relax][pressure][1]
\end{pmatrix},
\end{align}
\vspace{-1em}
\begin{align}
	(\source[relax][speed][fluid])^{t} =
	 -\epsilon_{\ite[fluid]} \left( 0 , \speed[s][diff], \speed[s][int] \speed[s][diff] \right), \
	(\source[relax][pressure][fluid])^{t} =
	-\epsilon_{\ite[fluid]} \left( 0 , 0 , \pressure[int] \pressure[diff] \right),
\end{align}
\end{subequations}
where we have introduced the pressure difference $\pressure[diff] = \pressure[2]-\pressure[1]$ and the slip velocity $\speed[sdiff] = \speed[s2]-\speed[s1]$. An expression for the relaxation parameters $\coeff[relax][speed]$ and $\coeff[relax][pressure]$ have been derived for example using the DEM technic in \citeay{Saurel_Gavrilyuk_2003}. Hereafter, we will consider the phase $1$ to be the gaseous phase and the phase $2$ to be the liquid phase. We will thus explicit the subscripts by using $g$ and $l$.

From this seven-equation model, the instantaneous relaxation of the pressures and the velocities leads to the \textit{five-equation} model \citeay{Kapila_2001} and relaxing also instantaneously the temperatures, one obtains the compressible multi-species Navier-Stokes equations referred as the \textit{four-equation} model. These three models define a hierarchy of diffuse interface model and each of them are hyperbolic and appropriate for reactive two-phase flow or interface problems. 

In cryogenic applications, the pressure of the phases may be considered to relax instantaneously, but due to the strong velocity and temperature gradients at the interface, it is unrealistic to assume hydrodynamic and thermal instantaneous relaxations. Whereas single velocity models are sufficient in the separated zone with an adapted resolution mesh, it is not the case in the mixed region and thus fail at predicting the atomization process. Therefore the Baer-Nunziato model~\eqref{eq:chap1_BNZ_model_anal_maths_with_relax_terms} appears as the best candidate for the present study.

Nonetheless, stemming from rational thermodynamics, the macroscopic set of equations can not be derived from physics at small scale of interface dynamics and thus require closure of interfacial pressure and velocity as well as to postulate the thermodynamics. The theory on the existence of a supplementary conservative equation to first-order order system of partial differential equation including non-conservative terms \citeay{Cordesse_2019_CMS} has brought about supplementary conservative equations together with constraints on the interfacial quantities and the definition of the thermodynamics of the mixture. In the present work, we use a non-miscible fluid thermodynamics, where the mixture entropy is defined by the mass averaged sum of the phasic entropies with no mixing effect, together with a closure proposed in \citeay{Furfaro_Saurel_2015} obtained through a discrete element method approach \citeay{Saurel_Gavrilyuk_2003}.
 Only mechanical and hydrodynamic relaxations are accounted for in the present study. They are modeled as in \citeay{Furfaro_Saurel_2015} through a pressure difference term and a drag term as in Equation~\eqref{eq:chap1_BNZ_model_anal_maths_relax_term}. Each term is driven by a characteristic time, which can be finite to account for disequilibrium. In the limit of an instantaneous relaxation, any disequilibrium disappear. As described in Section~\ref{sec:num_methods}, we have used an instantaneous pressure relaxation but a finite velocity relaxation.

The direct numerical simulation solves the incompressible Navier-Stokes equations and takes into account surface tension. Then, the interface is implicitly derived by the zero of a Level Set of a scalar function and its motion is captured by the transport of the Level Set function at the hydrodynamic speed, combined with a projection technique \citeay{sussmanredist}. To guarantee the conservation of mass, we also transport the volume-of-fluid function (VOF), defined as the liquid volume fraction expressed in terms of the level set function. The benefits of a VOF formulation coupled with a Level Set function are to conserve mass and to have access to geometrical properties of the interface. %
\section{Numerical methods}\label{sec:num_methods}
\subsection{CEDRE Solver}
The numerical methods employed to solve the Baer-Nunziato model~\eqref{eq:chap1_BNZ_model_anal_maths_with_relax_terms} are implemented in the multiphysics computational fluid dynamics software CEDRE \citeay{Gaillard_2016} working on general unstructured meshes and organized as a set of solver \citeay{Gaillard_2016} developed at the french aerospace lab ONERA. The solver \textit{SEQUOIA} is in charge of the diffuse interface model.

The Strang splitting technique applied on a HLLC solver proposed in \citeay{Furfaro_Saurel_2015} has been successfully implemented in the solver \textit{SEQUOIA} of the CEDRE CFD code. Second-order extension is achieved using a second-order Runge-Kutta scheme for the time discretization of each sub-system together with a multislope MUSCL second-order space discretization \citeay{Letouze_2014}. The technique consists in computing two dedicated slopes for each face of each mesh cell. It uses the vertex neighbors and ensures the $L^{\infty}$ norm stability on the scalar advection. The issue encountered when discretizing the non-conservative terms is tackled in \citeay{Furfaro_Saurel_2015} by assuming $(1)$ the interfacial quantities $\pressure[int]$ and $\speed[int]$ to be local constants in the Riemann problem, $(2)$ the volume fraction to vary only across the interfacial contact discontinuity $\speed[int]$. As a result, the non-conservative terms vanish, $\speed[int]$ and $\pressure[int]$ are determined locally by Discrete Equation Method (DEM) \citeay{Saurel_Gavrilyuk_2003} at each time step and stay constant during the update. Thus, phases are decoupled, the system splits into two conservative sub-systems to which we apply the multi-slope HLLC with hybrid limiter solver.

Depending on the application, the relaxations are assumed either instantaneous or finite in time. In the present test case, it is reasonable to assume a instantaneous pressure relaxation, $\coeff[relax][pressure] \rightarrow +\infty$, but one need to consider a finite velocity relaxation since the interface dynamic is mainly driven by the shear stress induced by a high velocity difference between the phases at the injection.

To obtain the relaxed pressure, since the characteristic time is taken to be infinitely small, the problem reduces to apply an iterative procedure as a Newton method to solve a second order ordinary differential equation (ODE). The iterative procedure returns a single equilibrium pressure. Detailed equations can be found in \citeay{Furfaro_Saurel_2015}. As for the velocities, since we want to account for finite relaxation time, the associated ODE takes the form
\begin{align}\label{eq:7eq_relax_speed}
	\fpdv{t} \speed[diff] - \frac{A^{o}}{ \epsilon_{\speed[s]}} \speed[diff]= 0, 
\	\text{with } A^{o} = \frac{\density^{o}}{\vol[frac][l]^{o} \density[l]^{o} \vol[frac][g]^{o} \density[g]^{o}},
\end{align}
where $\epsilon_{\speed[s]}$ is the characteristic relaxation time, $\epsilon_{\speed[s]} =  1/\coeff[relax][speed]$, superscript ${}^{o}$ denotes the state before relaxation and $\density$ is the mixture density, $\density = \vol[frac][l]\density[l]+\vol[frac][g]\density[g]$. A first numerical approach is to fix a remaining slip velocity ratio target at each computational time step $\Delta \Time$. It defines the characteristic relaxing time as
\begin{align}
	\frac{\epsilon_{\speed[s]}}{A^{o}} = \ln{\qty(\frac{\speed[sdiff]\qty(\Delta t)}{\speed[sdiff]^{o}})} \Delta t.
\end{align}
An instantaneous velocity relaxation is in practice also possible and manipulating the ODE leads to a unique relaxed velocity, which is the mass weighted average of the two velocities before relaxing. In the present simulation, we have however been using the finite velocity relaxation.

\subsection{ARCHER Solver}

As for the DNS, we have used the High-Performance-Computing code ARCHER developed at the CORIA laboratory. It was one of the first code worldwide, undertaking the simulation of liquid-jet atomization under a realistic diesel injection configuration \citeay{Menard_2007}. For in depth information about the ARCHER solver, the reader can also refer to e.g. \citeay{Menard_2007,Duret_2012,Vaudor_2017,Canu_2018}.

\subsubsection{Navier-Stokes equations}

It solves on a Cartesian mesh the one-fluid formulation of the incompressible Navier-Stokes equation, viz.
\begin{align}
\fpdv{t} (\density \vec{\speed}) 
+ \div{\qty( \density \vec{\speed} \otimes  \vec{\speed})} 
= - \grad{\pressure}  + \div{ \qty( 2 \mu \vec{\source[viscous]})} 
 +  \capillarity \meancurv \delta_s \vec{\normal[s]}, 
\label{eq:NS}
\end{align}
where $\vec{\speed}$ is the hydrodynamical velocity vector, $\pressure$ is the pressure field, $\vec{\source[viscous]}$ the strain rate tensor, $\mu$ the dynamic viscosity, $\density$ the density, $\capillarity$ the surface tension, $\vec{\normal[s]}$ the unit normal vector to the liquid-gas interface, $\meancurv$ its mean curvature and $\delta_s$ is the Dirac function characterizing the locations of the liquid gas interface. For solving Equation~\eqref{eq:NS}, the convective term is written in conservative form. The mixture density is calculated from the Volume Of Fluid approach as $\density= \density_l \alpha_l + \density_g(1-\alpha_l)$.

For transporting the interface, use is made of a coupled Level Set and volume-of-fluid (CLSVOF) solver, in which the Level Set function accurately describes the geometric features of the interface (its normal and curvature) and the VOF function ensures mass conservation. A detailed description of the interface capturing methods used is proposed hereafter.

\subsubsection{Interface capturing method}

The interface is captured using a Coupled Level Set/Volume Of Fluid (CLSVOF) algorithm. This method allows an accurate representation of the interface with the Level Set function and the mass conservation with the VOF method. The general algorithm is based on the one presented in Menard et al. \citeay{Menard_2007}. The volume of fluid transport equation is defined as
\begin{equation}
\fpdv{t} \alpha_{l} +\div \left(\alpha_{l}\vec{\speed}\right)= 0.
\label{eq:VOF}
\end{equation}
Equation~\eqref{eq:VOF} is derived from the continuity equation by considering $\rho=\alpha_l\rho_l+\alpha_g\rho_g$. 

In the incompressible CLSVOF algorithm, the liquid volume fraction and Level Set transport equations have the same formulation and are split in the three space directions. To improve the splitting algorithm accuracy, the divergence correction proposed in \citeay{Puckett_1997} has been applied to each splitting step. Consequently, the liquid volume fraction equation becomes
\begin{equation}
\fpdv{t} \alpha_{l} +\div \left(\alpha_{l}\vec{\speed}\right)=\alpha_{l}\div  \vec{\speed}.
\label{eq:VOFpuckett}
\end{equation}
For the sake of clarity, only the discretized VOF transport equation on a $(i,j,k)$ indexed cartesian grid is shown below
\begin{equation}
\left\{
\begin{array}{l}
	\frac{\tilde{\alpha_{l}}-\alpha_{l}^{n}}{\Delta t}
	+\frac{\left(\alpha_{l}^{n}u\right)_{i+\frac{1}{2}}
		-\left(\alpha_{l}^{n}u\right)_{i-\frac{1}{2}}}{\Delta x}
	=\tilde{\alpha_{l}}\frac{u_{i+\frac{1}{2}}-u_{i-\frac{1}{2}}}{\Delta x}, \\
\frac{\hat{\alpha_{l}}-\tilde{\alpha_{l}}}{\Delta t}
	+\frac{\left(\tilde{\alpha_{l}}v\right)_{j+\frac{1}{2}}
		-\left(\tilde{\alpha_{l}}v\right)_{j-\frac{1}{2}}}{\Delta y}
	=\hat{\alpha_{l}}\frac{v_{j+\frac{1}{2}}-v_{j-\frac{1}{2}}}{\Delta y},	\\
\frac{\bar{\alpha_{l}}-\hat{\alpha_{l}}}{\Delta t}+\frac{\left(\hat{\alpha_{l}}w\right)_{k+\frac{1}{2}}-\left(\hat{\alpha_{l}}w\right)_{k-\frac{1}{2}}}{\Delta z}=\bar{\alpha_{l}}\frac{w_{k+\frac{1}{2}}-w_{k-\frac{1}{2}}}{\Delta z}
\end{array}
\right.
,
\label{eq:alphal}
\end{equation}
where $u$, $v$ and $w$ are respectively the $x-$, $y-$ and $z-$ coordinates of the velocity vector $\vec{\speed}$ evaluated at the interfaces of a cell. The cell center is indexed $(i,j,k)$ and the cell faces $(i\pm 1/2,j\pm 1/2,k\pm 1/2)$. The grid size is $\Delta x \times \Delta y \times \Delta z$ and $\Delta t$ is the time step. Intermediate volume fractions $\tilde{\alpha_{l}}$, $\hat{\alpha_{l}}$ and $\bar{\alpha_{l}}$ are introduced to calculate the liquid volume fraction at time step $n+1$, $\alpha_{l}^{n+1}$, from its value at time step $n$, $\alpha_{l}^{n}$. A final equation couples the three directions to update the liquid volume fraction
\begin{equation}
\begin{split}
\frac{\alpha_{l}^{n+1}-\bar{\alpha_{l}}}{\Delta t}
	=-\bar{\alpha_{l}}\frac{w_{k+\frac{1}{2}}-w_{k-\frac{1}{2}}}{\Delta z}
	-\hat{\alpha_{l}}\frac{v_{j+\frac{1}{2}}-v_{j-\frac{1}{2}}}{\Delta y}
	-\tilde{\alpha_{l}}\frac{u_{i+\frac{1}{2}}-u_{i-\frac{1}{2}}}{\Delta x}	\\
	+\alpha_{l}^{n+1}\left(
		\frac{w_{k+\frac{1}{2}}-w_{k-\frac{1}{2}}}{\Delta z}
		+\frac{v_{j+\frac{1}{2}}-v_{j-\frac{1}{2}}}{\Delta y}
		+\frac{u_{i+\frac{1}{2}}-u_{i-\frac{1}{2}}}{\Delta x}
		\right).
\end{split}
\label{eq:alphatot1}
\end{equation}

The Level Set transport equation is defined as
\begin{align}
\fpdv{t} G + \vec{\speed} . \grad{G} = 0,
\label{eq:LS}
\end{align}
where $G$ is the level set function.

As mentioned earlier, the level transport equation is solved in the same way (splitting) as the liquid volume fraction to maintain consistency.
However, since the function $G$ no longer remains a distance when solving Equation \eqref{eq:LS} due to velocity gradients, a reinitialization step of the Level Set function is performed at each time step  \citeay{Sussman_1998,Tanguy_2007}. The reinitialization step allows to keep the Level Set function as a proper signed distance function, i.e. $| \grad{G} |=1$.
All the geometrical information of the interface are obtained with the Level Set function $G$. The mean curvature $H$ is calculated by $H=\nabla . \vec{n}$  with $\vec{n}=\ffrac{\nabla G}{|\nabla G|}$, the normal to the interface.

\subsubsection{Projection method}

Then, to compute velocity and pressure, the momentum Equation~\eqref{eq:NS} is solved using a standard projection method. This method allows to decouple velocity and pressure. An intermediate velocity is first calculated without the pressure term $-\vec{\nabla}p$ and the surface tension term
\begin{equation}
\vec{\speed}^{*}=\frac{\rho^{n}\vec{\speed}^{n}}{\rho^{*}}-\underbrace{\Delta t\frac{\div \left(\left(\rho \vec{\speed}\right)^{n}\otimes\vec{\speed}^{n}\right)}{\rho^{*}}}_\textrm{I}+\underbrace{\Delta t\frac{\div \left(2\mu \vec{\source[viscous]} \right)}{\rho^{*}}}_\textrm{II},
\label{eq:ustar}
\end{equation}
where $\rho^{*}$ corresponds to the density calculated with the new value of $\alpha_l$ obtained after the interface resolution, $\rho^{*}= \alpha_l^{n+1} \rho_l+ (1-\alpha_l^{n+1})\rho_g$. 

Regarding the convection term $I$, the method of Vaudor et al. \citeay{Vaudor_2017} has been used ensuring consistent mass and momentum fluxes computation. The mesh is an eulerian staggered grid so the velocity is computed on the faces of the cells and the other variables (pressure, density, liquid volume fraction, ...) are computed in the center of the cells. This method initially proposed in \citeay{Rudman_1998} consists in calculating the mass flux $\rho \speed$ in the center of the cell by using the continuity equation. The mass flux has to be known in the center of the cell in order to have a second order centered scheme for the divergence operator. This procedure allows mass and momentum to be transported in a consistent manner thereby enabling flows with large liquid/gas density ratios to be simulated accurately. A detailed explanation of the algorithm can be found in Vaudor et al. \citeay{Vaudor_2017}.

Following the procedure of Sussman et al. \citeay{Sussman_2007}, The dynamic viscosity used in the viscous term $II$ ($\mu_{l}$ or $\mu_{g}$) depends on the sign of the Level Set function. In mixed cells, a specific treatment is performed to evaluate the dynamic viscosity. This method takes into account directly the viscous tensor jump across the interface. 

Then, the momentum equation is discretized in the following way by using the intermediate velocity obtained previously,
\begin{equation}
\vec{\speed}^{n+1}=\vec{\speed}^{*}-\Delta t \frac{\nabla P}{\rho^{*}}
\label{eq:ufinal}
\end{equation}

By applying the divergence operator to Equation~\eqref{eq:ufinal}, an Helmholtz equation for the pressure is obtained 
\begin{equation}
\begin{split}
\div \left(\frac{\nabla P^{n+1}}{\rho^{*}}\right)=\frac{\nabla . \vec{\speed}^{*}}{\Delta t}.
\end{split}
\label{eq:pression}
\end{equation}
Equation~\eqref{eq:pression} is solved using a MultiGrid preconditioned Conjugate Gradient algorithm (MGCG) \citeay{Zhang_1996}.
The $\div \vec{\speed}^{*}$ term is solved with a second order centered scheme. A Ghost Fluid Method (GFM) \citeay{Fedkiw_1999} is used to apply the pressure jump due to the presence of surface tension \citeay{Menard_2007}. The advantage of the GFM is a more realistic representation of the interface (sharp, infinitely thin) : jump conditions are directly added at the interface position through a local modification of the numerical scheme. The distance of the interface is provided by the Level Set function. 
The final velocity is computed by Equation~\eqref{eq:ufinal} using a second order centered scheme for the pressure gradient. A second order predictor-corrector Runge Kutta scheme have been used for temporal integration. 

Concerning the time step calculation, a CFL condition similar to the one used by \citeay{Kang_2000} is used. For $CFL=1$, the CFL condition can be written as
\begin{subequations}\label{eq:CFL_Kang}
\begin{equation}
\Delta t \left(\frac{\left(C_{cfl}+V_{cfl}\right)+\sqrt{\left(C_{cfl}+V_{cfl}\right)^{2}+4\left(S_{cfl}\right)^{2}}}{2}\right)\leq 1
\end{equation}
with
\begin{align}
	C_{cfl}	= \frac{ \max{ \qty( \vert u \vert )} }{\Delta x} 
			+ \frac{\max{ \qty( \vert v \vert  )} }{\Delta y}
			+ \frac{\max{ \qty( \vert w \vert  )} }{\Delta z},\\
	V_{cfl}	= \left(\frac{2}{\Delta x^{2}}+\frac{2}{\Delta y^{2}}+\frac{2}{\Delta z^{2}}\right)
			\times \max{\qty(\frac{\mu_l}{\rho_l},\frac{\mu_g}{\rho_g})}, \\
	S_{cfl}	=\sqrt{\frac{\capillarity \vert \meancurv \vert}
					{\min{\qty( \rho_g, \rho_l )} \times \min{ \qty( \Delta x,\Delta y,\Delta z)}^{2}}}.
\end{align}
\end{subequations} %
\section{Description of the configuration}\label{sec:config}
To attest the reduced-order model introduced in Section~\ref{sec:math_modelling} with direct numerical simulation, we propose the study of the atomization of an air-assisted flat liquid sheet at high Reynolds and Weber number. The reason for choosing this test case are two folds: first it makes a complementary test case to the air-assisted coaxial atomization analyzed in \citeay{Cordesse_2018_NASATM_simu} in terms of injection type while still offering an atomized regime, second while the Reynolds and Weber numbers are farther away from cryogenic applications, they still are typical from the mid-range aeronautical engines.

\subsection{Air-assisted liquid sheet atomization}

Atomization of air-assisted flat liquid sheets have been widely experimentally investigated such as in \citeay{Lozano_2001,Carvalho_2002} and \citeay{Dumouchel_2008}. The behavior of such flow is driven by several parameters. The ratio of the norm of the gas velocity and the liquid velocity, $\speed[s][g]/\speed[s][l]$, and the momentum flux ratio $M$ are predominant to determine the breakup regime. Then the width of liquid injection, $\diam[l]$ and the relative gaseous Weber number, $We_{R}$, influence also the breakup length, the breakup frequency and the liquid sheet vibrating frequency \citeay{Dumouchel_2008}. $M$ and $We_{R}$ are defined as:
\begin{align}\label{eq:number}
	M = \frac{\density[g] \speed[s][g]^{2}}{\density[l] \speed[s][l]^{2}},  && We_{R} = \frac{\density[g] \qty(\speed[s][g]-\speed[s][l])^{2} \diam[l]}{\capillarity}.
\end{align}

Other flow parameters, that are used for liquid atomization of cylindrical jet can be adapted as well, such as the liquid Reynolds number, $Re_{L} = \density[l] \speed[s][l] \diam[l] / \mu_{L}$, the liquid Weber Number, $We_{L} = \density[l] \speed[s][l]^{2} \diam[l] / \capillarity$. However, they are less important for air-blast atomizer because, for high momentum ratio $M$, the effect of the dynamics of the liquid jet is very limited. The atomization process is then mostly driven by the shear that is induced by high gas velocity with respect to the liquid velocity. On the contrary, the dynamics of the gaseous flow is determinant in particular in the vicinity of liquid-gas layer. Thus, the gas longitudinal velocity profile normal to the issued liquid-gas surface is determinant for the development of the first instabilities \citeay{Lozano_2001}. For experiments, it is of primary importance to characterize the gas boundary layer that depend on the geometry of the air-blast injector upstream of the injection location.  

From these initial Kelvin-Helmholtz instabilities, if the liquid thickness is small, a coupling occurs between instabilities on both sides of the liquid sheet leading to the flag effect. From this feature shared by most of liquid sheet atomization systems based on air blast, several processes are involved in the disintegration of the liquid sheet. For instance, the liquid sheet flapping induces variation of the liquid thickness. In the stretched part, this thinness can induce perforation of the liquid sheet leading to the emergence of a hole surrounded by a liquid rim. The accumulation of liquid within these rims or in other part of the liquid sheet under the effect of surface tension can create ligaments, which are finally destabilized by Rayleigh-Plateau instabilities. In many industrial applications, to reach high injection rate with an efficient atomization, the flow conditions are such that turbulence is induced at least in the gas flow. The turbulent nature of the two-phase flow complicates then the previously described instabilities, adding a transient stretching effect and a possible collision between liquid elements. 
This is where the numerical simulation of the entire flow is supposed to help the design of injection systems. Accordingly, the numerical test case has been designed to be as close as possible of these typical conditions. The limitation comes from the mesh resolution that should be fine enough to capture most of these phenomena but still limited to offer an accessible CPU time. 

To design the appropriate test case we refer to previous study that have classified break-up mechanism with respect to velocity ratio \citeay{Lozano_2001,Carvalho_2002}. At low speed ratio, the liquid sheet oscillates with a potential low amplitude growth of the liquid core. Moreover the atomization happens following streamwise and spanwise ligament break-up and the spray angle is particularly low. It is called the \emph{cellular breakup regime}. This regime would be very challenging for computation requiring a very long computational domain together with high mesh resolution to capture the decreasing liquid thickness. Then at medium ratio, vorticity becomes more important and the breakup is controlled by streamwise liquid structure detaching from the principal liquid core. The high amplitude of the sinusoidal wave induces a high spray angle. This is the \emph{streamwise ligaments breakup} \citeay{Samuelsen_1990}. Finally, very high $\speed[s][g]/\speed[s][l]$ brings to a complete and immediate disintegration of the liquid sheet. In \citeay{Dumouchel_2008}, the authors also observed that high relative velocity usually leads to strong sinusoidal oscillations of the liquid flow inducing a high spray angle and short sheet breakup length, called \emph{flag-effect} \citeay{Dumouchel_2008}.

These last two observations indicate that a high velocity ratio is suitable to promote an efficient atomization with a quite large spray angle in order to reduce the required length of the computational domain.  High velocity ratio also increases the momentum ratio which is beneficial  to reduce the liquid core length at least in cylindrical liquid jet atomization \citeay{Lasheras_1998,Porcheron_2002,Leroux_2007}. In addition, having high velocity and momentum ratio will lead to simulations closer to most industrial application. The limitation comes from the turbulence that leads to smaller length scale as the Reynolds increases.  Similarly, it is expected that the final droplet size decreases as the velocity ratio increases, leading to higher Weber number. It has been chosen not to inject turbulent fluctuations of velocity at the inlet, though this may seem less realistic for real applications, it allows for more deterministic comparison between numerical approaches and postpones slightly the creation of small length scales. Finally, the density ratio has to be high enough to ensure a realistic representation of a liquid and a gas phase, typically of a thousandth for air and liquid water at atmospheric pressure. The lower is the density ratio, the less is the interaction between phases again leading to a less efficient atomization. In engines, the gas is very often pressurized leading to a density ratio of the order of a hundredth. 
Based on these principles, the present configuration has been set by adjusting flow parameters in order to promote a fast atomization and limit the liquid core penetration, while having a density ratio realistic for aircraft engines. A reasonably high momentum $M=18.7$ is obtained with a moderately high speed ratio, $\speed[s][g]/\speed[s][l] = 43$  and a density ratio $\ffrac{\rho_g}{\rho_l}=1/100$. To reduce the role played by the surface tension, since it is not solved by the diffuse interface model, we have chosen a high relative gaseous Weber number, $We_{R} = 403$. As we will see in Section~\ref{sec:result}, we obtain a regime resembling the \emph{streamwise ligaments breakup}, with a short liquid core, ligaments and an atomized spray.

\subsection{Geometry and description of various meshes}

\begin{figure}[h]
\centering
  \subfloat[Limits and initial boundary conditions.]{
\begin{overpic}[width=0.3\textwidth]{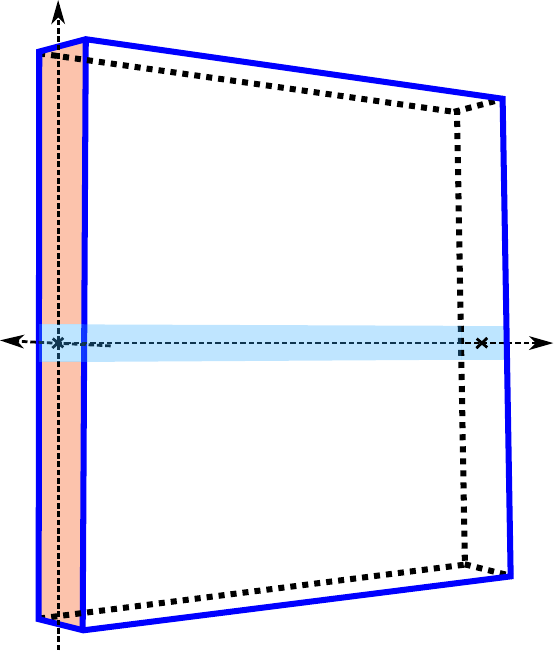} %
         \put(41,0){ \tikz[inner sep=.40ex,baseline=-.75ex] \node[circle,draw] {$4$};}
         \put(41,94){ \tikz[inner sep=.40ex,baseline=-.75ex] \node[circle,draw] {$5$};}
         \put(45,65){ \tikz[inner sep=.40ex,baseline=-.75ex] \node[circle,draw] {$3$};}
         \put(37,62){ \tikz[inner sep=.40ex,baseline=-.75ex] \node[circle,dashed, draw] {$3$};}
         \put(-10,46){ \tikz[inner sep=.40ex,baseline=-.75ex] \node[circle, draw] {$1$};}
         \put(-4,25){ \tikz[inner sep=.40ex,baseline=-.75ex] \node[circle, draw] {$2$};}
         \put(-4,75){ \tikz[inner sep=.40ex,baseline=-.75ex] \node[circle, draw] {$2$};}
         \put(79,65){ \tikz[inner sep=.40ex,baseline=-.75ex] \node[circle, draw] {$6$};}
         \put(11,96){$z$}
         \put(82,42){$x$}
         \put(0,42){$y$}
         \put(10,42){$0$}
       \end{overpic}
\label{fig:planarjet_domain}
}
\hspace{4em}
\subfloat[Injector schematic and velocity profiles.]{
\begin{overpic}[width=0.3\textwidth]{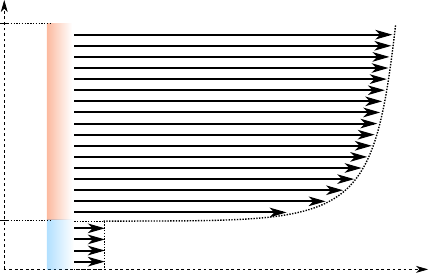} %
         \put(30,6){$\speed[s][l]$}
         \put(95,45){$\speed[s][g]$}
         \put(-10,-1){$ 0 $}
         \put(-11,10){$\dfrac{\diam[l]}{2}$}
         \put(-11,57){$\dfrac{\diam[g]}{2}$}
         \put(0,65){$z$}
         \put(102,0){$x$}
       \end{overpic}
\label{fig:velocity_profile}       
}
\caption{Geometry, limits and boundary conditions of the configuration.}
\end{figure}

The simulation domain described in Figure~\ref{fig:planarjet_domain} %
shows the boundary conditions of the simulation. \tikz[inner sep=.20ex,baseline=-.75ex] \node[circle, draw] {\footnotesize$1$}; is the liquid injection plan, %
\tikz[inner sep=.20ex,baseline=-.75ex] \node[circle, draw] {\footnotesize$2$}; is the gas injection plan, %
\tikz[inner sep=.20ex,baseline=-.75ex] \node[circle, draw] {\footnotesize$3$}; and %
\tikz[inner sep=.20ex,baseline=-.75ex] \node[circle, draw, dashed] {\footnotesize$3$}; are periodic plans, %
\tikz[inner sep=.20ex,baseline=-.75ex] \node[circle, draw] {\footnotesize$4$}; , %
\tikz[inner sep=.20ex,baseline=-.75ex] \node[circle, draw] {\footnotesize$5$};. %
\tikz[inner sep=.20ex,baseline=-.75ex] \node[circle, draw] {\footnotesize$6$}; are exit boundary conditions, which are implemented differently in CEDRE and ARCHER. For CEDRE, they are compressible subsonic outflow plans where the atmospheric pressure is imposed. Since no reliable theory of non-reflecting subsonic outflow boundary conditions for compressible two-phase flow exists in the spirit of compressible single phase flow \citeay{Poinsot_1992}, reflecting  waves can occur at the outflow plans. Nonetheless, as we will see in the next paragraph, the computational domain for the compressible solver has been extended and coarsened to impede these reflecting waves. For the ARCHER code, an outflow boundary condition has been applied on all the variables with a zero normal gradient condition except for the pressure, which is set to the constant atmospheric value. Moreover, for the incompressible solver, to confirm that the outflow boundary conditions we apply on the exit planes do not produce any bias on the results presented in the paper, supplementary computations on twice as long computational domain in the $x$ direction have been conducted and no noticeable modification of the results were to be found. The liquid injection height, $\diam[l]$, the gas injection height, $\diam[g]$, the x-y-z length of the box, $L_{x}$, $L_{y}$, $L_{z}$, are given in Table~\ref{table:box_dim}. Dimensionless lengths $x^*$, $y^*$, $z^*$ are defined as $x^*=\ffrac{x}{\diam[l]}$, $y^*=\ffrac{y}{\diam[l]}$ $z^*=\ffrac{z}{\diam[l]}$.

\begin{table}[h]
\renewcommand{\arraystretch}{1,5}
\centering
\subfloat[Dimensions]{
    \begin{tabular}{c|c|c|c|c|c}
     	units & $\diam[l]$ & $\diam[g]$   & $L_{x}$	& $L_{y}$ & $L_{z}$\\ \hline
     	$(mm)$ & 1   & 16	& 16 & 4 & 16 
    \end{tabular}
    }
\hspace{4em}
\subfloat[Fluids physical properties]{
    \label{table:physical_parameters}
    \begin{tabular}{c|c|c|c|c}
     	Phase   	& $\rho$ $(kg/m^{3})$ 	& $\pressure$ $(MPa)$	& $\capillarity$ $(N/m)$ 		& $\mu$ $(Pa.s)$ \\ \hline
        Liquid		& $100$	 			& 0.1					& $0.01$ 				& $0.0001$ \\ \hline
        Gas			& $1$ 					& 0.1					& $0.01$ 				& $0.0001$
    \end{tabular}
}
    \caption{Dimensions and physical properties of the configuration.}
    \label{table:box_dim}
\end{table}

Figure~\ref{fig:velocity_profile} defines the velocity profile of the gas and the liquid, which are symmetric with respect to the $x$-axis and $y$-invariant.
The gas velocity profile $\speed[s][g]$ given in Equation~\eqref{eq:velocity_profile} is typical for turbulent pipe flow \citeay{Schlichting_1979}.
\begin{align}\label{eq:velocity_profile}
    \speed[s][g] = \speed[s][g]^{avg} \frac{7}{6} \qty( \frac{2\abs{z}-\diam[l]}{\diam[g]-\diam[l]})^{\frac{1}{6}}+\speed[s][l].
\end{align}
The average gas velocity $\speed[s][g]^{avg}$ is $65\, m/s$. An offset equal to $\speed[s][l] = 1.5 \, m/s$ ensures the continuity of the velocity profile at the injection plan. The domain is initially filled with a liquid sheet in the $x-y$ plan, as thick as the liquid slit as shown in Figure~\ref{fig:planarjet_domain}. Table~\ref{table:physical_parameters} states the fluid properties in terms of density $\rho$, surface tension coefficient, $\capillarity$, and viscosity, $\mu$.

The ARCHER simulations are performed on a Cartesian mesh $512 \, \times \, 128  \, \times \,  512$ with a cell size equal to $\Delta x =3.125 \, 10^{-5} \, m$, so a total of $101$M faces, $33.6$M cells and $32$ cells in the liquid slit. In terms of degrees of freedom, which is defined as the product of the number of variables solved and the number of cells, ARCHER solves $42.0$M. CEDRE simulations have been performed on two meshes composed of tetrahedral cells. The first, referred later on as \emph{CEDRE (MR)}, proposes a medium refinement level with $148$k faces, $71.7$k cells and $788$k degrees of freedom, and the second, referred later on as \emph{CEDRE (HR)}, a high refinement level with $1.11$M faces, $546$k cells, thus $6.01$M degrees of freedom. \emph{CEDRE (MR)} has only $10$ cells in the slit, while \emph{CEDRE (HR)} has $20$ cells. The data are summarized in Table~\ref{table:mesh_data} and~\ref{table:CPU_info}.
\begin{table}[h]
\renewcommand{\arraystretch}{1,5}
\centering
    \begin{tabular}{c|c|c|c}
     				& Face number	& Cell number	& $\min{(\Delta x)}/\diam[l]$ 		\\ \hline
        CEDRE (MR)	&	$0.148$M	& $71.7$k		& $10$   	\\ \hline
        CEDRE (HR)	&	$1.11$M 	& $546$k		& $20$		\\ \hline
        ARCHER		&	$101$M		& $33.6$M		& $32$ 	
    \end{tabular}
    \caption{Mesh statistics of the three simulations.}
    \label{table:mesh_data}
\end{table}
In terms of degrees of freedom, there is a factor $7.6$ between the two CEDRE simulations, and similar factor, $7.0$, between the high refined CEDRE simulation and the DNS. The three simulations all together encompass thus a large level of refinements. 

\begin{figure}[h]
\centering
\subfloat[CEDRE mesh overview.]{%
       \centering
  \includegraphics[width=0.25\textwidth]{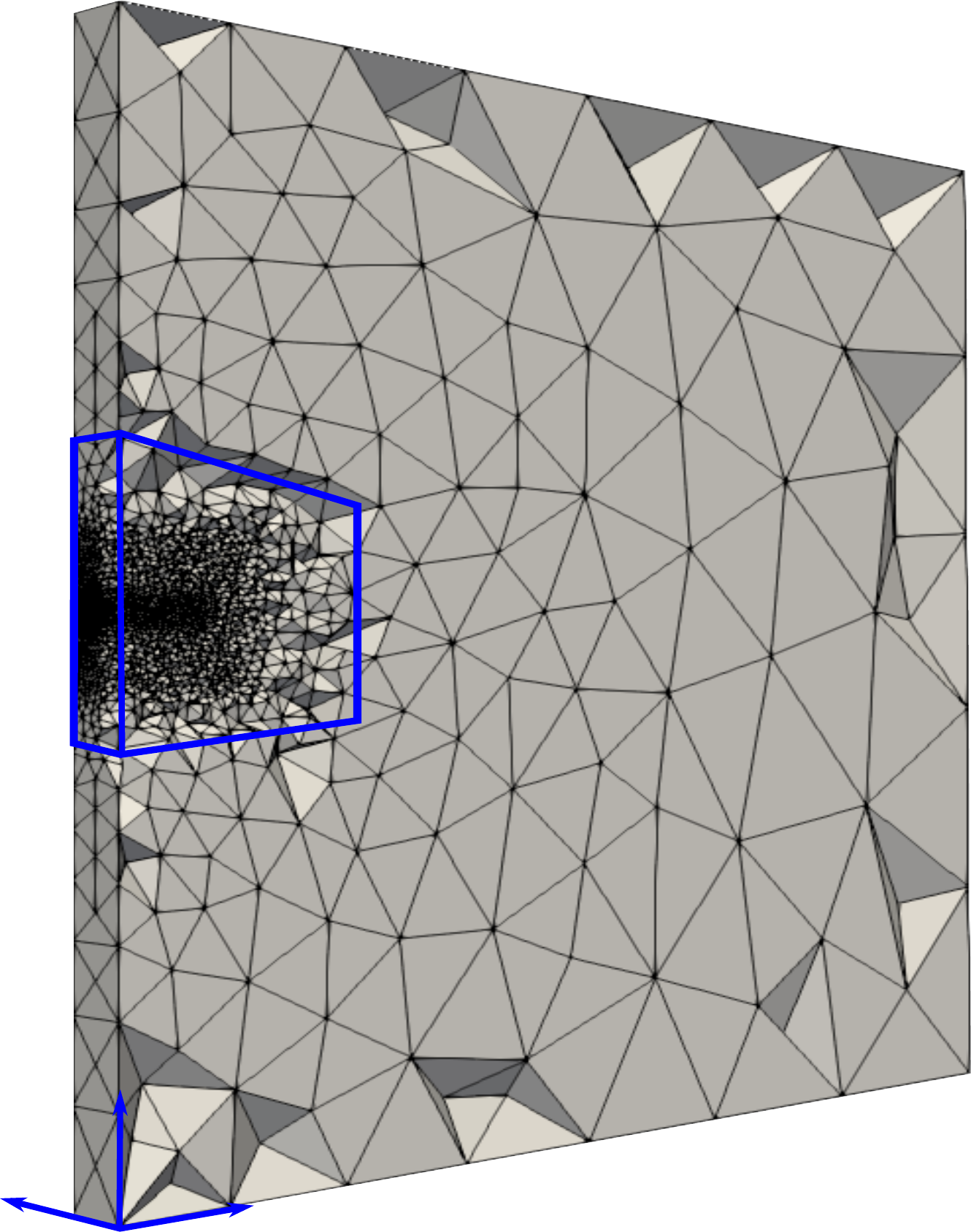} %
  \label{fig:cedre_mesh_all}
}
\hspace{6em}
\subfloat[Zoom on the refined DNS box.]{%
       \centering
  \includegraphics[width=0.25\textwidth]{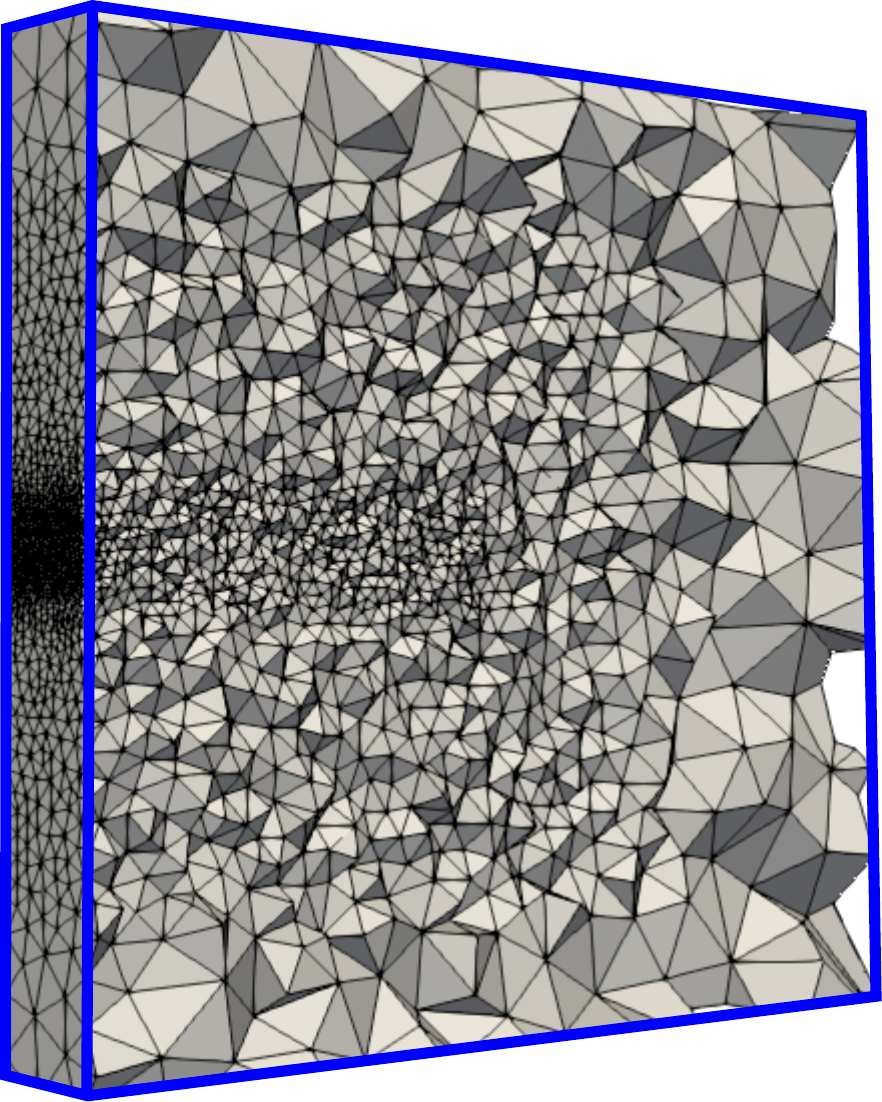} %
  \label{fig:cedre_mesh_DNS}
}
\caption{Clip of the CEDRE mesh on the plane $y=0$ but keeping whole elements.}
\label{fig:cedre_mesh}
\end{figure}

In order to compare the results of the DNS to the results obtained with CEDRE, one must consider the fact that the DNS solver is incompressible, thus there is no acoustic impacting the liquid sheet and its density is constant. To restrain the acoustic role in the CEDRE compressible solver, we have enlarged the computational domain by a factor $5$ in the $x$ and $z$ direction and meshed it with a very coarse mesh to avoid any reflecting waves as shown in Figure~\ref{fig:cedre_mesh}.
The minimum cell size is located along the liquid slit as one notices on Figure~\ref{fig:cedre_mesh_DNS}.

\begin{figure}[!h]\centering
\begingroup
  \makeatletter
  \providecommand\color[2][]{%
    \GenericError{(gnuplot) \space\space\space\@spaces}{%
      Package color not loaded in conjunction with
      terminal option `colourtext'%
    }{See the gnuplot documentation for explanation.%
    }{Either use 'blacktext' in gnuplot or load the package
      color.sty in LaTeX.}%
    \renewcommand\color[2][]{}%
  }%
  \providecommand\includegraphics[2][]{%
    \GenericError{(gnuplot) \space\space\space\@spaces}{%
      Package graphicx or graphics not loaded%
    }{See the gnuplot documentation for explanation.%
    }{The gnuplot epslatex terminal needs graphicx.sty or graphics.sty.}%
    \renewcommand\includegraphics[2][]{}%
  }%
  \providecommand\rotatebox[2]{#2}%
  \@ifundefined{ifGPcolor}{%
    \newif\ifGPcolor
    \GPcolortrue
  }{}%
  \@ifundefined{ifGPblacktext}{%
    \newif\ifGPblacktext
    \GPblacktextfalse
  }{}%
  \let\gplgaddtomacro\g@addto@macro
  \gdef\gplbacktext{}%
  \gdef\gplfronttext{}%
  \makeatother
  \ifGPblacktext
    \def\colorrgb#1{}%
    \def\colorgray#1{}%
  \else
    \ifGPcolor
      \def\colorrgb#1{\color[rgb]{#1}}%
      \def\colorgray#1{\color[gray]{#1}}%
      \expandafter\def\csname LTw\endcsname{\color{white}}%
      \expandafter\def\csname LTb\endcsname{\color{black}}%
      \expandafter\def\csname LTa\endcsname{\color{black}}%
      \expandafter\def\csname LT0\endcsname{\color[rgb]{1,0,0}}%
      \expandafter\def\csname LT1\endcsname{\color[rgb]{0,1,0}}%
      \expandafter\def\csname LT2\endcsname{\color[rgb]{0,0,1}}%
      \expandafter\def\csname LT3\endcsname{\color[rgb]{1,0,1}}%
      \expandafter\def\csname LT4\endcsname{\color[rgb]{0,1,1}}%
      \expandafter\def\csname LT5\endcsname{\color[rgb]{1,1,0}}%
      \expandafter\def\csname LT6\endcsname{\color[rgb]{0,0,0}}%
      \expandafter\def\csname LT7\endcsname{\color[rgb]{1,0.3,0}}%
      \expandafter\def\csname LT8\endcsname{\color[rgb]{0.5,0.5,0.5}}%
    \else
      \def\colorrgb#1{\color{black}}%
      \def\colorgray#1{\color[gray]{#1}}%
      \expandafter\def\csname LTw\endcsname{\color{white}}%
      \expandafter\def\csname LTb\endcsname{\color{black}}%
      \expandafter\def\csname LTa\endcsname{\color{black}}%
      \expandafter\def\csname LT0\endcsname{\color{black}}%
      \expandafter\def\csname LT1\endcsname{\color{black}}%
      \expandafter\def\csname LT2\endcsname{\color{black}}%
      \expandafter\def\csname LT3\endcsname{\color{black}}%
      \expandafter\def\csname LT4\endcsname{\color{black}}%
      \expandafter\def\csname LT5\endcsname{\color{black}}%
      \expandafter\def\csname LT6\endcsname{\color{black}}%
      \expandafter\def\csname LT7\endcsname{\color{black}}%
      \expandafter\def\csname LT8\endcsname{\color{black}}%
    \fi
  \fi
    \setlength{\unitlength}{0.0500bp}%
    \ifx\gptboxheight\undefined%
      \newlength{\gptboxheight}%
      \newlength{\gptboxwidth}%
      \newsavebox{\gptboxtext}%
    \fi%
    \setlength{\fboxrule}{0.5pt}%
    \setlength{\fboxsep}{1pt}%
\begin{picture}(5102.00,2834.00)%
    \gplgaddtomacro\gplbacktext{%
      \csname LTb\endcsname
      \put(0,550){\makebox(0,0)[r]{\strut{}\small 94}}%
      \put(0,845){\makebox(0,0)[r]{\strut{}\small 95}}%
      \put(0,1139){\makebox(0,0)[r]{\strut{}\small 96}}%
      \put(0,1434){\makebox(0,0)[r]{\strut{}\small 97}}%
      \put(0,1729){\makebox(0,0)[r]{\strut{}\small 98}}%
      \put(0,2024){\makebox(0,0)[r]{\strut{}\small 99}}%
      \put(0,2318){\makebox(0,0)[r]{\strut{}\small 100}}%
      \put(0,2613){\makebox(0,0)[r]{\strut{}\small 101}}%
      \put(132,330){\makebox(0,0){\strut{}\small 0}}%
      \put(669,330){\makebox(0,0){\strut{}\small 1}}%
      \put(1207,330){\makebox(0,0){\strut{}\small 2}}%
      \put(1744,330){\makebox(0,0){\strut{}\small 3}}%
      \put(2282,330){\makebox(0,0){\strut{}\small 4}}%
      \put(2819,330){\makebox(0,0){\strut{}\small 5}}%
      \put(3357,330){\makebox(0,0){\strut{}\small 6}}%
      \put(3894,330){\makebox(0,0){\strut{}\small 7}}%
      \put(4432,330){\makebox(0,0){\strut{}\small 8}}%
      \put(4969,330){\makebox(0,0){\strut{}\small 9}}%
    }%
    \gplgaddtomacro\gplfronttext{%
      \csname LTb\endcsname
      \put(-374,1581){\rotatebox{-270}{\makebox(0,0){\strut{}$\text{\small $\density[l] \, [kg/m^{3}]$ }$}}}%
      \put(2550,176){\makebox(0,0){\strut{}\small $\Time[sim]$ $[$ms$]$}}%
    }%
    \gplbacktext
    \put(0,0){\includegraphics{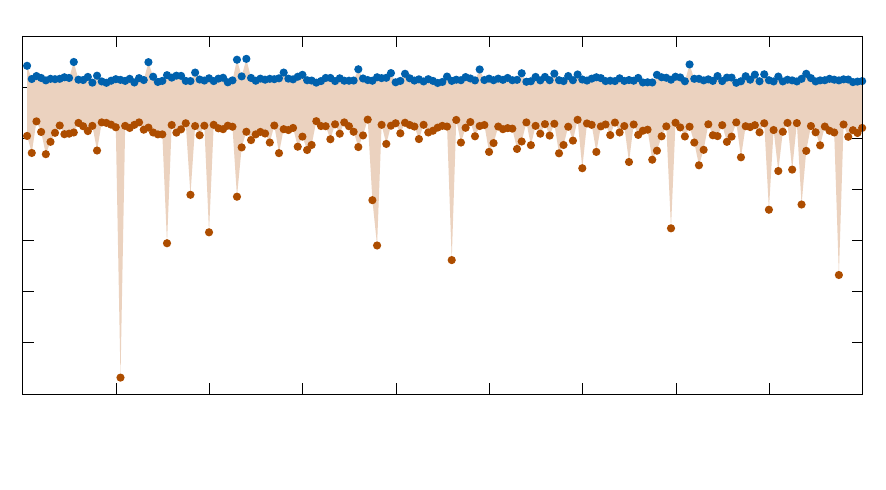}}%
    \gplfronttext
  \end{picture}%
\endgroup
\caption{CEDRE (MR) instantaneous liquid density $\density[l]$ over time, \fillcircle{Blueone} maximum value, \fillcircle{Redone} minimum value.}
\label{plt:liquid_density}
\end{figure}

Furthermore the use of a compressible thermodynamics in CEDRE through the Stiffened-Gas equation of state makes it impossible to maintain the liquid density constant. The temperature of the phases have been modified to obtain the same initial pressure and density conditions as in Table~\ref{table:physical_parameters} and in practice, the liquid density almost stays constant as shown on Figure~\ref{plt:liquid_density}.

The convection time $t_{conv}$ of the system is defined as $t_{conv} = \ffrac{L_{x}}{\speed[s][l]}$. The liquid is quickly accelerated and flows at an averaged speed of $\speed[s][l] \approx 15 \, m/s$ at $x= 3\, mm$ as we will see in Section~\ref{sec:result} in Figure~\ref{fig:ux_0.03t}. Hence the convection time is approximately equals $t_{conv} \approx 1 \, ms$. The minimum simulation time of the three simulations is $8 \, ms$, which corresponds to approximately height convective times, yielding a priori satisfying statistical convergences for each simulation. The simulation information are summarized in Table~\eqref{table:CPU_info}.
\begin{table}[ht]
\renewcommand{\arraystretch}{1,5}
\centering
    \begin{tabular}{c|c|c|c|c}
     				& Total $\Time[sim]$	& CPU cost for $\Time[sim] = 9 \, ms$  $[h]$ 	& Degrees of freedom	\\ \hline
        CEDRE (MR)	&	18				 	& $8.54 \, 10^{3}$  	& $0.788$M 	 \\ \hline
        CEDRE (HR)	&	10.5 				& $69.3 \, 10^{3}$		& $6.01$M 	 \\ \hline
        ARCHER		&	14					& $245 \, 10^{3}$ 		& $42.0$M	
    \end{tabular}
    \caption{Simulation time and costs comparisons.}
    \label{table:CPU_info}
\end{table}
The total CPU cost is defined as the product of the number of processors, $N_{proc}$, times the computational time for a given simulation time, $\Time[sim]$. The three simulations have not run in overall the same amount of simulation time, due to time, ressource and statistical convergence constraints. Therefore, only partial comparison in terms of CPU costs can be proposed. For $\Time[sim] = 9 \, ms$, we see that ARCHER is $3.5$ times more costly than CEDRE (HR), the latter being $8.1$ times more costly than CEDRE (MR). However, as we will see in Section~\ref{sec:result}, time simulation needed to gain statistical convergence differs from one simulation to another. In overall $CEDRE$ offers quicker stastical convergence than ARCHER with a significant reduction of degrees of freedom, highlighting again the interest of reduced-order model simulations for industrial configurations.

\begin{figure}[h]
\centering
\includegraphics[width=1.0\textwidth]{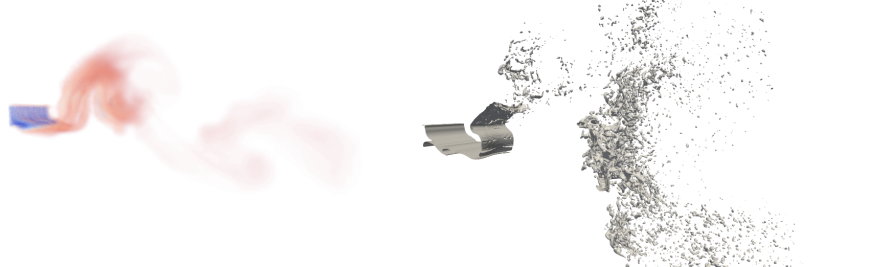} %
\caption{Instantaneous overview of the simulations: volume rendering of the liquid volume fraction, $\vol[frac][l]$ high \protect \includegraphics[height=0.7em, width=5em]{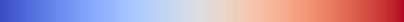} $\ $ low (CEDRE (HR) - left) and level set in grey (ARCHER - right).}
\label{fig:instant_volfrac_3D_comparison}
\end{figure}

Before moving on to Section~\ref{sec:result}, where we analyze and discuss the results obtained with the DNS and the diffuse interface model, Figure~\ref{fig:instant_volfrac_3D_comparison} offers a global overview of the simulations performed with ARCHER and CEDRE. On the left hand side of Figure~\ref{fig:instant_volfrac_3D_comparison}, we have drawn a volume rendering of the liquid volume fraction $\vol[frac][l]$ obtained with the CEDRE (HR) simulation. On the right hand side is shown the level set solved by ARCHER, indicating the position of the interface.

\section{Results and discussion}\label{sec:result}

We propose a qualitative and quantitative comparison of the results of the simulations performed with CEDRE and ARCHER in order to, first, analyze the behavior of each model, second, emphasize the main conceptual differences between them and, finally, evaluate to which extent CEDRE reduced-order model is predictive and to what extent it is departing from the DNS results.

This section is organized as follows: we start by assessing the DNS reliability to identify the most relevant region where the DNS can stand as a trustworthy benchmark. Then, we provide a comparison of the atomization global characteristics obtained by each simulations through the volume fraction time evolution and averaging. We further detail the comparison with statistical analyses of the flow through first and second order moments based on the liquid volume fraction and velocities.
\subsection{Identification of the DNS region of validity}

The problem of under resolution is a real issue in general for simulation of turbulent gas-liquid interfacial flows. The resolution required to solve completely most of atomization processes, even with moderate Weber number, is often skyrocketing. Solving entirely a two-phase flow of an atomization process would require several mesh cell per diameter of any droplet, a condition that is hardly met in practice and leads to the question of mesh convergence. Even with the most intensive computational effort \citeay{Shinjo_2010}, the full resolution is not necessarily guaranteed and statistical analysis is often intractable. The strategy developed in the present work is to reduce the zone of analysis to a restricted area where the mesh resolution is high enough to accurately capture the two-phase flow. This is among the first attempts to determine which part of the flow is well-resolved based on numerical criteria characterizing the interface resolution quality (IRQ). This approach is based on previous work, on a simplified problem \citeay{Demoulin_2013} where the statistical analysis of this IRQ criteria has been carried out by comparing several level of grid refinement leading to mesh convergence. The main advantage of this criterion is to evaluate the quality of complex configuration simulations, such as the present one, without running the simulation with several levels of mesh resolution to achieve a proper mesh convergence study. This criterion is defined as $IRQ_K=\ffrac{1}{( \Delta x \, 2 \, \meancurv)}$, where $\Delta x$ is the grid size and $\meancurv$ the mean curvature. This criterion highlights the under-resolved regions of the simulation where its value diminishes. For instance, a droplet described with 4 mesh cells along its radius has an $IRQ_K=2$. Therefore, the lower the IRQ is, the lower is the resolution of the liquid (or gas) structure highlighted. Note that the threshold value of $2$ chosen here is arbitrary and the user is expected to choose a value that is the most adequate to a given simulation. Instead of investigating the individual $IRQ_K$ for each structure, the $IRQ_K$ PDF can be studied to obtain a better picture of the resolution level of the overall simulation.

\begin{figure}[h]
\centering
\subfloat[{$x^* \in \qty[ 0,4]$ region.}]{%
       \centering
\begin{overpic}[width=0.39\textwidth]{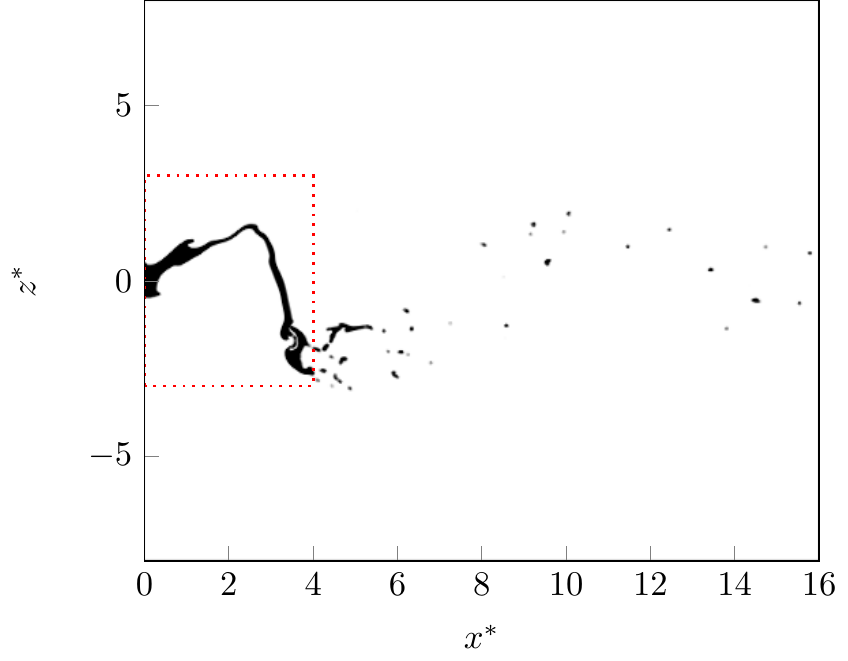} %
       \end{overpic}  
  \label{fig:seq_curv_05}
}
\hspace{4em}
\subfloat[{$x^* \in \qty[ 4,8]$ region.}]{%
       \centering
\begin{overpic}[width=0.35\textwidth]{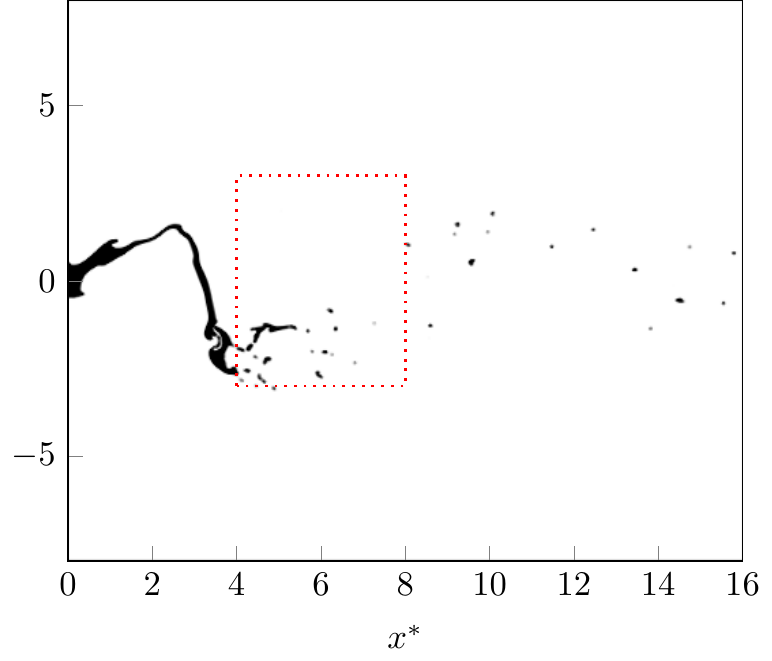} %
       \end{overpic}  
  \label{fig:seq_curv_510}
}
\\

\subfloat[{$IRQ_K$ PDF in the $x^* \in \qty[ 0,4]$ region.}]{%
\begin{overpic}[width=0.35\textwidth]{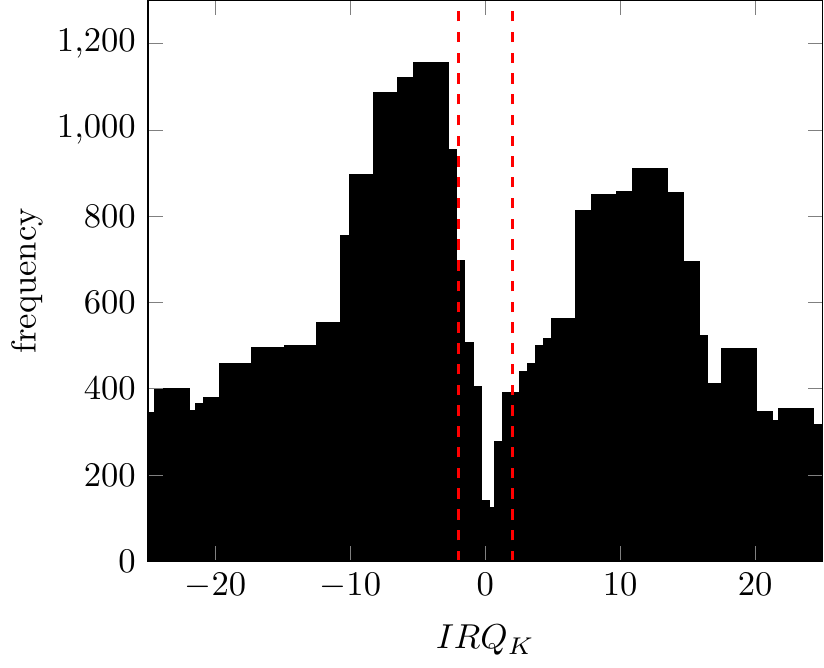} %
       \end{overpic}  
  \label{fig:IRQ_curv_05}
}
\hspace{4em}
\subfloat[{$IRQ_K$ PDF in the $x^* \in \qty[ 4,8]$ region.}]{%
\begin{overpic}[width=0.32\textwidth]{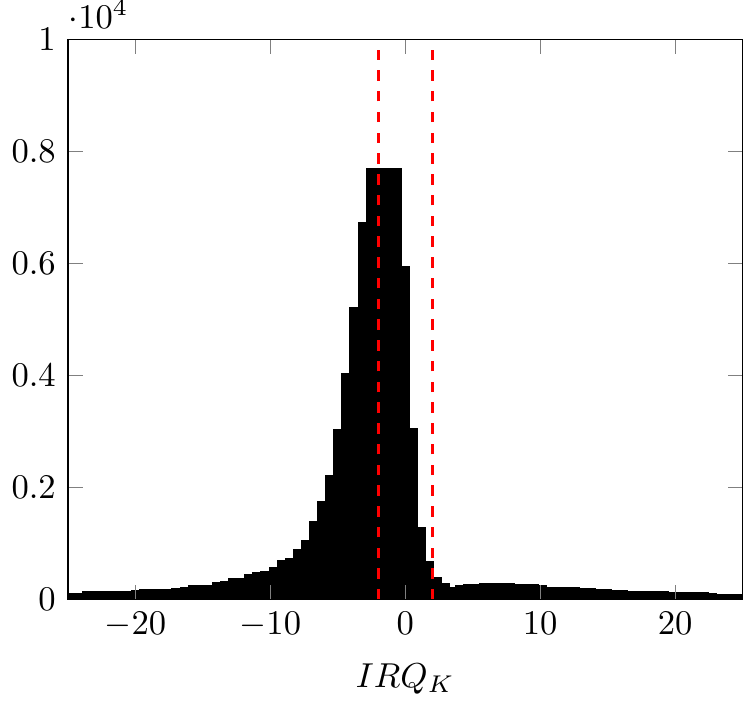} %
\end{overpic}  
  \label{fig:IRQ_curv_510}
}
\caption{{$IRQ_k$ PDF analysis at time $t=12 \, ms$ (bottom), in two regions identified by the red dashed rectangles drawn on the plan $y^*=0$ showing liquid volume fraction $\vol[frac][l]$ (top).}}
\end{figure}

Since the DNS results have to be reliable to assess the predictiveness of the reduced order model, an analysis of the $IRQ_K$ criterion is first performed. The $IRQ_K$ is extracted from two specific regions of the jet: in the main liquid core, where the criterion should behave perfectly since most wrinkling of the interface are expected to be captured, and in the most critical secondary atomization region, where liquids structures encounter fragmentation/breakup processes that can lead to under-resolved liquids structures. Theses two regions of interest are illustrated in Figures~\ref{fig:seq_curv_05}, \ref{fig:seq_curv_510}. 

The PDF of $IRQ_K$, shown in Figures~\ref{fig:IRQ_curv_05}, \ref{fig:IRQ_curv_510}, gives useful information about the quality of the simulation. In the first region of the liquid sheet, most of the IRQ values of the PDF (around $95\%$) are located beyond the critical zone of low IRQ, $IRQ_K \in [-2:2]$, as shown in Figure~\ref{fig:IRQ_curv_05}. This range is indicated by two red dotted lines in the figure. As expected, the resolution is sufficient here to describe the main physical phenomena such as instabilities on the liquid surface and the flapping mechanism. However, in the second region, Figure~\ref{fig:IRQ_curv_510}, the PDF is compressed close to the critical $IRQ_K$ zone, hence only around $60\%$ of the interface is well resolved. It indicates that the DNS results should be taken with caution for large $x^*$. This can be explained by the production of droplets of small scale due to atomization of the flapping sheet. Note that this kind of simulations at high Weber and Reynolds numbers are quite challenging in term of computational costs, explaining the difficulty to reach higher IRQ in the most atomized or dispersed region. Besides, we specifically have chosen two different regions with opposite behavior to assess the pertinence of the criterion.

Consequently, the DNS appears sufficiently trustworthy until the jet become more dispersed, due to the presence of very small droplets. In addition, in the present work, CEDRE simulation has not been coupled with any specialized reduced-order model, such as KBMM, to better describe the dispersed flow. It is unlikely the diffuse interface model alone will capture properly this part of the flow. The following analysis will therefore be conducted in the zone $x^* \in \qty[ 0,4]$ along the streamwise axis where the DNS can be considered as the reference simulation.

\subsection{Evaluation of the atomization global characteristics of both numerical approaches}

To begin with, we would like to evaluate the macroscopic features of the investigated flow. We first analyze the time evolution of the liquid volume fraction in order to reveal the expected flag effect mentioned in Section~\ref{sec:config}. Then we interest ourselves to the liquid core penetration as it is usually a key feature measured in experiments. 

\subsubsection{Time evolution}

\captionsetup[subfloat]{labelformat=empty}
\begin{figure}[h]
\centering
\begin{minipage}{.25\textwidth}
\subfloat[$t_0$]{%
	\centering
\begin{overpic}[width=0.4\textwidth, frame, trim={6cm 1.4cm 1.6cm 1.5cm}, clip]{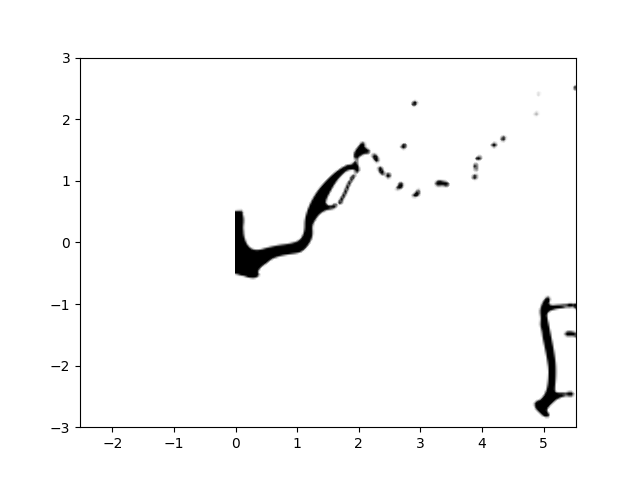}
       \end{overpic}  
  \label{fig:volfrac_ARCHER_t2375}
}
\subfloat[]{%
       \centering
\begin{overpic}[width=0.4\textwidth, frame, trim={6cm 1.4cm 1.6cm 1.5cm}, clip]{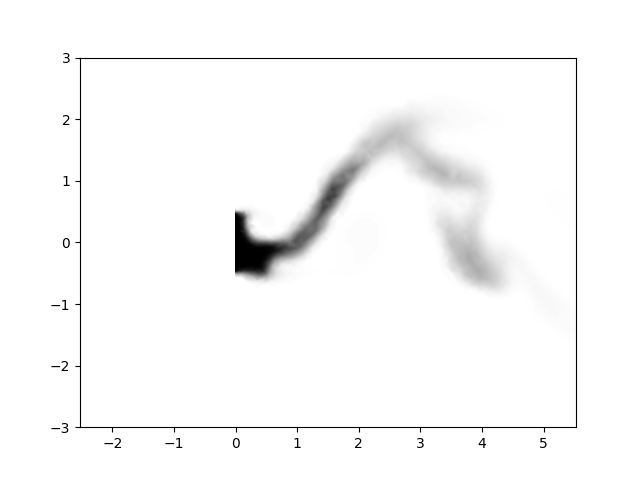}
       \end{overpic}  
  \label{fig:volfrac_CEDRE_t23.90}
}
\end{minipage}%
\begin{minipage}{.25\textwidth}
\subfloat[$t_0+\delta t$]{%
       \centering
\begin{overpic}[width=0.4\textwidth, frame, trim={6cm 1.4cm 1.6cm 1.5cm}, clip]{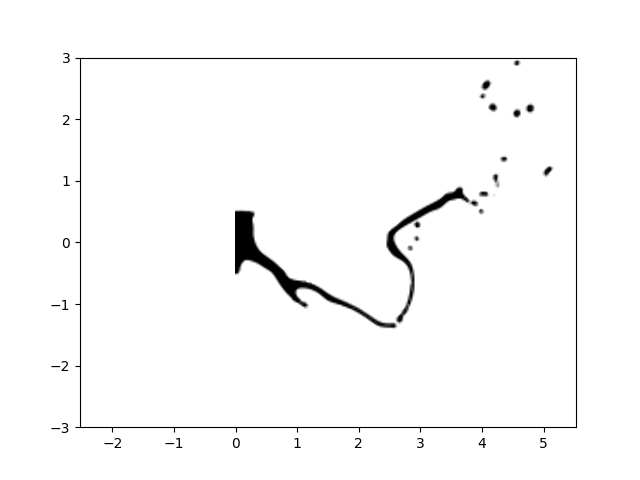}
       \end{overpic}  
  \label{fig:volfrac_ARCHER_t2380}
}
\subfloat[]{%
       \centering
\begin{overpic}[width=0.4\textwidth, frame, trim={6cm 1.4cm 1.6cm 1.5cm}, clip]{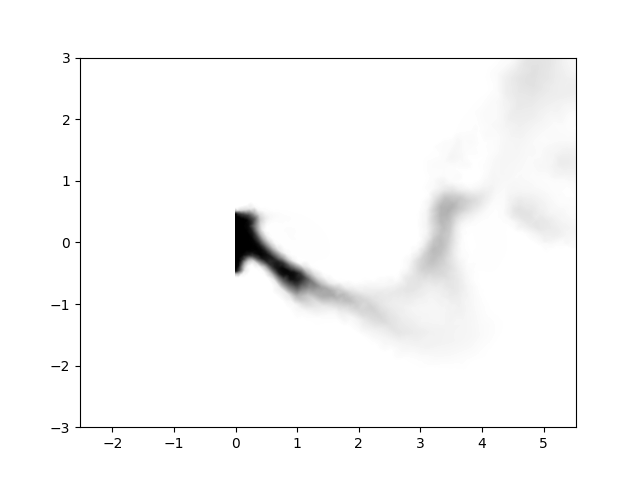}
       \end{overpic}  
  \label{fig:volfrac_CEDRE_t23.95}
}
\end{minipage}%
\begin{minipage}{.25\textwidth}
\subfloat[$t_0+2\delta t$]{%
       \centering
\begin{overpic}[width=0.4\textwidth, frame, trim={6cm 1.4cm 1.6cm 1.5cm}, clip]{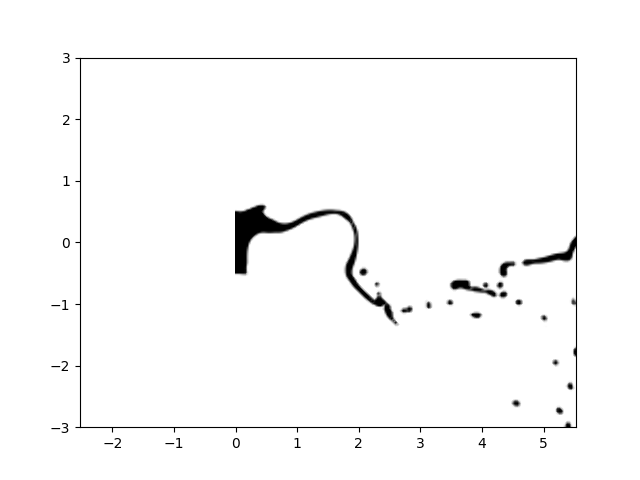}
       \end{overpic}  
  \label{fig:volfrac_ARCHER_t23.85}
}
\subfloat[]{%
       \centering
\begin{overpic}[width=0.4\textwidth, frame, trim={6cm 1.4cm 1.6cm 1.5cm}, clip]{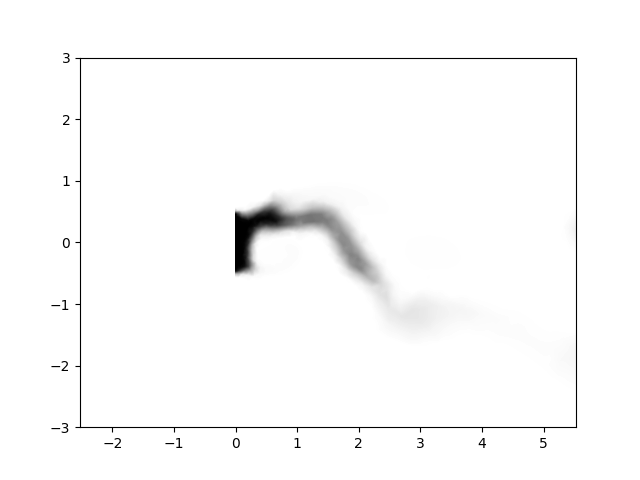}
       \end{overpic}  
  \label{fig:volfrac_CEDRE_t24.00}
}
\end{minipage}
\begin{minipage}{.25\textwidth}
\subfloat[$t_0+3\delta t$]{%
       \centering
\begin{overpic}[width=0.4\textwidth, frame, trim={6cm 1.4cm 1.6cm 1.5cm}, clip]{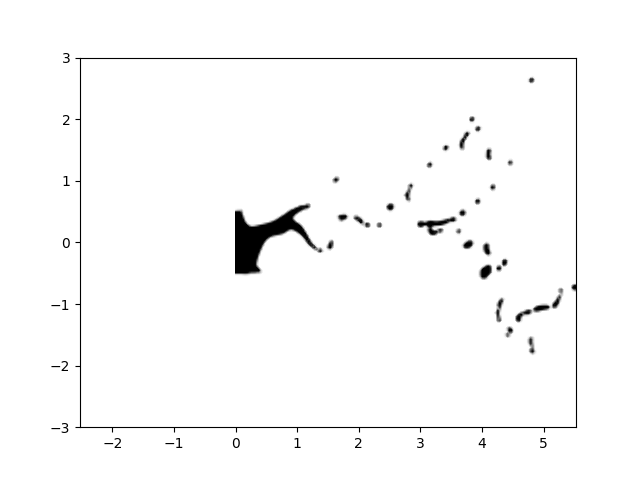}
       \end{overpic}  
  \label{fig:volfrac_ARCHER_t23.90}
}
\subfloat[]{%
       \centering
\begin{overpic}[width=0.4\textwidth, frame, trim={6cm 1.4cm 1.6cm 1.5cm}, clip]{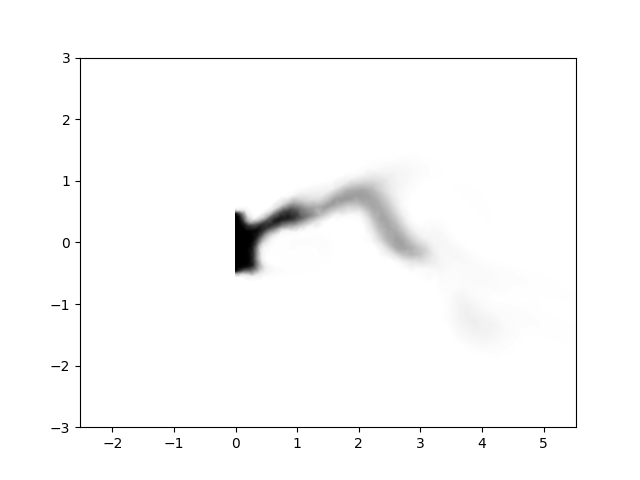}
       \end{overpic}  
  \label{fig:volfrac_CEDRE_t24.05}
}
\end{minipage}%
\begin{minipage}{.25\textwidth}
\subfloat[$t_0+4\delta t$]{%
       \centering
\begin{overpic}[width=0.4\textwidth, frame, trim={6cm 1.4cm 1.6cm 1.5cm}, clip]{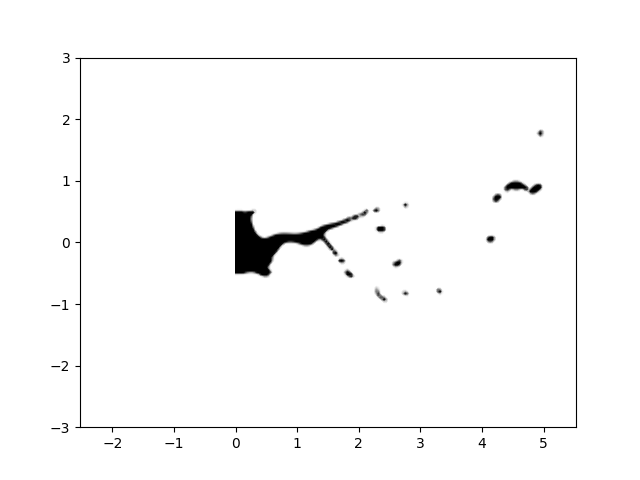}
       \end{overpic}  
  \label{fig:volfrac_ARCHER_t23.95}
}
\subfloat[]{%
       \centering
\begin{overpic}[width=0.4\textwidth, frame, trim={6cm 1.4cm 1.6cm 1.5cm}, clip]{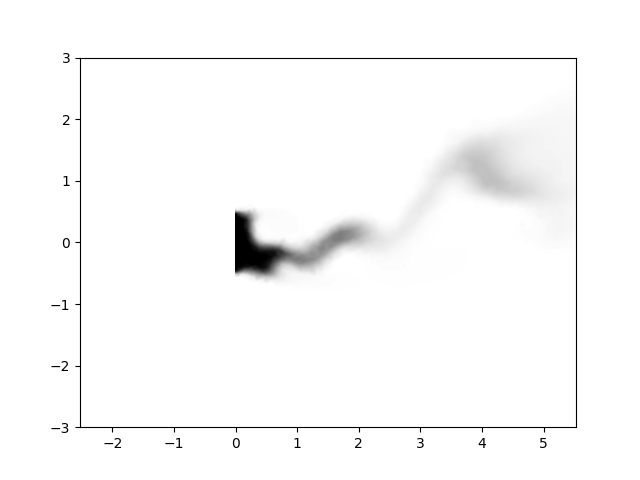}
       \end{overpic}  
       \label{fig:volfrac_CEDRE_t24.10}
}
\end{minipage}%
\begin{minipage}{.25\textwidth}

\subfloat[$t_0+5\delta t$]{%
       \centering
\begin{overpic}[width=0.4\textwidth, frame, trim={6cm 1.4cm 1.6cm 1.5cm}, clip]{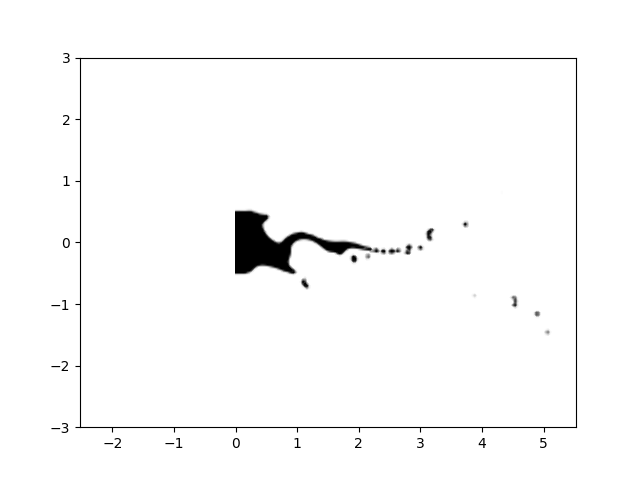}
       \end{overpic}  
  \label{fig:volfrac_ARCHER_t24.00}
}
\subfloat[]{%
       \centering
\begin{overpic}[width=0.4\textwidth, frame, trim={6cm 1.4cm 1.6cm 1.5cm}, clip]{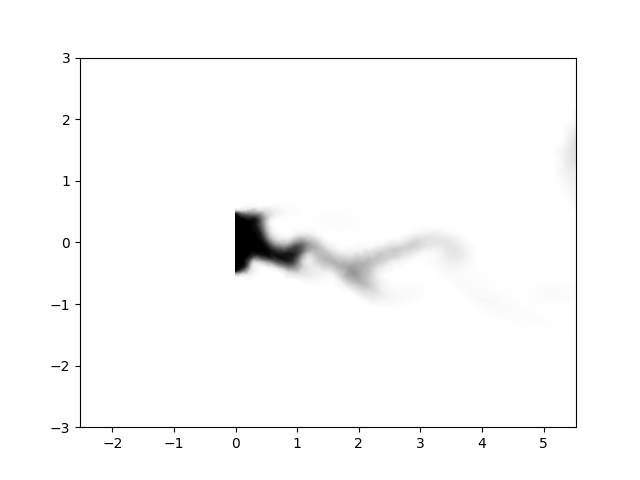}
       \end{overpic}  
  \label{fig:volfrac_CEDRE_t24.15}
}
\end{minipage}
\begin{minipage}{.25\textwidth}
\subfloat[$t_0+6\delta t$]{%
       \centering
\begin{overpic}[width=0.4\textwidth, frame, trim={6cm 1.4cm 1.6cm 1.5cm}, clip]{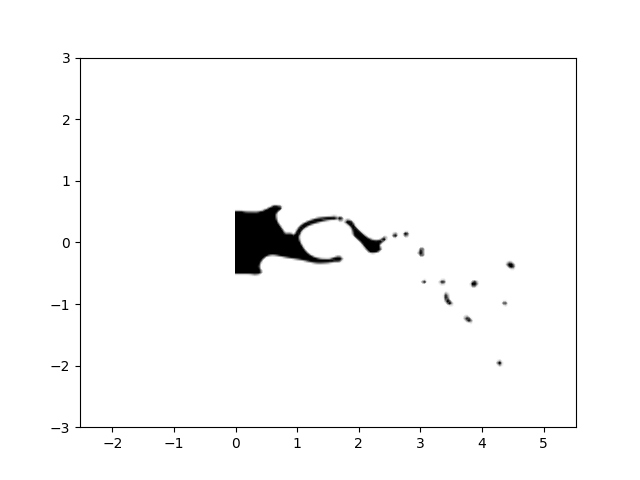}
       \end{overpic}  
  \label{fig:volfrac_ARCHER_t24.05}
}
\subfloat[]{%
       \centering
\begin{overpic}[width=0.4\textwidth, frame, trim={6cm 1.4cm 1.6cm 1.5cm}, clip]{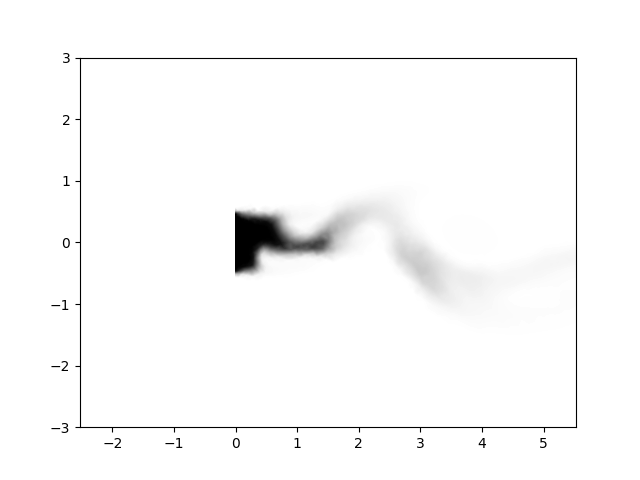}
       \end{overpic}  
  \label{fig:volfrac_CEDRE_t24.20}
}
\end{minipage}%
\begin{minipage}{.25\textwidth}
\subfloat[$t_0+7\delta t$]{%
       \centering
\begin{overpic}[width=0.4\textwidth, frame, trim={6cm 1.4cm 1.6cm 1.5cm}, clip]{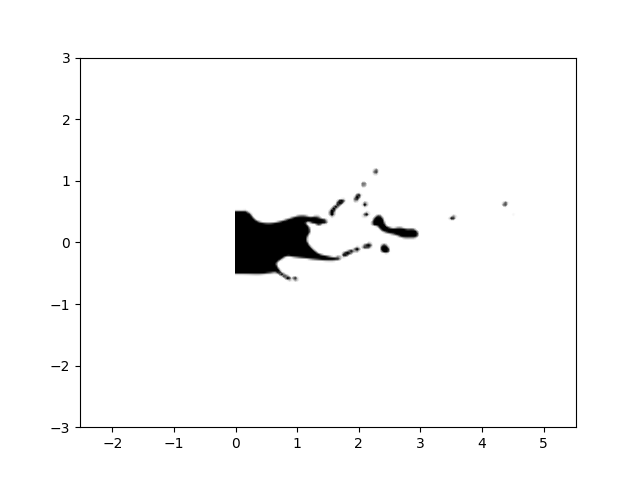}
       \end{overpic}  
  \label{fig:volfrac_ARCHER_t24.10}
}
\subfloat[]{%
       \centering
\begin{overpic}[width=0.4\textwidth, frame, trim={6cm 1.4cm 1.6cm 1.5cm}, clip]{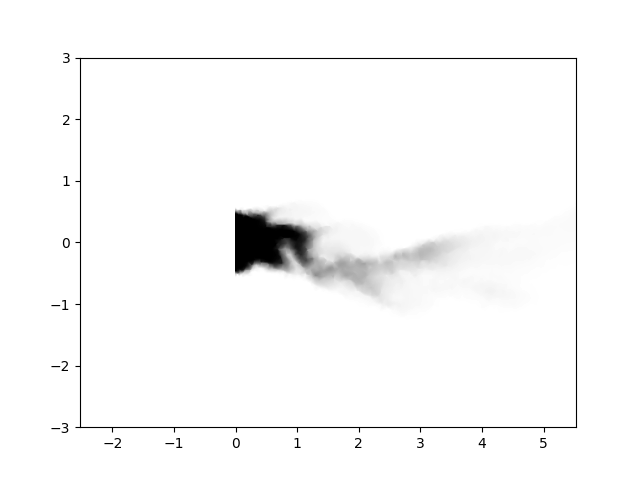}
       \end{overpic}  
  \label{fig:volfrac_CEDRE_t24.25}
}
\end{minipage}%
\begin{minipage}{.25\textwidth}
\subfloat[$t_0+8\delta t$]{%
       \centering
\begin{overpic}[width=0.4\textwidth, frame, trim={6cm 1.4cm 1.6cm 1.5cm}, clip]{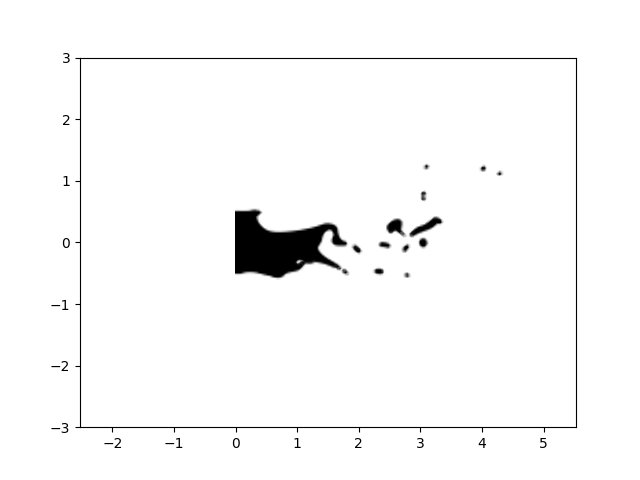}
       \end{overpic}  
  \label{fig:volfrac_ARCHER_t24.15}
}
\subfloat[]{%
       \centering
\begin{overpic}[width=0.4\textwidth, frame, trim={6cm 1.4cm 1.6cm 1.5cm}, clip]{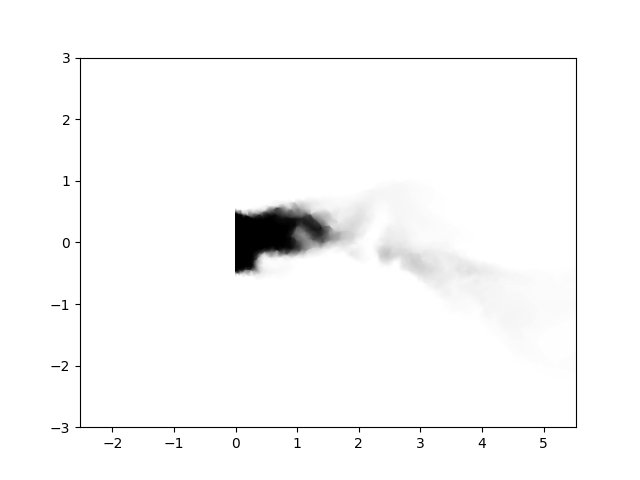}
       \end{overpic}  
  \label{fig:volfrac_CEDRE_t2370}
}
\end{minipage}
\caption{Time lapse of liquid volume fraction near the injector (ARCHER), $\volfrac[l] = 0$ \protect \includegraphics[width=0.07\textwidth, height=0.014\linewidth]{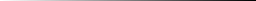}1. Left: ARCHER, right: CEDRE. We define $t_0$ as the starting time and $\delta t$ as the time step.}
\label{fig:inst_volfrac_period}
\end{figure}

\captionsetup[subfloat]{labelformat=parens}

Figure~\ref{fig:inst_volfrac_period} shows the temporal evolution of the liquid volume fraction obtained with the interface capturing technique (left) and the diffuse interface model (right). The time-frame starts at a given simulation time $\Time[sim] = t_{0}$ with a $\delta \Time$ frame rate and stops after nine iterations.

In the first place some inherent properties linked to these numerical methods are clearly observable: results obtained with the CLSVOF method exhibit a sharp interface between liquid and gas, i.e. a one cell transition between $\vol[frac][l]=0$ or $\vol[frac][l]=1$ at the interface. On the contrary, the liquid volume fraction obtained with the seven equation model allows a range of $\vol[frac][l]$ to be found across the interface, corresponding to a smooth transition between liquid and gas.

When comparing the instantaneous snapshots, the local representation of the interface is clearly different due to the intrinsic modeling gap between a sharp and a diffuse interface modeling. Nevertheless, these differences are less obvious when considering that the diffuse interface approach may be interpreted in terms of some statistical probability to find the liquid phase, i.e. $0.5$ value of liquid volume fraction does not actually mean that there is actually $50\%$ of liquid phase and gas phase but that pure liquid may be found there with a probability of $50\%$.

Focusing on large scale motions, Figure~\ref{fig:inst_volfrac_period} reveals a sinusoidal flapping. Measuring this flapping frequency could be done by Fourier transform of a signal characteristic of the flapping varying over time. However, while this procedure is conducted in experiments since operators can record the signal in time for a large number of flapping, it is nonetheless hardly applicable to the present numerical simulations including only few flapping events. Instead, we have chosen an arbitrary initial time $\Time_{0}$ at which both simulations are synchronized. The global flapping evolution of both numerical approaches are much the same inferring that, despite the different representation of the interface, the liquid sheet flapping frequency is similar. One may have also noticed at the end of Figure~\ref{fig:inst_volfrac_period}, that both simulations exhibit a transient regime during which we observe compaction of the thin flapping liquid sheet starting to be thicker and shorter. Then the liquid sheet stretches again at the very end of the time-lapse for both simulations.

\subsubsection{Liquid core penetration}

\begin{figure}[h]
\centering
\includegraphics[width=0.35\textwidth]{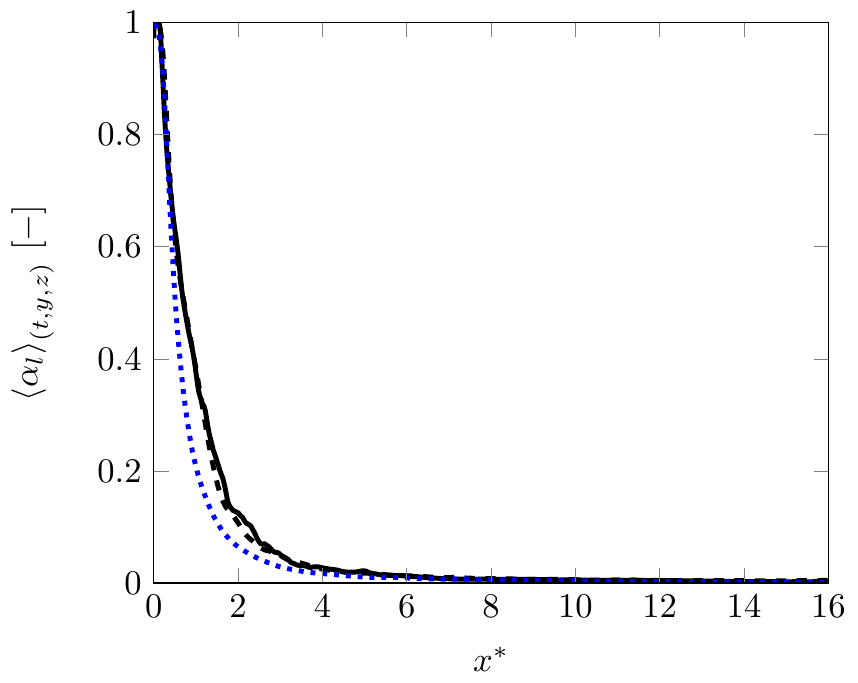} %
\caption{Averaged liquid volume fraction $\average{\volfrac_l}_{(t,y,z)}$ along streamwise axis.\protect\lineplain{black} ARCHER, \protect\linedot{black} CEDRE (HR), \protect\linedot{blue} CEDRE (MR).}
\label{fig:Av_alpha_streamwise_liquidx}
\end{figure}

One key quantity in the study of liquid injection is the liquid penetration length. For transient injection, for which no liquid is initially present, the liquid spray grows from the injector outlet and penetrates into the chamber. The liquid core penetration length is then easily obtained at the first break-up event. Since the present case focuses on the established state of the flow, we cannot monitor such event. Worst, the flow has undergone multiple break-ups through time and the spray has reached the outlet of the computational domain. Experimentally, some apparatus may give a picture of the spray with high contrast between a zone where the spray is present and a zone where there is nearly no liquid and average it over time. Since this procedure is hard to apply on numerical simulations, we choose to study the time and $y-$axis and $z-$axis averaged liquid volume fraction, $\average{\vol[frac][l]}_{(t,y,z)}$, along the streamwise direction (x-axis) in Figure~\ref{fig:Av_alpha_streamwise_liquidx}.  

The liquid profile along the axis obtained by the DNS ranges from unity close to the injector slit towards zero further downstream, once atomization and dispersion have occurred. The high resolution profile CEDRE (HR) of the diffuse interface model matches nicely the DNS data. It is a noticeable result that despite of the difference on the representation of the interface the evolution of the liquid concentration is accurately captured. The CEDRE (MR) profile on the contrary shows a certain discrepancy with the reference DNS result. This suggests a correlation with the mesh resolution, which can be explained as follows: the diffuse interface approach integrates a part of the interface statistics by representing the phase transition by a smooth function rather than by a discontinuity as for the DNS. The diffuse interface model contains thus two averaging procedures, one  related to the model derivation itself and a second due to the mesh resolution, causing the observed discrepancy with the medium refined mesh.

\subsubsection{Discussion on small-scale representations}

Turbulence is intrinsically impossible to be reproduced and thus to be predicted locally in space and time by simulations. Accordingly, only statistical characteristics of turbulent flows can be predicted. With this in mind, DNS simulation has the meaning of a numerical procedure that can be used directly without any modeling assumptions to generate flow samples having the same statistical characteristics as the real investigated flow, such as the averaged velocity. DNS instantaneous results can not be considered as the true realization of the flow, but rather possible realizations. Our understanding of turbulence in liquid-gas flow and in particular for this test case is not complete, but it may be a reasonable hypothesis that these unpredictable velocity fluctuations are accounted for. It is even possible that the complex interaction between phases across the interface enforces the unpredictability of the exact flow. Thus, it is possible to expect that a small perturbation of the DNS can lead to a significant position change of some ligaments and droplets later on.

On the contrary, as explained in Figure~\ref{fig:inst_volfrac_period}, the diffuse interface model makes explicit this non-determination of the interface real position by estimating the probability of the liquid phase presence in each cell. The diffuse interface approach is then potentially representing correctly the real flow, but it is highly relying on the closure of the models, such as the velocity relaxation time introduced in Equation~\eqref{eq:7eq_relax_speed} or the interfacial pressure and speed.

To conclude, on these preliminary comparisons, while the two models show a very different representation of small-scales due to the distinct interface modeling, large-scale motions seems reasonably in good agreements and sufficient to recover global features of the atomization process such as liquid penetration and spray angle.

Beyond the global characteristic of the atomization, numerical simulations give also access to more complete data set of results. In particular, for this established flow, the time averaging procedure provides local statistics of the flow. The following part is thus dedicated to statistical analysis of results obtained by diffuse interface method compared to the reference DNS data. To ensure a fair comparison the domain is limited to the well-resolved part $x^* \in \qty[0,4]$.

\subsection{Statistical analysis}

To push further the comparison of the simulations, we now provide statistical analyses of the flow. The first paragraph evaluates first-order moments with the time averaged liquid volume fraction and the time averaged hydrodynamical velocity components. Then, the second paragraph discusses about the second-order moments obtained from the liquid volume fraction and velocity fluctuations.

\subsubsection{First-order moments}
\begin{figure}[h]
\centering
\subfloat[ARCHER]{%
       \centering
\begin{overpic}[width=0.33\textwidth]{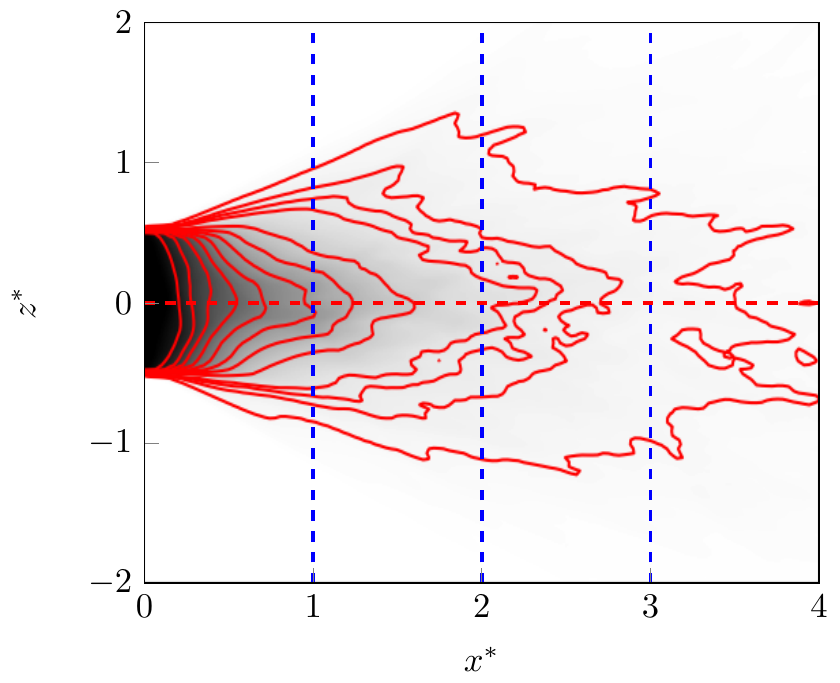} %
       \end{overpic}  
  \label{fig:vof_archer}
}
\quad
\subfloat[CEDRE (HR)]{%
       \centering
\begin{overpic}[width=0.30\textwidth]{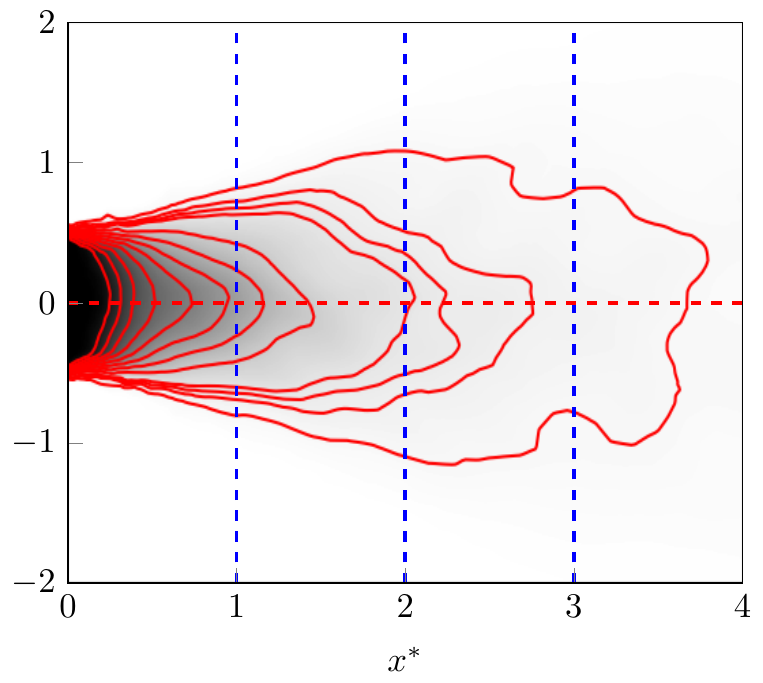} %
       \end{overpic}  
  \label{fig:VOF_CEDRE}
}
\quad
\subfloat[CEDRE (MR)]{%
       \centering
\begin{overpic}[width=0.30\textwidth]{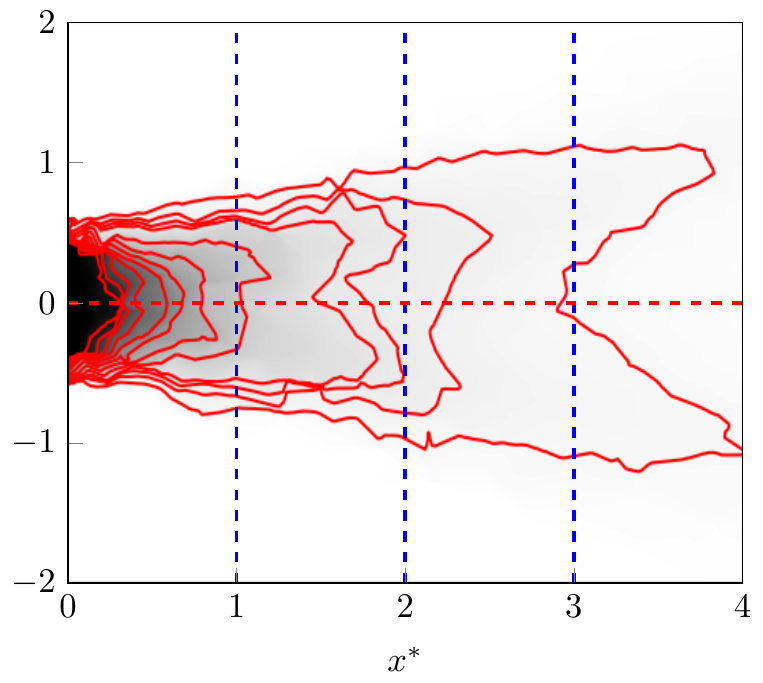} %
       \end{overpic}  
  \label{fig:VOF_CEDRE_mr}
}
\caption{Averaged liquid volume fraction isolines $\average{\vol[frac][l]}_{t} = 0.1$\protect \includegraphics[width=0.07\textwidth, height=0.014\linewidth]{./gray.png}$1$ at the $y^*=0$ plane. Dashed located at $x^*\in \left\lbrace 1, 2, 3 \right\rbrace$.} 

\label{fig:VOF_penelenght}
\end{figure}

\paragraph{Time averaged liquid volume fraction}
Isolines of the time averaged liquid volume fraction $\average{\vol[frac][l]}_{t}$ with its spatial evolution on the $y^*=0$ plane are drawn in Figure~\ref{fig:VOF_penelenght}. This figure also give an overview of the spray angle obtained with both formalisms, which is a key quantity in atomization studies. %
The averaging time is $\Time[sim] = 9 \, ms$. 

A good agreement is observed qualitatively for both methods: the isolines almost match each other and contours of liquid volume fraction are similar. However, the isolines extend a bit further along the $x-$ axis for the DNS than those of the diffuse interface model. This effect is probably due to the numerical dissipation, since the DNS mesh is about four times refined in each direction compared to CEDRE (HR) mesh. The isolines extend slightly more for the DNS case, by a small margin. Interestingly, from the lower value of isolines, we can clearly see qualitatively the spray angle of the simulations. The angle is a little wider for the DNS confirming that DNS is more subjected to \emph{dispersion} than the diffuse interface model. This effect is nearly noticeable from the HR diffused interface simulation but clearly visible from the MR simulation. 

\begin{figure}[h]
\centering
\subfloat[{$x^*=1$.}]{%
       \centering
\begin{overpic}[width=0.3\textwidth]{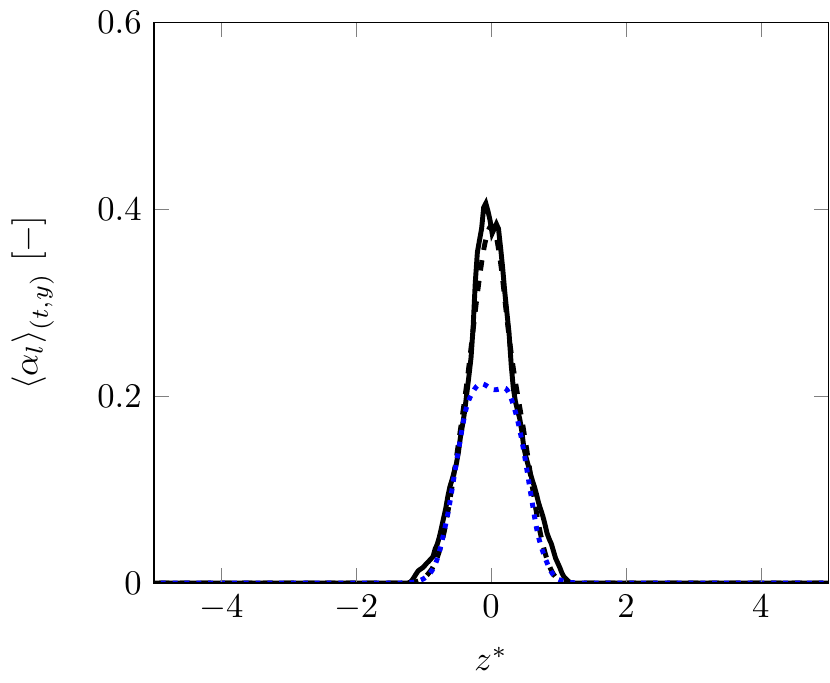} %
       \end{overpic}  
  \label{fig:vof_0.0005t}
}
\quad
\subfloat[{$x^*=2$.}]{%
       \centering
\begin{overpic}[width=0.27\textwidth]{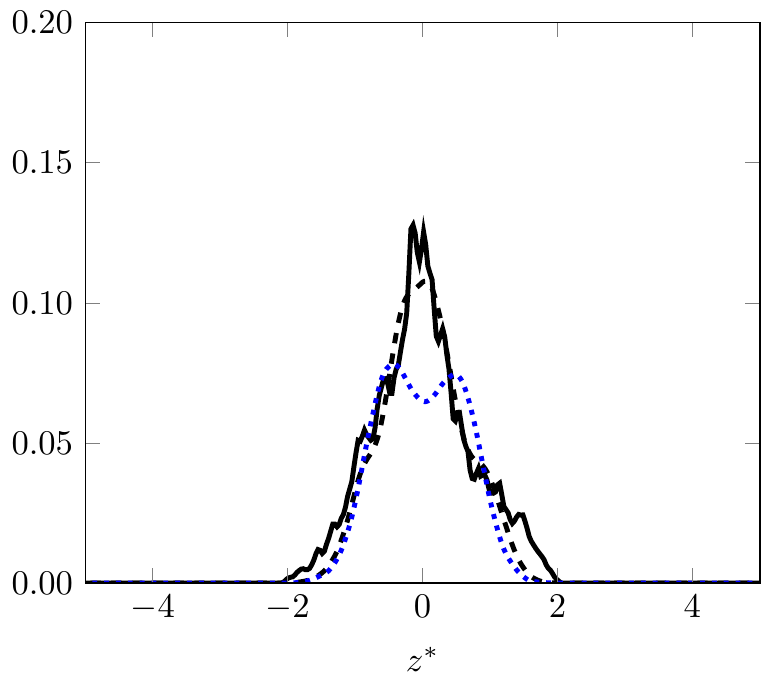} %
       \end{overpic}  
  \label{fig:VOF_0.001t}
}
\quad
\subfloat[{$x^*=3$.}]{%
       \centering
\begin{overpic}[width=0.27\textwidth]{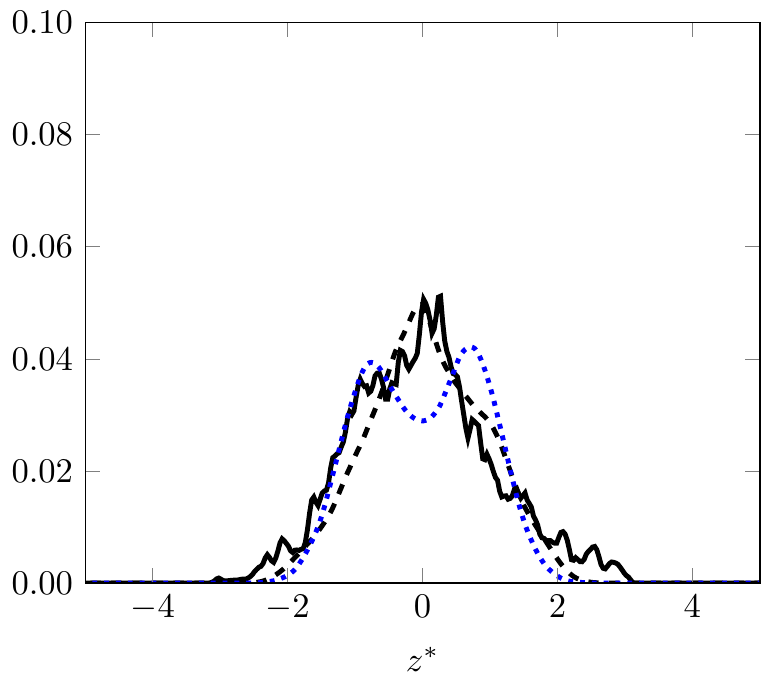} %
       \end{overpic}  
  \label{fig:VOF_0.01t}
}
\caption{Averaged liquid volume fraction $\average{\volfrac_l}_{(t,y)}$ along Figure~\ref{fig:VOF_penelenght} dashed lines.\protect\lineplain{black} ARCHER, \protect\linedot{black} CEDRE (HR), \protect\linedot{blue} CEDRE (MR). Reader must take care to the fact that the $\average{\vol[frac][l]}_{(t,y)}$ range is different for the three slices.}
\label{fig:av_alpha_profiles}
\end{figure}

To obtain a quantitative comparison, time and $y-$axis averaged liquid volume fraction profiles along the transverse direction $(z)$ are depicted for three different $x$ positions in Figure~\ref{fig:av_alpha_profiles}. The positions of transverse profiles are shown on Figure~\ref{fig:VOF_penelenght}. 
The diffuse interface model is very close to the DNS reference simulation, but confirms the slightly higher dispersion in the reference DNS, visible at least on the CEDRE (MR) case. A possible explanation, involving the representation of the interface, could be that droplets formed during the atomization process are thrown away by the flapping motion of the liquid sheet. 
Once they are ejected, they have a relative transverse velocity component with respect to the surrounding gas phase. The relaxation of this slip velocity has not the same mechanism in the DNS that aims at solving the force acting at the liquid surface, whereas the diffuse interface models it (see Equation~\eqref{eq:7eq_relax_speed}). The correct resolution of these forces by the DNS depends on the mesh resolution, hence is difficult to achieve for droplets having a radius of the order of the mesh resolution. This may lead to these slight differences on the dispersion, the DNS approach estimating a longer time of relaxation than the diffused interface approach.

As a concluding remark, in future work, we could characterize the relaxation time of the velocity relaxation based on the dispersion observed in the results to close the diffuse interface model.

\begin{figure}[h]
\centering
\subfloat[ARCHER]{%
       \centering
\begin{overpic}[width=0.33\textwidth]{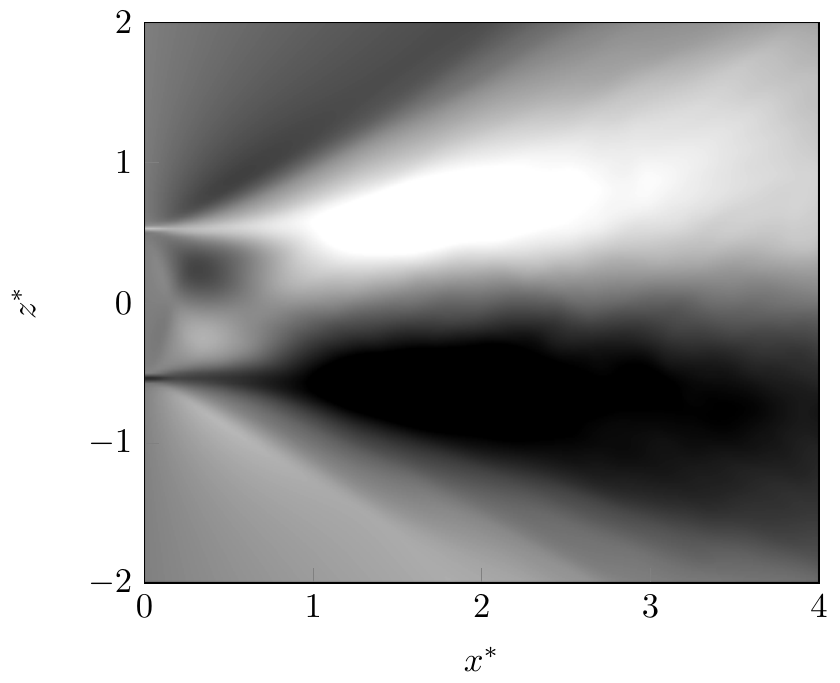} %
       \end{overpic}  
  \label{fig:uz_archer}
}
\quad
\subfloat[CEDRE (HR)]{%
       \centering
\begin{overpic}[width=0.3\textwidth]{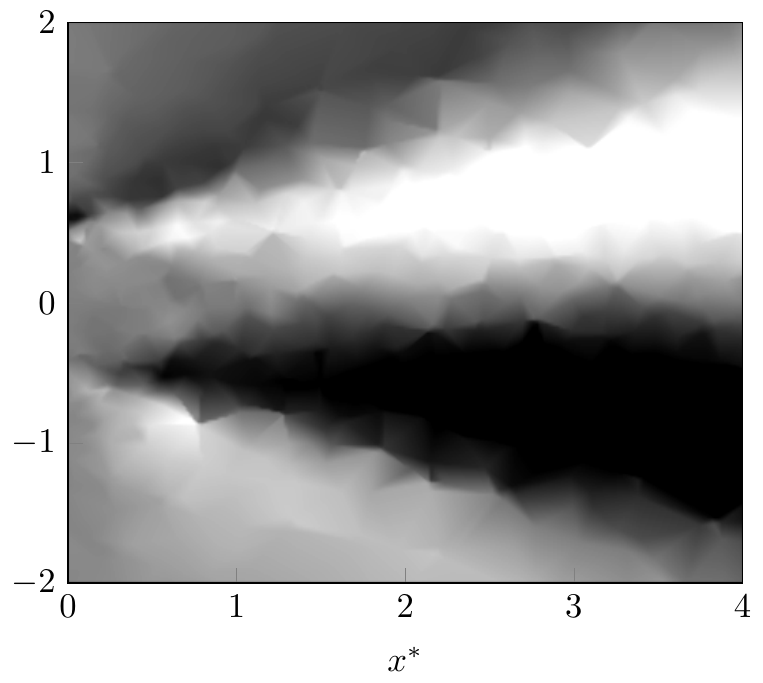} %
       \end{overpic}  
  \label{fig:uz_CEDRE}
}
\quad
\subfloat[CEDRE (MR)]{%
       \centering
\begin{overpic}[width=0.3\textwidth]{./UZ_CEDRE.pdf} %
       \end{overpic}  
  \label{fig:uz_CEDRE_MR}
}
\caption{Averaged tansverse velocity $\average{\speed[z]}_{t} = -3$ \protect \includegraphics[width=0.07\textwidth, height=0.014\linewidth]{./gray.png} $3 \, m/s$ at the $y^*=0$ plane.} 
\label{fig:uz}
\end{figure}

\begin{figure}[h]
\centering
\subfloat[{$x^*=1$.}]{%
       \centering
\begin{overpic}[width=0.3\textwidth]{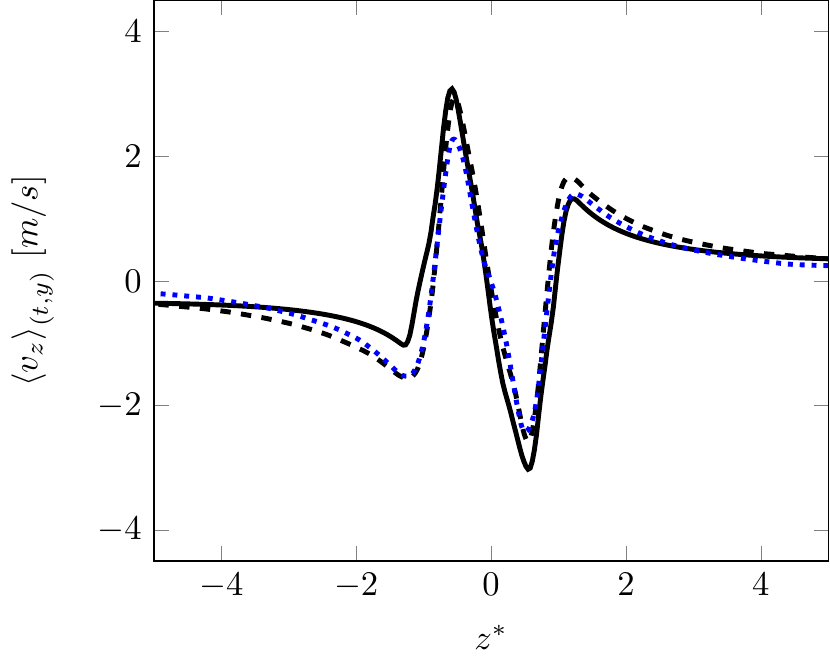} %
       \end{overpic}  
  \label{fig:uz_0.0005t}
}
\quad
\subfloat[{$x^*=2$.}]{%
       \centering
\begin{overpic}[width=0.27\textwidth]{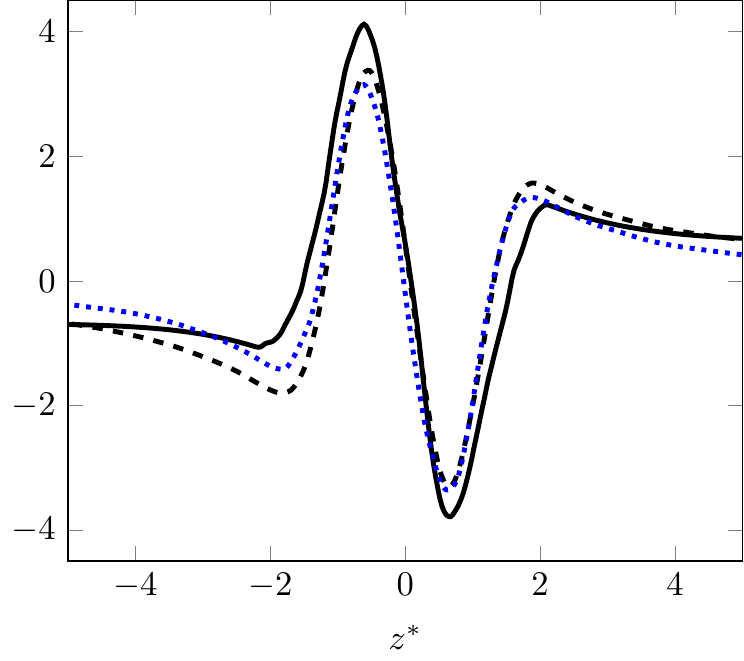} %
       \end{overpic}  
  \label{fig:uz_0.001t}
}
\quad
\subfloat[{$x^*=3$.}]{%
       \centering
\begin{overpic}[width=0.27\textwidth]{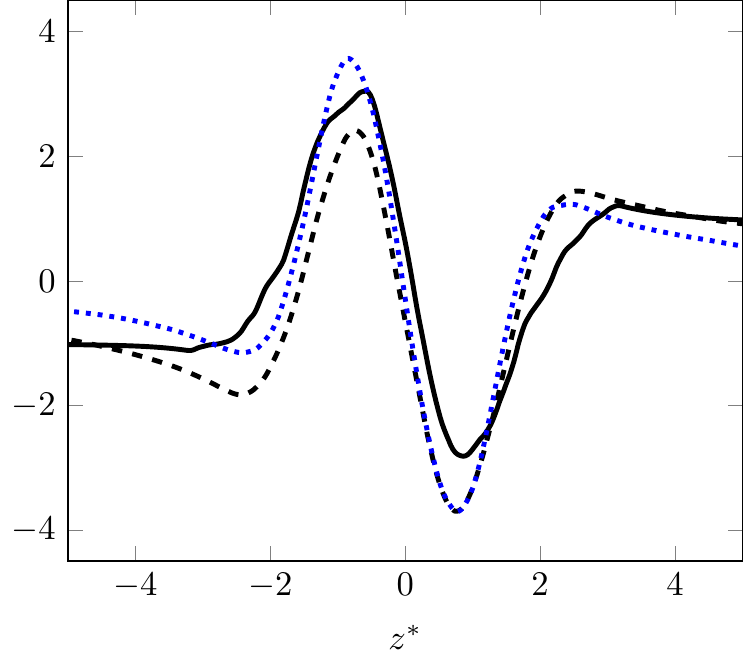} %
       \end{overpic}  
  \label{fig:uz_0.01t}
}
\caption{Averaged tansverse velocity $\average{\speed[z]}_{(t,y)}$ along Figure~\ref{fig:VOF_penelenght} dashed lines. \protect\lineplain{black} ARCHER, \protect\linedot{black} CEDRE (HR), \protect\linedot{blue} CEDRE (MR).}
\label{fig:uzprofiles}
\end{figure}

\paragraph{Mean velocities}
Time average transverse velocity contours are illustrated in Figure~\ref{fig:uz} at slice $y^* = 0$.
Qualitatively the global behaviour is the same between DNS and diffuse interface approach for a resolution high enough. Medium resolution case leads to an over dissipation of the velocity variations, most probably due to the numerical diffusion.  In particular, just at the exit of the injection plane the longitudinal acceleration of the liquid sheet induce a contraction that can be seen on a small area ($x^*<1$ and $\abs{z^*}<0.5$) where the transverse liquid velocities are directed towards the centre plane ($z^*=0$). This effect is completely damped for the less resolved mesh.  Outside this zone, the velocity is always directed outward due to the liquid dispersion and jet spreading. To achieve a more quantitative analysis of the dispersion, the transverse velocity profiles have been extracted and averaged in time and along the $y$-direction in Figure~\ref{fig:uzprofiles}. 

For high  enough resolutions both methods leads to very similar profile: in the zone dominated by the liquid  phase close to the center plane $z^*=0$, the  transverse velocity is always directed outward due to the liquid dispersion process. The magnitude of the velocities are mostly similar even slightly higher values are achieved through the DNS approach with respect to the diffuse interface approach. It should be, like for the liquid concentration profile, an effect of the relaxation time of the liquid droplet and structure that are initially ejected from the main liquid stream with high velocity. Then it takes more time for this velocity to relax to the ambient gas velocity than for the diffuse interface approach.  On the external zone the transverse velocity is mostly directed toward the centre and the differences between DNS and diffuse interface approach may be due to the numerical treatment of the external boundary conditions.

\begin{figure}[h]
\centering
\subfloat[ARCHER]{%
       \centering
\begin{overpic}[width=0.33\textwidth]{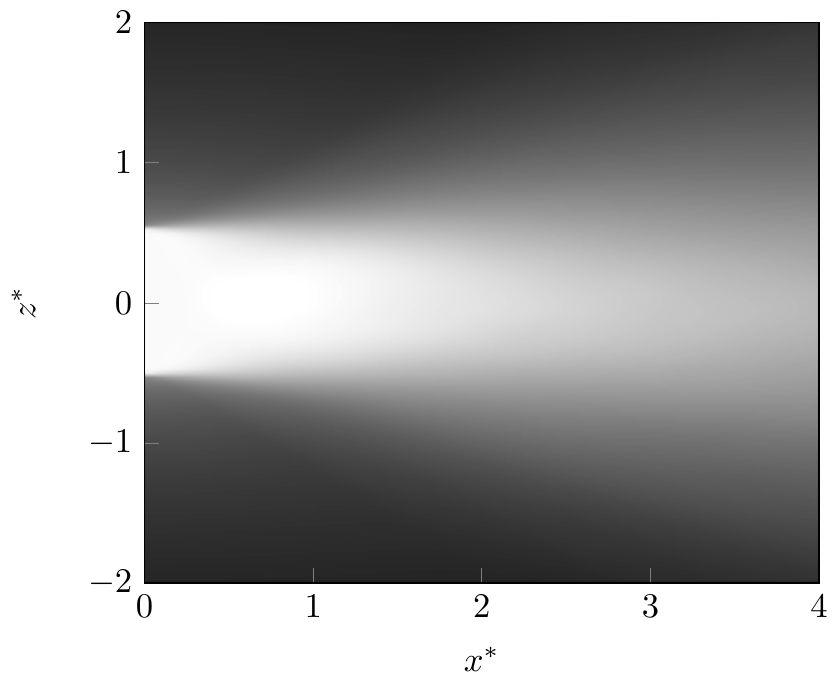} %
       \end{overpic}  
  \label{fig:ux_archer}
}
\quad
\subfloat[CEDRE (HR)]{%
       \centering
\begin{overpic}[width=0.3\textwidth]{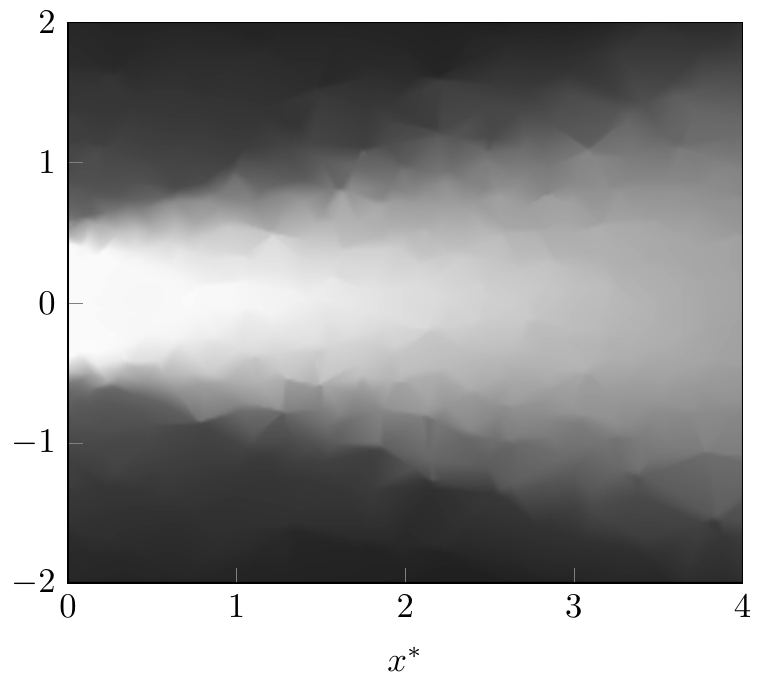} %
       \end{overpic}  
  \label{fig:ux_CEDRE}
}
\quad
\subfloat[CEDRE (MR)]{%
       \centering
\begin{overpic}[width=0.3\textwidth]{./UX_CEDRE.pdf} %
       \end{overpic}  
  \label{fig:ux_CEDRE_MR}
}
\caption{Averaged axial velocity $\average{\speed[x]}_{t} = 0$ \protect \includegraphics[width=0.07\textwidth, height=0.014\linewidth]{./gray.png} $70 \, m/s$ at the $y^*=0$ plane.}
\label{fig:ux2d}
\end{figure}

\begin{figure}[h]
\centering
\subfloat[{$x^*=1$.}]{%
       \centering
\begin{overpic}[width=0.3\textwidth]{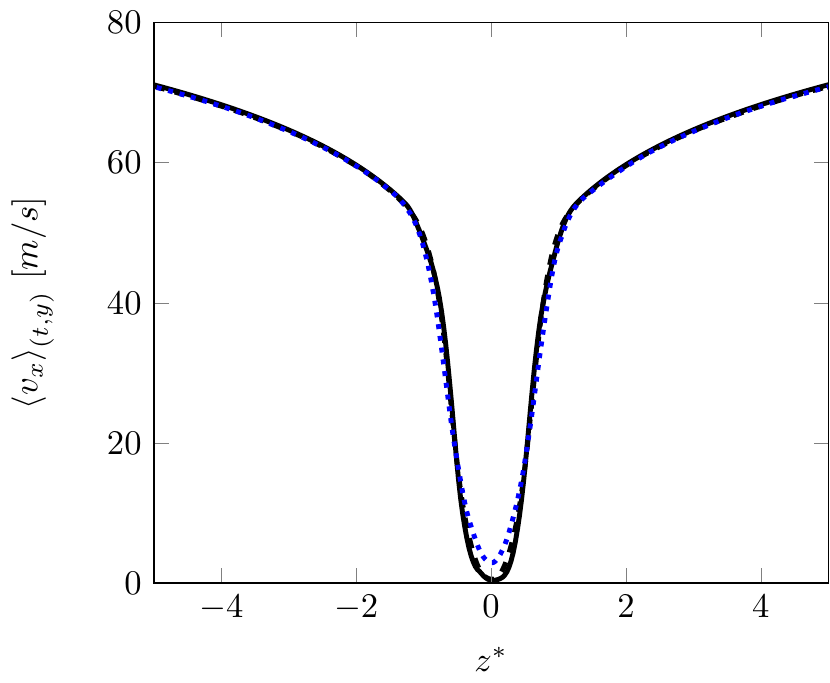} %
       \end{overpic}  
  \label{fig:ux_0.0001}
}
\quad
\subfloat[{$x^*=2$.}]{%
       \centering
\begin{overpic}[width=0.27\textwidth]{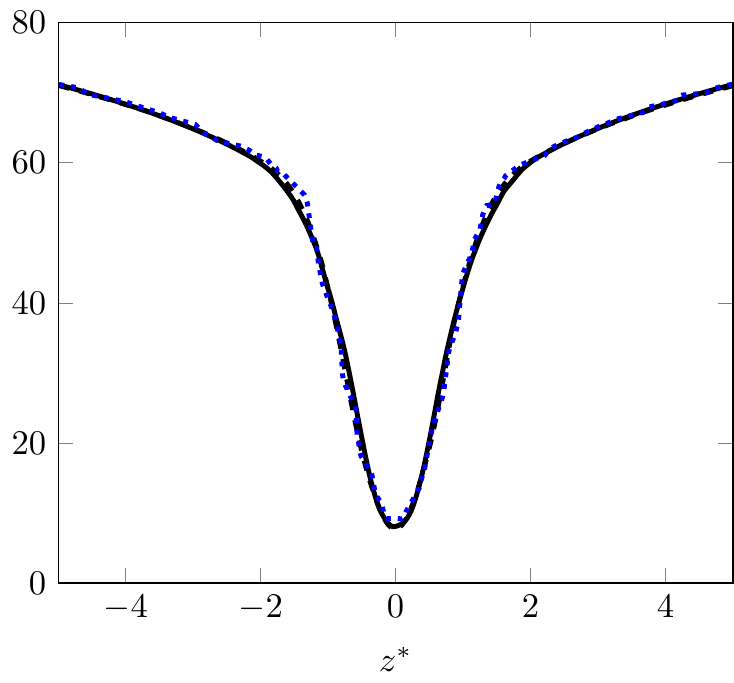} %
       \end{overpic}  
  \label{fig:ux_0.002t}
}
\quad
\subfloat[{$x^*=3$.}]{%
       \centering
\begin{overpic}[width=0.27\textwidth]{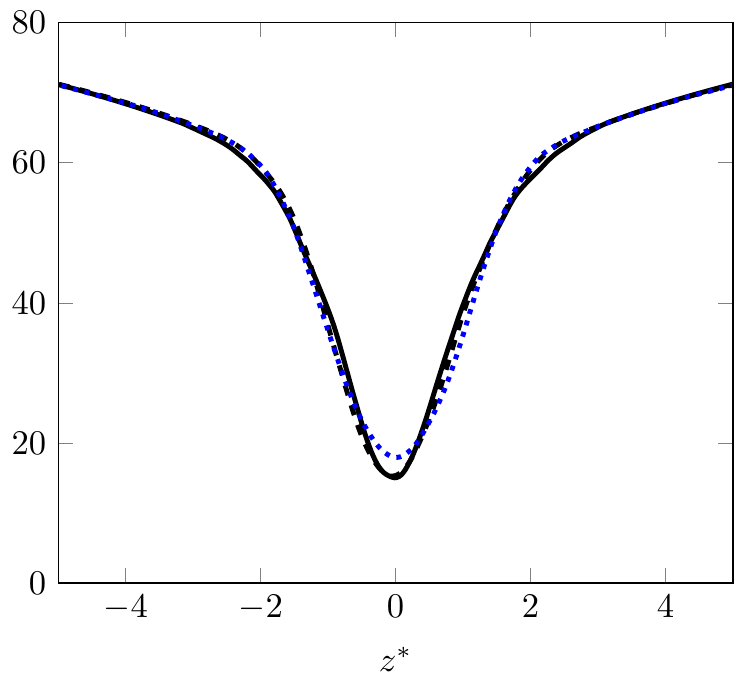} %
       \end{overpic}  
  \label{fig:ux_0.03t}
}
\caption{Averaged axial velocity $\average{\speed[x]}_{(t,y)}$ along Figure~\ref{fig:VOF_penelenght} dashed lines. \protect\lineplain{black} ARCHER, \protect\linedot{black} CEDRE (HR), \protect\linedot{blue} CEDRE (MR).}
\label{fig:uxprofiles}
\end{figure}

Similar comparisons are performed on the time averaged streamwise velocity, $\average{\speed[x]}_{t}$, in Figure~\ref{fig:ux2d}. An excellent agreement is observed with respect to the diffuse interface model results. Again, the intensity of the velocity is slightly smaller due to numerical dissipation inherent to the diffuse interface model. 
The agreement is confirmed by analyzing the time and $y-$axis averaged velocity profiles at different locations shown in Figure~\ref{fig:uxprofiles}.
\FloatBarrier

The next paragraph is dedicated to the study of turbulent statistics through the analysis of the second order moments of the flow using the liquid volume fraction and velocity components fluctuations.

\subsubsection{Second-order moments}

\paragraph{Mean components of the Reynolds stress tensor}
To study more deeply the statistics of the flow, we have computed the second order moments based on the liquid volume fraction and velocity fields. We introduce $\vol[frac][l][fluctu]$ is the fluctuating part of the liquid volume fraction, defined as $\vol[frac][l][fluctu] =  \vol[frac][l] - \average{\vol[frac][l]}_{t}$ and $\speed[x][fluctu]$, $\speed[y][fluctu]$ and $\speed[z][fluctu]$ the fluctuating velocity components. Three kinds of second order moment emanate from these quantities: the liquid volume fraction variance, $\average{\vol[frac][l][fluctu] \vol[frac][l][fluctu]}_{t}$, nine components of the Reynolds stress tensor, $\average{\speed[i][fluctu] \speed[j][fluctu]}_{t}$, and the three components of the liquid turbulent flux $\average{\vol[frac][l][fluctu] \speed[i][fluctu]}_{t}$. 

\begin{figure}[h]
\centering
\subfloat[ARCHER]{%
       \centering
\begin{overpic}[width=0.33\textwidth]{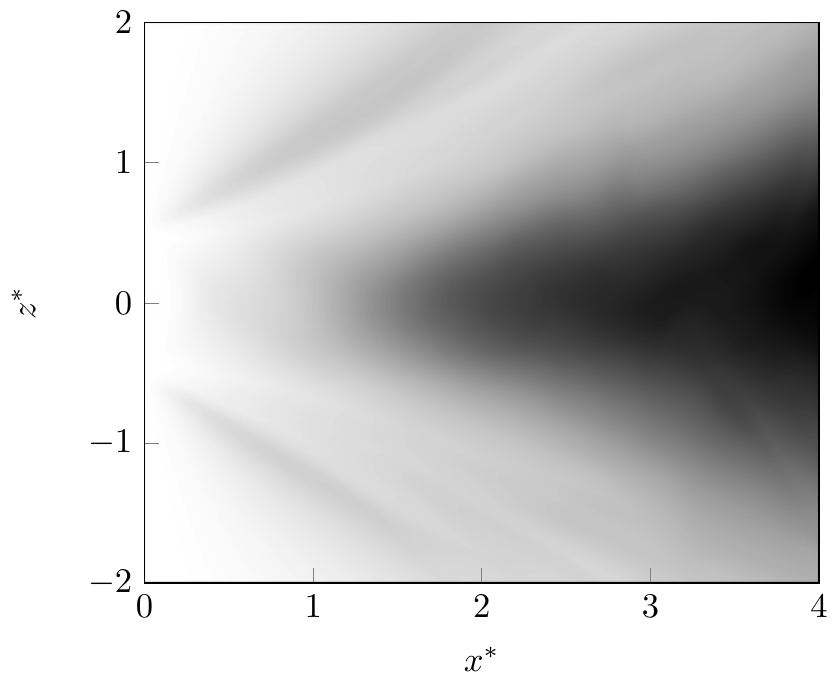} %
       \end{overpic}  
  \label{fig:uzuz_archer}
}
\quad
\subfloat[CEDRE (HR)]{%
       \centering
\begin{overpic}[width=0.3\textwidth]{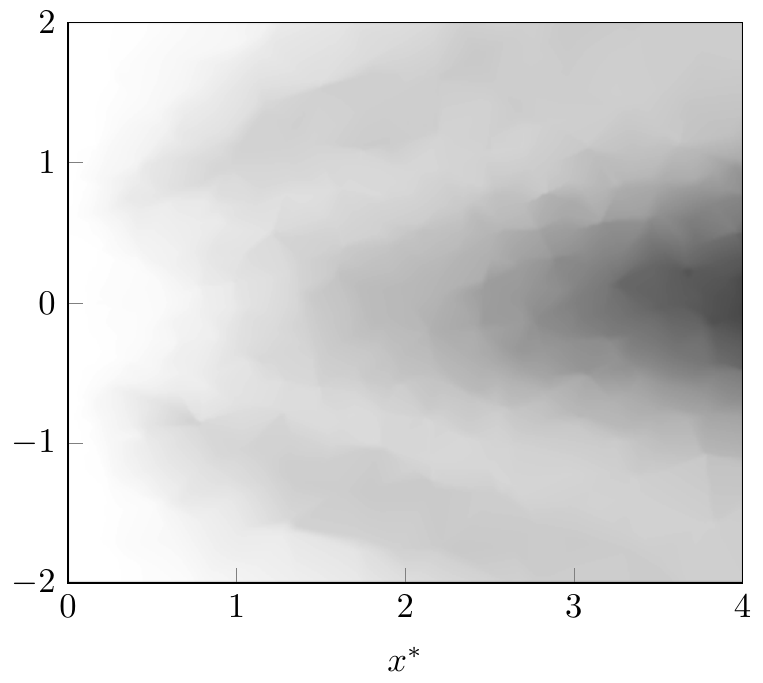} %
       \end{overpic}  
  \label{fig:uzuz_CEDRE}
}
\quad
\subfloat[CEDRE (MR)]{%
       \centering
\begin{overpic}[width=0.3\textwidth]{./UZUZ_CEDRE.pdf} %
       \end{overpic}  
  \label{fig:uzuz_CEDRE_MR}
}

\caption{Transverse Reynolds stress tensor component $\average{\speed[z][fluctu]\speed[z][fluctu]}_{t}  = 0$ \protect \includegraphics[width=0.07\textwidth, height=0.014\linewidth]{./gray.png} $350 \, m^2/s^2$ at the $y^*=0$ plane.}
\label{fig:ret_uzuz2d}
\end{figure}

The analysis of the Reynolds stress tensor allows a validation of the turbulence characteristics observed in the diffuse interface model. Figure~\ref{fig:ret_uzuz2d} shows the transverse Reynolds stress tensor component $\average{\speed[z][fluctu] \speed[z][fluctu]}_{t}$ contour in the $y^*=0$ plane. 
Turbulence intensity is higher downstream in both formalisms. Again, the intensity obtained with the seven equation model is slightly smaller but the overall results are close to the one obtained with the DNS.

\begin{figure}[h]
\centering
\subfloat[{$x^*=1$.}]{%
       \centering
\begin{overpic}[width=0.3\textwidth]{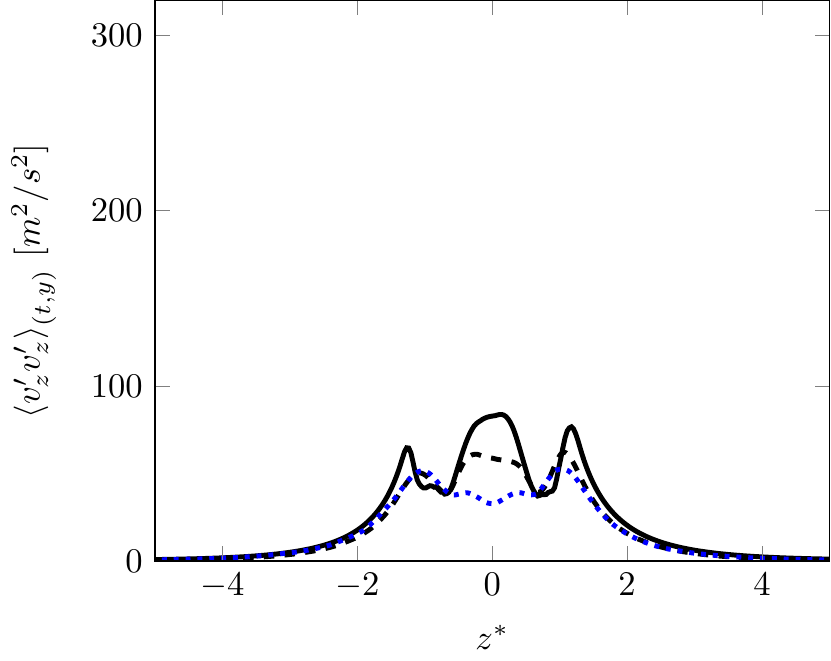} %
       \end{overpic}  
  \label{fig:uzuz_0.0005t}
}
\quad
\subfloat[{$x^*=2$.}]{%
       \centering
\begin{overpic}[width=0.27\textwidth]{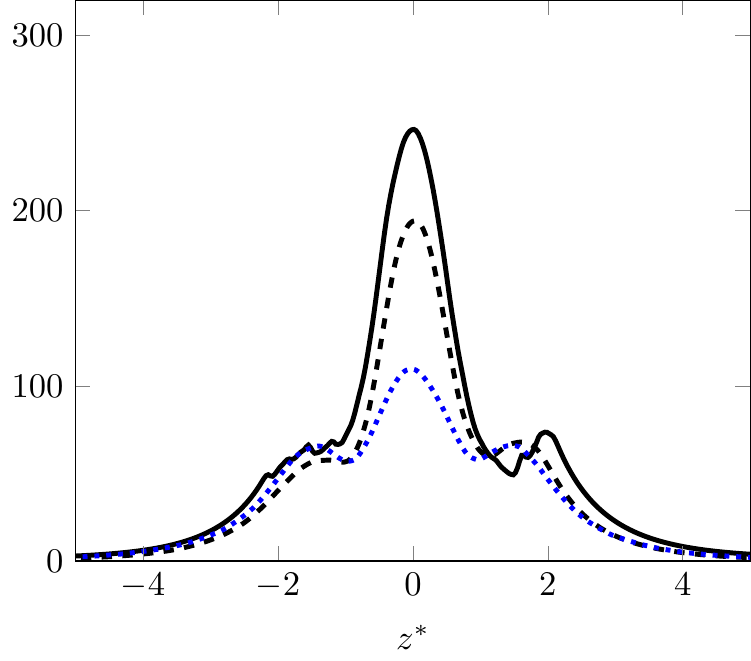} %
       \end{overpic}  
  \label{fig:uzuz_0.001t}
}
\quad
\subfloat[{$x^*=3$.}]{%
       \centering
\begin{overpic}[width=0.27\textwidth]{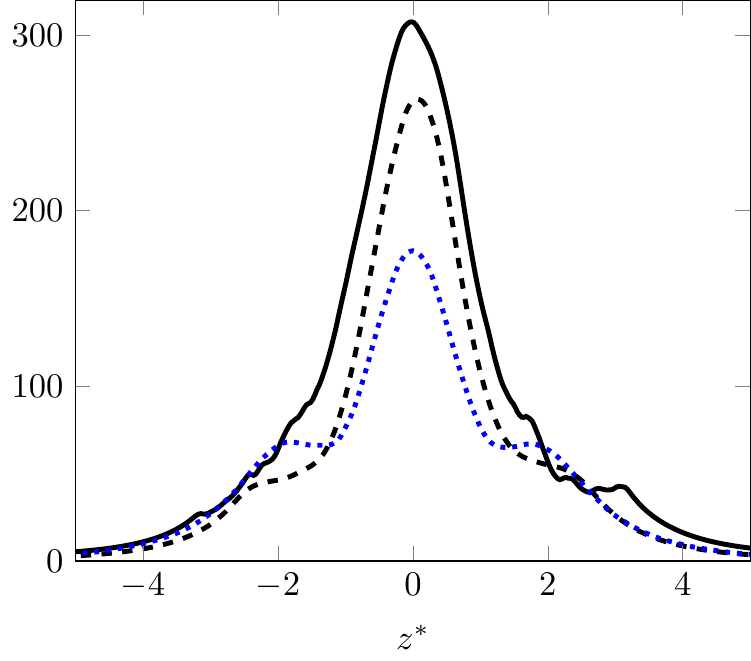} %
       \end{overpic}  
  \label{fig:uzuz_0.01t}
}
\caption{Transverse Reynolds stress tensor component $\average{\speed[z][fluctu]\speed[z][fluctu]}_{(t,y)}$ along Figure~\ref{fig:VOF_penelenght} dashed lines. \protect\lineplain{black} ARCHER, \protect\linedot{black} CEDRE (HR), \protect\linedot{blue} CEDRE (MR).}
\label{fig:ret_uzuz}
\end{figure}

Figure~\ref{fig:ret_uzuz} illustrates the $y-$axis averaged transverse Reynolds stress tensor component for different locations in the streamwise direction. 
As mentioned before, the maximum turbulence intensity is lower than the DNS but the trend of the curve are very similar in both methods, confirming the ability of the diffuse interface model to represent the fluctuating velocities.

\begin{figure}[h]
\centering
\subfloat[ARCHER]{%
       \centering
\begin{overpic}[width=0.33\textwidth]{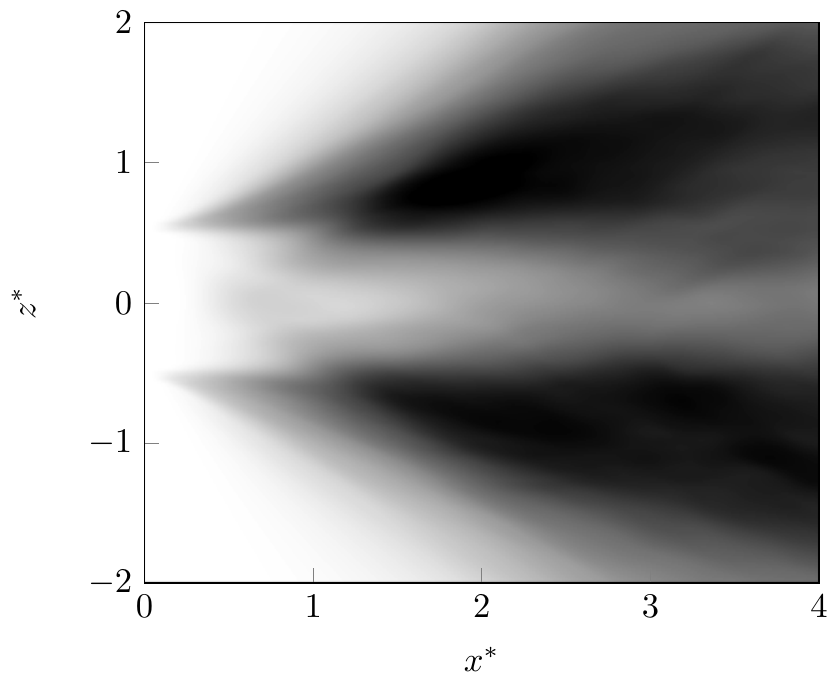} %
       \end{overpic}  
       \label{fig:uxux_archer}
}
\quad
\subfloat[CEDRE (HR)]{%
       \centering
\begin{overpic}[width=0.3\textwidth]{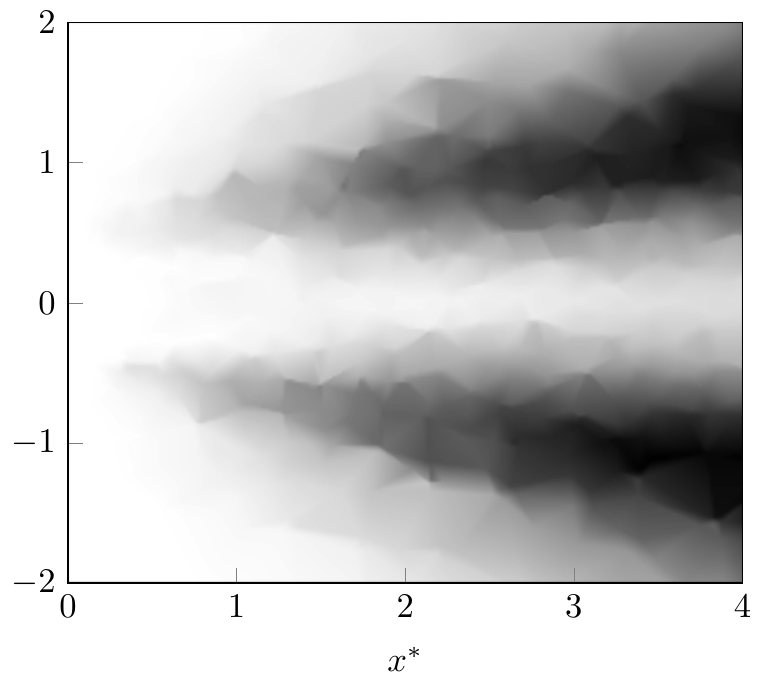} %
       \end{overpic}  
       \label{fig:uxux_CEDRE}
}
\quad
\subfloat[CEDRE (MR)]{%
       \centering
\begin{overpic}[width=0.3\textwidth]{./UXUX_CEDRE.pdf} %
       \end{overpic}  
       \label{fig:uxux_CEDRE_MR}
}
\caption{Longitudinal Reynolds stress tensor component $\average{\speed[x][fluctu]\speed[x][fluctu]}_{t}= 0$  \protect \includegraphics[width=0.07\textwidth, height=0.014\linewidth]{./gray.png} $300 \, m^2/s^2$ at the $y^*=0$ plane.}
\label{fig:ret_uxux2d}
\end{figure}

The same analysis is performed on the longitudinal Reynolds stress component $\average{\speed[x][fluctu] \speed[x][fluctu]}_{t}$, shown in Figure~\ref{fig:ret_uxux2d} . Here, some differences are observed: turbulent intensity in the DNS is higher, but also more spread vertically compared to the results obtained with the diffuse interface model, for which the turbulence fluctuations are located closer to the main axis. 

\begin{figure}[h]
\centering
\subfloat[{$x^*=1$.}]{%
       \centering
\begin{overpic}[width=0.3\textwidth]{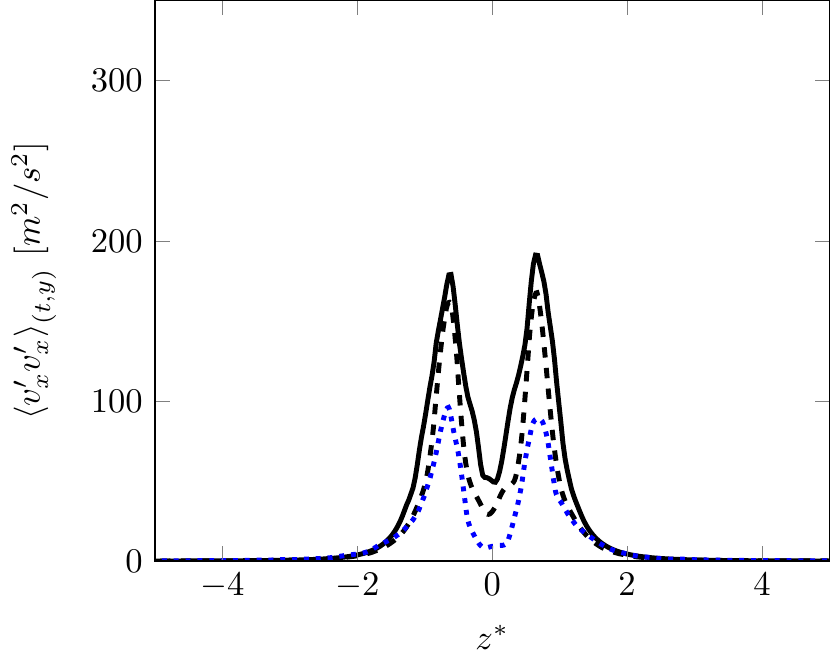} %
       \end{overpic}  
       \label{fig:uxux_0.0005t}
}
\quad
\subfloat[{$x^*=2$.}]{%
       \centering
\begin{overpic}[width=0.27\textwidth]{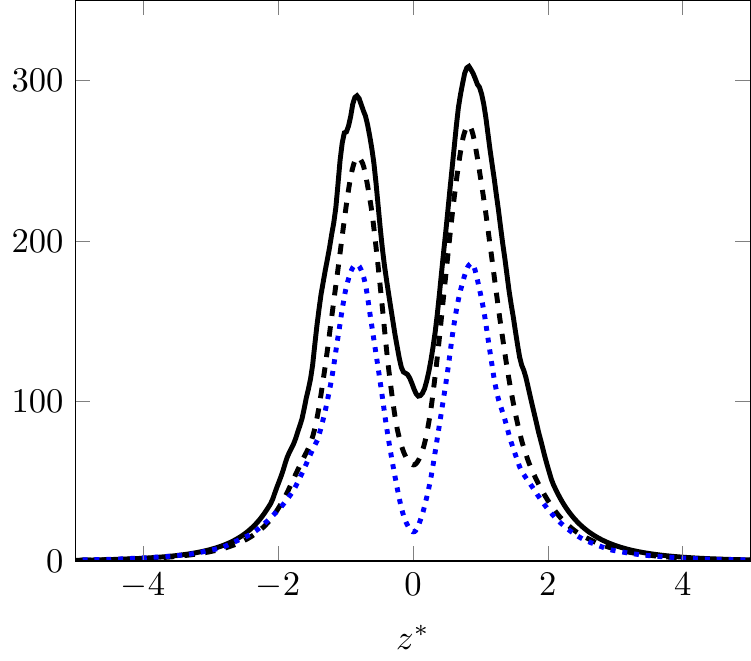} %
       \end{overpic}  
       \label{fig:uxux_0.001t}
}
\quad
\subfloat[{$x^*=3$.}]{%
       \centering
\begin{overpic}[width=0.27\textwidth]{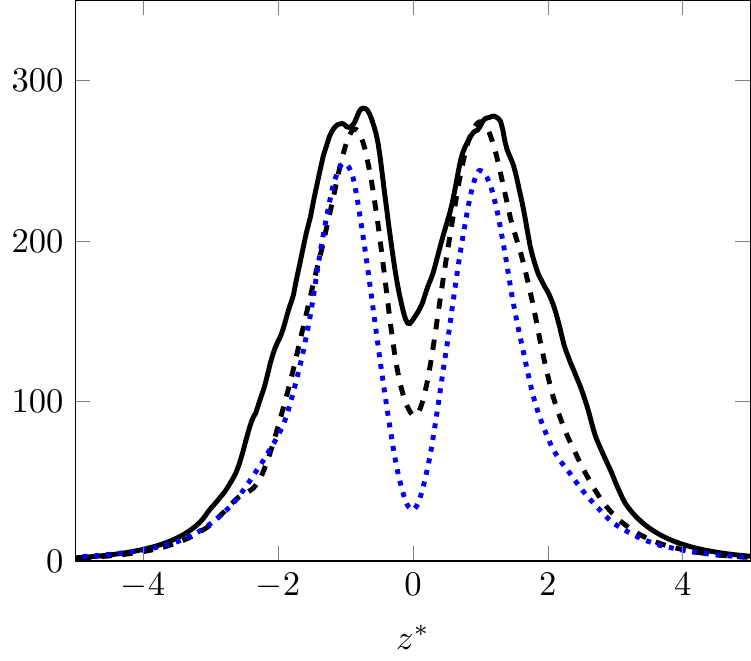} %
       \end{overpic}  
       \label{fig:uxux_0.01t}
}
\caption{$y-$axis averaged longitudinal Reynolds stress tensor component $\average{\speed[x][fluctu]\speed[x][fluctu]}_{(t,y)}$ along Figure~\ref{fig:VOF_penelenght} dashed lines. \protect\lineplain{black} ARCHER, \protect\linedot{black} CEDRE (HR), \protect\linedot{blue} CEDRE (MR).}
\label{fig:ret_uxux}
\end{figure}

Concerning the $y-$axis averaged longitudinal Reynolds stress tensor profiles shown in Figure~\ref{fig:ret_uxux}, an overall good agreement is found between the DNS and the diffuse interface model results.
Still for $x^*=2$ the peak values obtained through the DNS are noticeably larger than for the one obtained with the diffuse interface model. At the contrary at $x^*=3$ the peak values are almost the same but the spreading of the DNS is greater than that of the diffuse interface approach.

\paragraph{Transversal turbulent liquid flux}

The transversal turbulent liquid flux $\average{\vol[frac][l][fluctu] \speed[z][fluctu]}_{t}$ represents the transport of the liquid volume fraction induced by velocity fluctuations. This quantity is important for atomization modeling \citeay{Anez_2019}, since it has to be well captured in order to recover an accurate dispersion of the liquid.

\begin{figure}[h]
\centering
\subfloat[ARCHER]{%
       \centering
\begin{overpic}[width=0.33\textwidth]{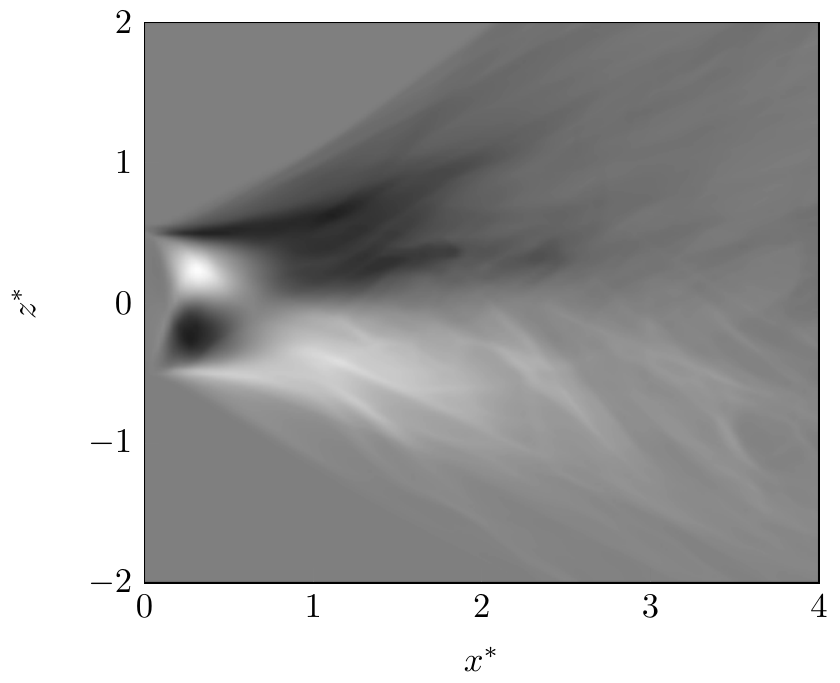} %
       \end{overpic}  
       \label{fig:alphav_archer}
}
\quad
\subfloat[CEDRE (HR)]{%
       \centering
\begin{overpic}[width=0.3\textwidth]{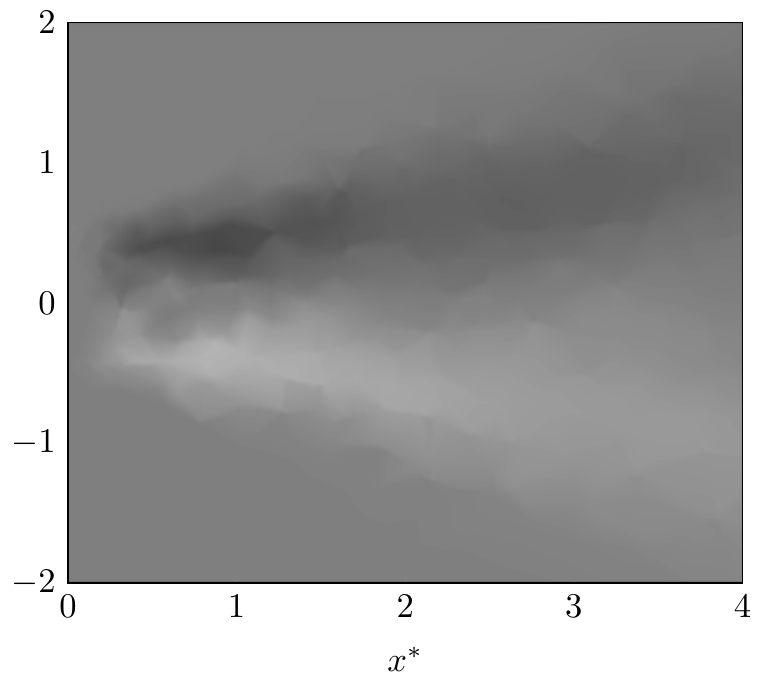} %
       \end{overpic}  
       \label{fig:alphav_CEDRE}
}
\quad
\subfloat[CEDRE (MR)]{%
       \centering
\begin{overpic}[width=0.3\textwidth]{./ALPHAV_CEDRE.pdf} %
       \end{overpic}  
       \label{fig:alphav_CEDRE_MR}
}

\caption{Transversal turbulent liquid flux tensor component $\average{\vol[frac][l][fluctu] \speed[z][fluctu]}_{t} = -1.5$ \protect \includegraphics[width=0.07\textwidth, height=0.014\linewidth]{./gray.png} $1.5 \, m.s^{-1}$ at the $y^*=0$ plane.}
\label{fig:turbliqflux}
\end{figure}

Using the plane $y^*=0$ as in previous paragraphs, $\average{\vol[frac][l][fluctu] \speed[z][fluctu]}_{t}$  contours are summarized in Figure~\ref{fig:turbliqflux}.

\begin{figure}[h]
\centering
\subfloat[{$x^*=1$.}]{%
       \centering
\begin{overpic}[width=0.3\textwidth]{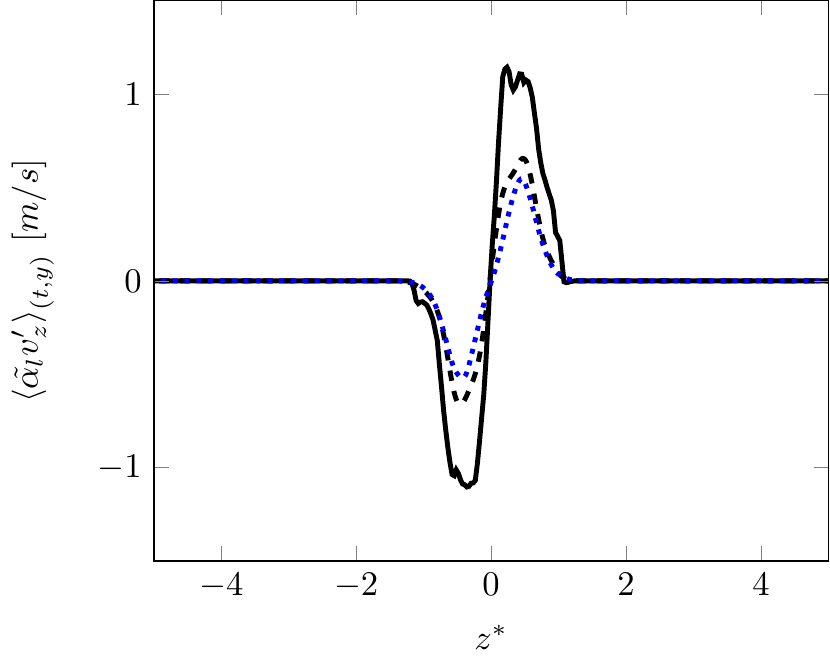} %
       \end{overpic}  
       \label{fig:alphav_0.001t}
}
\quad
\subfloat[{$x^*=2$.}]{%
       \centering
\begin{overpic}[width=0.27\textwidth]{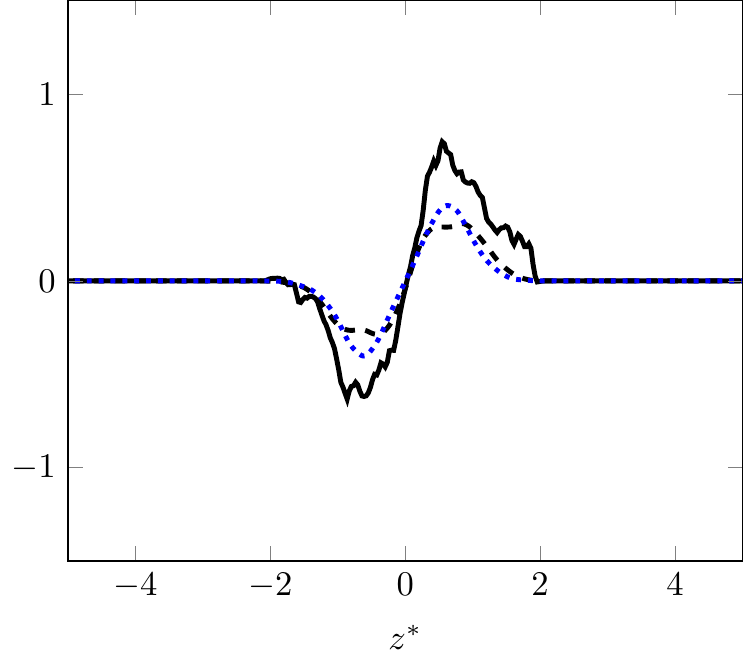} %
       \end{overpic}  
       \label{fig:alphav_0.002t}
}
\quad
\subfloat[{$x^*=3$.}]{%
       \centering
\begin{overpic}[width=0.27\textwidth]{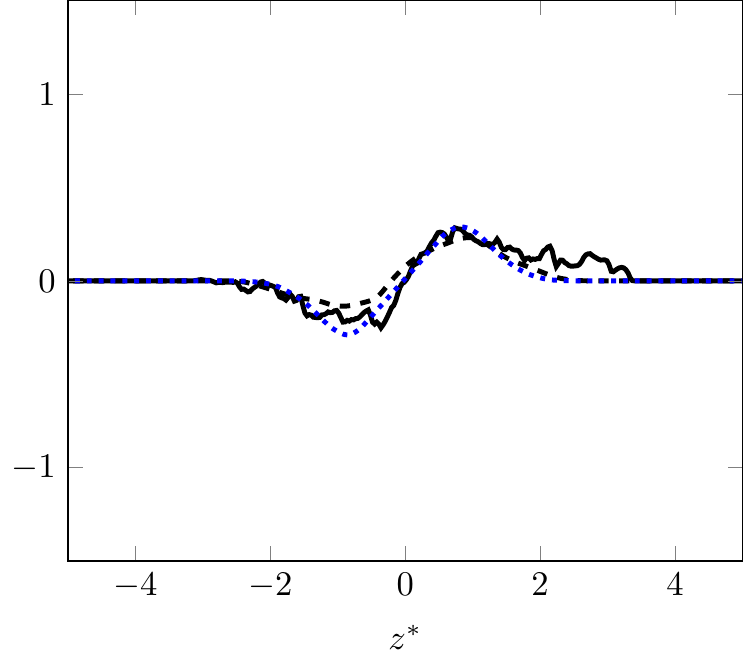} %
       \end{overpic}  
       \label{fig:alphav_0.003t}
}
\caption{$y-$axis averaged transversal liquid flux tensor component $\average{\vol[frac][l][fluctu] \speed[z][fluctu]}_{(t,y)}$ along Figure~\ref{fig:VOF_penelenght} dashed lines.\protect\lineplain{black} ARCHER, \protect\linedot{black} CEDRE (HR), \protect\linedot{blue} CEDRE (MR).}
\label{fig:ret_alphau}
\end{figure}

Figure~\ref{fig:ret_alphau} illustrates the $y-$axis averaged transversal turbulent liquid flux for different locations in the streamwise direction. 

This variable behavior combines the effects pointed out earlier on the liquid volume fraction and the transverse velocity fluctuation: the turbulent liquid flux is more spread in the DNS case due to the higher spray angle. The turbulent liquid flux intensity is lower in the seven equation model. These results confirms that the liquid dispersion is higher in the DNS results.

\FloatBarrier

\paragraph{Discussion on second-order moments}
The second-order moment statistics exhibit some differences between both numerical approaches that are most probably related to the different representation of the small scale features of the flows, and more particularly to the representation of the liquid-gas transition as already attested with the global feature flow comparison at the beginning of the section.

In addition, it can be pointed out that it is more difficult to achieve the statistical convergence, as shown by remaining fluctuations in all the second-order moment figures. To show the effect of small scale structure on the second-order fluctuations, let us first consider the variance of the liquid volume fraction, defined as
\begin{align}
  \average{\vol[frac][l][fluctu] \vol[frac][l][fluctu]}=\average{\vol[frac][l] \vol[frac][l]}-\average{\vol[frac][l]}\,\average{\vol[frac][l]}.
\end{align}
Since at a given location $\mathrm{x}$  there can either be liquid or gas, the admissible values of $\alpha_l$ are only one and zero respectively, thus the probability density function, at this location, reads 
$P(\vol[frac][l])=(1-\average{\vol[frac][l]}) \delta({\vol[frac][l]}) +\average{\vol[frac][l]} \delta({\vol[frac][l]}-1)$, thus yielding
\begin{align}
  \average{\vol[frac][l][fluctu] \vol[frac][l][fluctu]}=
  \average{\vol[frac][l]}-\average{\vol[frac][l]}~\average{\vol[frac][l]}=\average{\vol[frac][l]}(1-\average{\vol[frac][l]}).
\end{align}
The sharp representation of the interface preserves this feature as shown in Figure~\ref{fig:instant_volfrac_3D_comparison}, since the product ${\vol[frac][l]}(1-{\vol[frac][l]})$ is always null, whereas the diffuse interface model gives necessarily lower resolved variance level since ${\vol[frac][l]}(1-{\vol[frac][l]})$ is not null in zones where the interface has been diffused (see Figure~\ref{fig:inst_volfrac_period}), yielding a lower variance. This variance gap has not disappeared in the diffuse interface approach. If the construction of the diffuse interface approach is based for instance on some probability to find the liquid, the same statistical approach induces that a certain amount of variance is related to this probability, this part is somehow included as a model and hidden at the resolved scale.

Furthermore, the turbulent liquid flux includes also this kind of effect, by considering at one location and time there can be only liquid and gas. It can be shown that
\begin{align}
\average{\vol[frac][l][fluctu] \speed[s][fluctu]}=\average{\vol[frac]}(1-\average{\vol[frac]})(\average{\speed[s][l]}-\average{\speed[s][g]})= \average{\vol[frac][l][fluctu] \vol[frac][l][fluctu]} (\average{\speed[s][l]}-\average{\speed[s][g]}).
\end{align}
Thus, the turbulent liquid flux is related to the variance and the difference of velocity between phases.  Since the resolved part of the diffuse interface underestimates the variance, it penalizes the turbulent liquid flux. Here again, it is due to the fact that second-order moments are sensitive to the small-scales feature of the flow. 

On the same line not all the features of the flow are perfectly represented by the sharp interface approach (DNS), since for low mesh resolution the precision on the determination of the interface position will decrease leading to other kind of numerical artifacts. For low resolution, the methods that preserve the sharp interface transition to keep also the total amount of each phase forces somehow the phase inclusion to remain at the resolution level. This has been shown on Figure~\ref{fig:IRQ_curv_510}, for the low resolution part of the simulation, an accumulation of surface that has a curvature related to $\Delta x $ appears showing that the numerical method start to limit the atomization process to preserve droplet that can be captured with the present mesh resolution. 

Finally, the implication of the reduction of variance can also be demonstrated for Reynolds stress correlation, as an example the same PDF of $\alpha_l$ implies
\begin{align}
\average{\speed[s][fluctu] \speed[s][fluctu]}=\average{\alpha_l}~\average{\speed[s][l]' \speed[s][l]'}+
(1-\average{\alpha_l})~\average{\speed[s][g]' \speed[s][g]'}+\average{\alpha_l'~\alpha_l'}(\average{\speed[s][l]}-\average{\speed[s][g]})^2.
\end{align}
The Reynolds stress contains three contributions: a part in the liquid phase, a part in a gas phase and a last part due to the velocity difference between gas and liquid. This last contribution is proportional to the variance of $\alpha_l$. Thus the under estimation of the variance by the diffuse interface approach reduces also the Reynolds stress. This effect can be observed on Figure~\ref{fig:ret_uxux} and it is more pronounced for the MR case than for the HR case.

To conclude this discussion, the fact that the diffuse interface model underestimates the variance does not mean that the diffuse interface model produces wrong result, but that a part of second-order moments is included in the statistical representation of the transition between the gas and liquid.

\section*{Conclusions}

In the present work, we have proposed to pursue the evaluation of reduced-order models to perform predictive simulations of the primary atomization in engines ranging from cryotechnic to aeronautic applications. As a baseline comparator, we have relied on the DNS results of a hierarchy of specific test-cases. The original test-case  consisted in an air-assisted water atomization using a coaxial injector with experimental results from the LEGI test bench \citeay{Cordesse_2018_NASATM_simu}. In the present paper, we have selected a second test-case, an air-assisted water atomization using this time a planar injector rather than a co-axial injector. This planar injector reproduces in terms of Weber and Reynolds numbers the liquid sheet flowing out a swirling atomizer used in agricultural applications \citeay{Belhadef_2012}, $Re_{L} =1.5\, 10^{3}$ and $We_{R} = 4.0 \, 10^{2}$ and $p=1 \, bar$. The Reynolds and Weber numbers are also typical from the mid-range aeronautical engines. As analyzed in the present contribution, this test-case offers also an atomization regime, which makes it complementary with the first test case in order to eventually evaluate and assess our reduced-order models, the final aim being a cryogenic coaxial injection.

We have thus qualitatively and quantitatively compared the results of simulations performed with CEDRE for the diffuse interface reduced-order model and ARCHER for the DNS. %
We have first assessed the DNS reliability through the use of the IRQ criterion to identify the most relevant region where the DNS can stand as a trustworthy benchmark. %
Then, the main result of this work concerns the comparisons of two models that have a different modelling of small-scales due to the distinct interface representation. With the DNS code ARCHER an interface reconstruction technique is used representing thus the interface like a sharp transition. On the other hand, the diffuse interface approach spreads the phase transition over few mesh cells, inducing a statistical representation of the probability of phase presence. Despite this intrinsic difference of representation of flow at small scale of interface deformation, the mean behavior of the flow (average velocity) and atomisation (average liquid volume fraction) are preserved and well recovered for both approaches.  Looking at the average slice of liquid volume fraction field, global features of the atomization process such as liquid penetration and spray angle are comparable even if a slightly higher dispersion of the reference DNS with respect to the diffuse interface model may be observed. To go further, quantitative comparison have been realized by comparing directly transverse average profile of liquid volume fraction and velocity component. The agreement is within the statistical noise for all quantities except for the transverse velocity comparison where  few percent of difference may be observed. It is important to stress that the diffuse interface results depend on the mesh resolution up to a certain level where the DNS based results can be recovered. This findings suggest further work to build a clear theoretical derivation of the statistical length scale induced by the diffuse interface representation. We have further detailed the comparison with statistical analysis of the flow through first and second-order moments based on the liquid volume fraction and velocities limited to the well-resolved part $\left\langle 0, 4\diam[l] \right\rangle$ of the computational domain. %
The second-order moments have confirmed the liquid higher dispersion for the DNS. We finally have pointed out why the diffuse interface model underestimates the variance and emphasized that it does not mean that the diffuse interface model produces wrong result, but that a part of second-order moments may be included in the representation of the transition between the gas and liquid. To summarize, we have understood the behavior of each numerical method and identified  the main conceptual differences between them.

Even if both numerical formalisms (i.e. diffuse interface method and interface capturing method) are quite different, their comparison has shown good agreement for key quantities such as spray angle, liquid penetration length, averaged velocities profiles, averaged volume fraction profiles, averaged Reynolds stress tensor and mean turbulent liquid flux.  Important CPU gains between the \textit{seven equation model} implemented in the CEDRE code and the DNS results from the ARCHER code have been obtained. 

The diverging modeling of the small-scale between the two approaches has brought out the need to examine in future works the underlying averaging process coming from the diffuse interface modeling, which is linked to the mesh resolution and produces numerical dissipation leading to a lost of amplitude of the fields. It also suggests to add a sub-scale description of the interface dynamics through geometric variables such as the interfacial area density, the mean and Gaussian curvatures as proposed in \citeay{Cordesse_2018_NASATM_LAP}. It is current research area part of the PhD thesis \citeay{Cordesse_PhD}.
\section*{Acknowledgments}
The support of the French Government Space Agency (CNES) and the French Aerospace Lab (ONERA) with the help of L.H. Dorey (ONERA) and M. Th\'eron (CNES) are gratefully acknowledged. Simulations have been successfully conducted using the CEDRE computational fluid dynamics software on the ONERA cluster. 
Regarding ARCHER computations, this work was granted access to the HPC resources of IDRIS, TGCC and CINES under the allocation A0052B10101 and A0072B10101 attributed by GENCI (Grand Equipement National de Calcul Intensif) and resources of CRIANN (Project number 2006011). We would like to thank Julien R\'eveillon for his help and numerous and fruitful discussions. The help of H. Deneuville and P. Tomov from SAFRAN Aircraft Engines (SAE) is gratefully acknowledged.

\section*{Compliance with Ethical Standards} This work has been partially funded by the French Government Space Agency (CNES) and the French Aerospace Lab (ONERA) through a PhD grant for P. Cordesse, and by SAFRAN Aircraft Engines (SAE) through a PhD grant for A. Remigi. %

\vfill %
\end{document}